\documentclass[fleqn,usenatbib, twocolumn]{mnras}

\usepackage{newtxtext,newtxmath}
\usepackage{array}
\usepackage{longtable}
\usepackage{lmodern}
\usepackage{rotating}
 \usepackage{xcolor}

\usepackage[T1]{fontenc}
\usepackage{ae,aecompl}
\usepackage{natbib}
\usepackage[normalem]{ulem}

\usepackage{graphicx}
\usepackage{float}
\usepackage{amsmath}	

\usepackage{amssymb}	
\usepackage{subcaption}

\title[GX 339--4]{The odyssey of the black hole low mass X-ray binary GX\,339--4:\\ 
Five years of dense multi-wavelength monitoring. }

\author[E. Tremou et al.]{E. Tremou $^{1}$\thanks{E-mail: etremou@nrao.edu},
S. Corbel $^{2,3}$,
R. Fender $^{4,5}$,
P. Woudt $^{5}$,
J.C.A  Miller-Jones $^{6}$,
\newauthor
I. Heywood$^{4}$,
F. Carotenuto $^{4, 16}$,
S. Motta $^{4}$,
A. Tzioumis $^{7}$,
P. J. Groot $^{5,8,15}$,
\newauthor
D. M. Russell $^{9}$,
J. Crook-Mansour $^{4}$,
P. Saikia $^{9}$,
W. Yu $^{10}$,
J. van den Eijnden $^{11}$,
\newauthor
A. J.  van der Horst $^{12}$,
D. R. A. Williams-Baldwin $^{13}$,
X. Zhang $^{10,14}$
\\
$^{1}$National Radio Astronomy Observatory, P.O. Box O, Socorro, NM 87801, USA\\
$^{2}$Universit\'e Paris Cit\'e and Universit\'e Paris Saclay, CEA, CNRS, AIM, F-91190 Gif-sur-Yvette, France\\
$^{3}$ ORN, Observatoire de Paris, Universit\'e PSL, Univ Orl\'eans, CNRS, 18330 Nan\c{c}ay, France \\
$^{4}$Department of Physics, University of Oxford, Keble Road, Oxford OX1 3RH, UK\\
$^{5}$Department of Astronomy and Interuniversity Institute for Data Intensive Astronomy, University of Cape Town, Private Bag X3, Rondebosch 7701, South Africa\\
$^{6}$International Centre for Radio Astronomy Research-Curtin University, GPO Box U1987, Perth, WA 6845, Australia\\
$^{7}$Australia Telescope National Facility, CSIRO, PO Box 76, Epping, New South Wales 1710, Australia\\
$^{8}$Department of Astrophysics/IMAPP, Radboud University Nijmegen, PO Box 9010, NL-6500 GL Nijmegen, the Netherlands \\
$^{9}$Center for Astrophysics and Space Science (CASS), New York University Abu Dhabi, PO Box 129188, Abu Dhabi, UAE\\
$^{10}$Shanghai Astronomical Observatory, Chinese Academy of Sciences, 80 Nandan Road, Shanghai 200030, China\\
$^{11}$Department of Physics, University of Warwick, Coventry CV4 7AL, UK\\
$^{11}$Anton Pannekoek Institute for Astronomy, Universiteit van Amsterdam, Science Park 904, 1098, XH, Amsterdam, The Netherlands\\
$^{12}$Department of Physics, The George Washington University, 725 21st Street NW, Washington, DC 20052, USA\\
$^{13}$Jodrell Bank Centre for Astrophysics, School of Physics and Astronomy, The University of Manchester, Manchester M13 9PL, UK\\
$^{14}$University of Chinese Academy of Sciences, 19A Yuquanlu, Beijing 100049, China\\
$^{15}$South African Astronomical Observatory, P.O. Box 9, 7935 Observatory, South Africa\\
$^{16}$INAF-Osservatorio Astronomico di Roma, Via Frascati 33, I-00076, Monte Porzio Catone (RM), Italy}

\date{Accepted XXX. Received YYY; in original form ZZZ}

\pubyear{2022}

\begin{document}
\label{firstpage}
\pagerange{\pageref{firstpage}--\pageref{lastpage}}

\maketitle

\begin{abstract}
We present the longest and the densest quasi-simultaneous radio, X-ray and optical campaign of the black hole low mass X-ray binary GX\,339--4, covering five years of weekly GX\,339--4 monitoring with MeerKAT, \textit{Swift}/XRT and MeerLICHT, respectively. Complementary high frequency radio data with the Australia Telescope Compact Array are presented to track in more detail the evolution of GX\,339--4 and its transient ejecta. During the five years, GX\,339--4 has been through two ``hard-only" outbursts and two ``full" outbursts, allowing us to densely sample the rise, quenching and re-activation of the compact jets. Strong radio flares were also observed close to the transition between the hard and the soft states. Following the radio flare, a transient optically thin ejection was spatially resolved during the 2020 outburst, and was observed for a month. We also discuss the radio/X-ray correlation of GX\,339--4 during this five year period, which covers several states in detail from the rising phase to the quiescent state. This campaign allowed us to follow ejection events and provide information on the jet proper motion and its intrinsic velocity. With this work we publicly release the weekly MeerKAT L-band radio maps from data taken between September 2018 and October 2023.
\end{abstract}




\section{Introduction}
Low mass X-ray binaries (LMXBs) are interacting binary stars that contain a low mass star ($<$3M$_{\odot}$; the companion or donor star), which feeds an accretion disk around a stellar remnant (black hole or neutron star).
The companion star usually fills its Roche lobe and therefore transfers mass to the compact star. 
The infalling mass forms a rotating accretion disk \citep{1973pringle} and is accreted on to the compact object. The rest of the matter is ejected back to the interstellar medium through powerful energetic outflows \citep[relativistic transient jets;][]{1992mirabel,1991hughes,1999fender,2002corbel} or disk winds \citep[e.g.][]{2012ponti}. 
The compact jets are tightly connected to the accretion flow \citep[disk-jet coupling;][]{1998hannikainen,2003corbel,2013corbel}, as revealed by the non-linear radio/X-ray correlation which extends to  Active Galactic Nuclei (AGN) through the fundamental plane of black hole activity \citep{2003merloni, 2003gallo, 2004falcke, 2015saikia, 2018saikia}. Analogous to AGN, the steady jets of black hole low mass X-ray binaries display a flat or inverted radio to infrared spectrum associated with their emission \citep{1979blandford,1988hjellming,2001markoff,2001fender,2003fender,2006fender} while the relativistic discrete/transient jets have spectra which evolve from optically thick to optically thin \citep{1995Hjellming,2004fender,2009fender}. 
Unlike AGNs, X-ray binaries evolve through their duty cycles on short timescales, typically rising from their quiescent state and undergoing entire outbursts over periods of days to months. Hence, LMXBs are ideal targets to study the physics of the accreting material within the system but also its interaction with the interstellar medium via winds and discrete/transient jets.  

During the episodic phases of outbursts, the luminosities of LMXBs may increase by orders of magnitude, sometimes reaching close to the Eddington limit. A power-law component that arises from inverse Compton emission dominates the spectrum of the quiescent state and the early rise of the outburst (``hard-state"). This emission is associated with a  geometrically thick, optically thin, radiatively inefficient accretion flow \citep[RIAF; e.g.][]{1995narayan} or alternatively from the jet base \citep[e.g.][]{2005markoff}. In this state, the radio emission originates from a steady, partially-self-absorbed compact jet \citep[e.g.][]{2001fender,2004fender} that displays a flat to inverted radio spectral index \citep[e.g.][]{2000corbel,2000dhawan,2014russell,2017plotkin,2020tremou}.

Matter in the outer part of the accretion disk flows inwards toward the black hole, while the X-ray spectrum softens as it becomes dominated by a multi-temperature blackbody component from the hot inner regions of an optically thick, geometrically thin accretion disk. The system enters the ``soft" state through the hard-intermediate \citep[HIMS;][]{2006remillard,2010belloni} and subsequently the soft-intermediate state (SIMS). Prior to the the ejections, in the intermediate states and in the soft state after the ejections, the radio emission from the compact steady jets is usually quenched by $>$3.5 orders of magnitude \citep{1999fender, 2011coriat, 2011russell, 2019russell,2021carotenutoa}. 
The system is often observed to be flaring in radio, as the powerful outflows are launched and travel away from the black hole at relativistic speeds. The luminosity of the system will then drop, and the outburst will decay until reaching the quiescent level.

However, not all outbursts follow the complete outburst track (``full" outburst) but instead spend their outburst only in the hard-state and eventually return to quiescence without completing a successful transition to the soft state \citep[``failed" or ``hard-only" outburst;][]{2004brocksopp,2011coriat,2016tetarenko,2021alabarta}.

\subsection{The black hole low mass X-ray binary, GX\,339--4}

GX\,339--4 is a Galactic LMXB hosting a black hole. It was discovered over 50 years ago, in 1972, by the MIT X-ray detector on board the Orbiting Solar Observatory (OSO) 7 satellite \citep{1973markert}.  

GX\,339--4 is one of the major X-ray binary targets that has been monitored for more than $\sim$50 years and both kinds of outbursts occurred in the past \citep{2013corbel,2016tetarenko,2021alabarta}. Thanks to a large sample of quasi-simultaneous radio and X-ray data, it is a perfect system to allow us to understand the correlation in the radio/X-ray plane and the differences in the context of jet production between the two types of outbursts \citep[``full" or ``hard-only" outburst;][]{2021dehaas}.

The spectral properties of the GX\,339--4 system display similarities to other black hole X-ray binaries indicating that GX\,339--4 is powered by a central black hole \citep{1998zdziarski,2000sunyaev}. 

Despite being a well-studied source, the main system properties of GX\,339--4 are not well constrained due to its faint companion star. Upper limits on the luminosity of its companion star allowed the identification of its low mass X-ray binary nature \citep{2001shahbaz}. The orbital period has been, initially, estimated to be 14.8 to 16.8 hours using optical spectroscopy \citep{2002cowley,1992callanan} with an upper limit on its inclination angle of $<78$ degrees \citep{2017heida}. \cite{2003hynes} later constrained the orbital period to be 42.14 $\pm$ 0.01 hours. The mass of the black hole has been debated \citep{2003hynes,2008munoz,2011shidatsu} with the most recent estimate by \cite{2017heida} being $\sim$ 9.8 M$_{\odot}$. \cite{2004hynes} used optical spectroscopy to determine the system's distance which gave a lower limit of 6 kpc, while \cite{2004zdziarski} placed it at 8 kpc using optical and infrared data. A similar value was also obtained more recently by \cite{2016parker} who used Nuclear Spectroscopic Telescope Array (NuSTAR) and \textit{Swift} data, deriving a distance of 8.4 $\pm$ 0.9 kpc.  

The radio counterpart of the system was discovered in 1994 (most likely in the hard to soft transition state) by \cite{1994sood}, and \cite{1999wilms} argued that the radio emission is associated with a compact self-absorbed jet \citep{1999fender,2000corbel}. 
Over the past decades ($\sim$ 40 years), GX\,339--4 has undergone numerous outbursts ($>20$), enabling every time a better understanding of the radio compact jet during quiescence \citep{2020tremou}, the hard states and its quenching during its transition to the soft state \citep{1999fender}, and the relation between the accretion disk and the jet \citep[disk-jet coupling;][]{1998hannikainen,2000corbel,2003corbel,2003markoff,2013homan,2013corbel}. Furthermore, GX\,339--4 was the first source to show X-ray correlations with an infrared jet, and optical disc \citep[e.g.:][]{2005homan,2009coriat,2012buxton}.
 Therefore, GX\,339--4 is an ideal target for dense monitoring in order to help us to better understand the launching mechanisms of compact jets in systems rising out of quiescence, how they are connected to the accretion signatures and what is their interaction with the surrounding medium when it goes through a ``hard-only" or ``full" outburst. 

In this paper, we present the results of a five years long,  weekly radio, X-ray and optical monitoring of the key X-ray binary source, GX\,339--4, as of \textbf{T}he \textbf{HUN}t for \textbf{D}ynamic and \textbf{E}xplosive \textbf{R}adio transients with Meer\textbf{KAT}
   \footnote{\url{http://www.thunderkat.uct.ac.za/}} \citep[ThunderKAT,][]{2017fender} project.

\section{Observations}
\subsection{MeerKAT radio observations}
\begin{figure*}
\centering
\includegraphics[width=\textwidth]{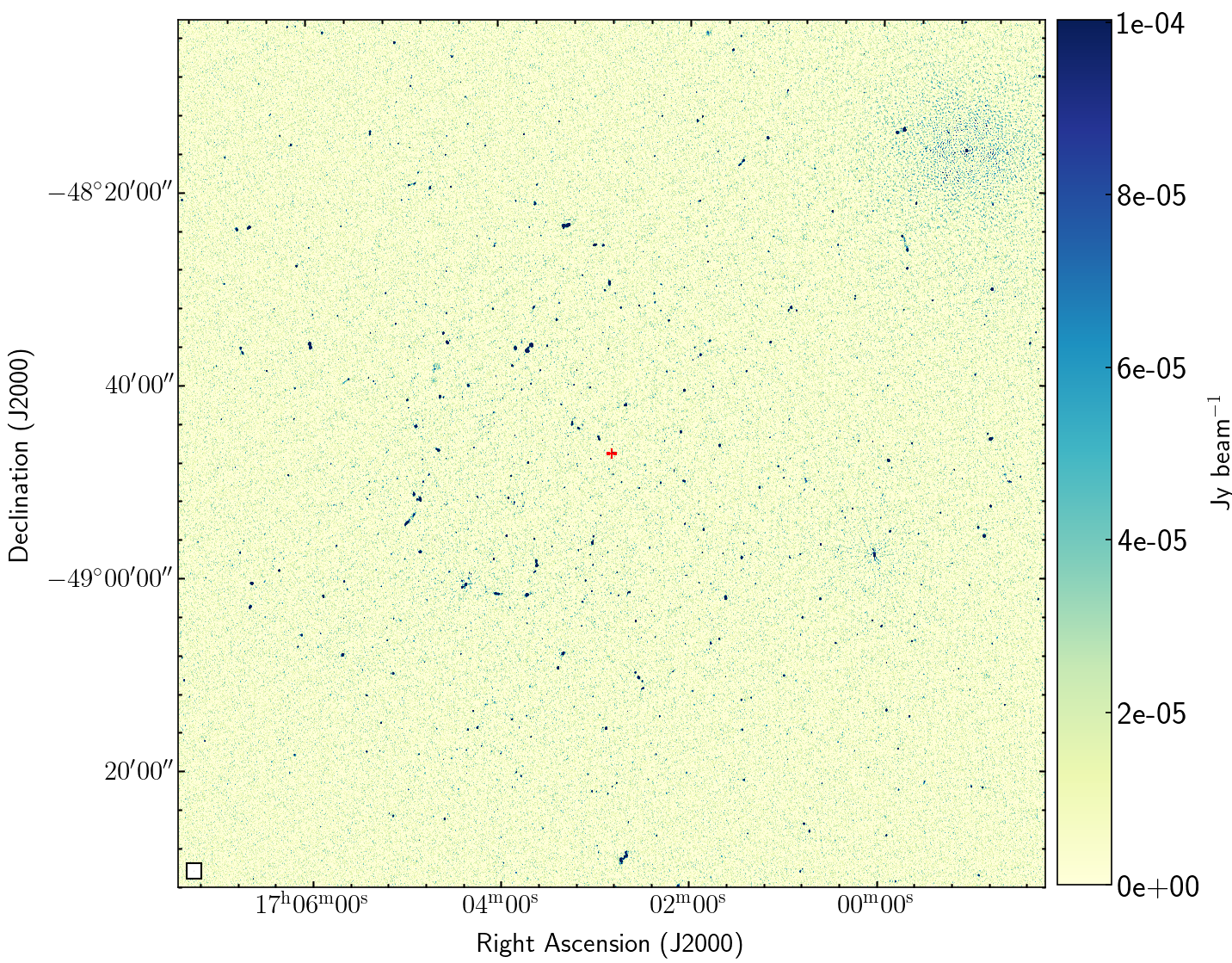}
\caption{Example of a single-epoch L-band MeerKAT radio map showing 1.5$\times$1.5 degrees of the field centered on GX\,339--4 (red cross mark). The rms noise is 32.6 $\mu$Jy\,beam$^{-1}$ enabling the detection of weak sources down to 100 $\mu$Jy. It was taken on June 19, 2021 (MJD 59384), with an integration time of 10 minutes on source. The circular synthesized beam size is 5$\arcsec$.} \label{fig:GXepoch}
\end{figure*}

As part of the ThunderKAT monitoring program, we visited the field of GX\,339--4 once per week from September 2018 until March 2019, with a 15-minute integration time. Afterwards, we decreased the integration time by 5 minutes (10 minutes on source in total) due to our existing knowledge of the field and the ability to reach the desired sensitivity. Here, we present 252 epochs of weekly data taken between November 2018 and October 2023. 
The MeerKAT radio telescope \citep{2018camilo1} is located in the Karoo desert in South Africa and comprises 64 antennas, of 13.5 meters diameter each, with a maximum baseline of 8 km. Observations were made using the L-band (900 - 1670 MHz) receiver, centered at 1284 MHz with 856 MHz bandwidth.  Observations typically alternated between the target and phase calibrator (J1744--5144), while a bandpass / flux calibrator (J1939--6342) was also observed. All observations were obtained in full polarization mode (I, Q, U, V stokes parameters recorded), however given the polarization properties were beyond the scope of this project, no polarization calibrator was observed to allow for cross-hand (XY and YX) calibration.  The data were flagged using \textsc{Tricolour} \footnote{\url{https://github.com/ratt-ru/tricolour}} \citep{2022hugo} and were calibrated using the Common Astronomy Software Application\footnote{\url{https://casa.nrao.edu/}} \citep[CASA;][]{2022casa}. The \textsc{Oxkat} \footnote{\url{https://github.com/IanHeywood/oxkat}} \citep{2020oxkat} pipeline was used for the 1GC calibration (phase correction, antenna delays, and band-pass correction). Imaging, self-calibration and direction-dependent calibration of the data were carried out with the wide-band, wide-field imager, \textsc{DDFacet} \citep{2018tasse} and the \textsc{killMS} package \citep{2015smirnov} for direction-dependent calibration. For the imaging we set the image size to be 8192 $\times$ 8192 pixels,  1.5$\arcsec$ cell size, while we used Briggs weighting with a robust parameter of -0.7  in order to achieve a compromise between angular resolution and the level of sidelobes of the synthesized beam. 
The average rms noise of a single epoch is 34 $\mu$Jy/beam and the single epoch images are publicly available at \url{https://doi.org/10.48479/4fpq-sd16}. An example of a single-epoch radio map at L-band is shown in Figure \ref{fig:GXepoch}. We used the \textsc{pyBDSF}\footnote{\url{https://www.astron.nl/citt/pybdsf/}}
 source finder software to extract the position and flux density of GX\,339--4 in each epoch. Furthermore an extensive blind search for variables and transients in the field of GX\,339--4 revealed the discovery of the first transient with MeerKAT, as well as several long-term variable sources \citep[see][for details]{2020driessen,2022driessen}.

\subsection{ATCA radio observations}
GX\,339--4 was observed with the Australia Telescope Compact Array (ATCA) for eight epochs between April and August 2020 (project code: C1199, PI: S. Corbel). The array's configuration varied. Three observations were taken with the 6A configuration, three with the 1.5C configuration, one in the H214 configuration, and the last one with the EW352 configuration. The integration time per observation spanned between 2 and 4 hours depending on the expected flux density of GX\,339--4, ensuring a solid detection. 
All of the observations were taken in the 4cm band; central frequencies of 5.5 and 9 GHz with a total bandwidth of
2 GHz in each of these two basebands. Similarly to MeerKAT observations, J1939--6342 was used as the flux calibrator and for bandpass corrections and PMN J1650--5044 was used as the phase calibrator for gain corrections. The data were calibrated following the standard procedures with CASA and were imaged and self-calibrated using \textsc{killMS} and \textsc{DDFacet} with a robust parameter of -0.3 for an optimal trade-off between angular resolution and sensitivity. We also imaged using different parameters such as  uniform weighing in order to achieve the sharpest possible resolution. We also used the source finder software \textsc{pyBDSF} to extract the positions and flux densities of GX\,339--4 and the large-scale outflow.
A full list of observations is presented in Table \ref{tab:obs}.

\subsection{\textit{\textit{Swift}} XRT X-ray observations}

We used data taken by the \textit{Swift} X-Ray Telescope (XRT) instrument \citep{2000burrows} as part of a dedicated monitoring program (SwiftKAT, PI: S. Motta) associated with the
ThunderKAT project. Data were taken any time the source was not observationally constrained due to the proximity to the Sun. For our analysis,  we include measurements made from January 2019 through September 2023 where the source position was not close to the Sun. We used 145 observations from XRT in both photon counting (PC) and Windowed Timing (WT) mode that are close in time to our radio observations. Photon pile-up was negligible at low photon count rates, while for higher count rates ($>$0.2 counts, PC mode), we filtered grade 0 events and used annuli of variable inner and outer radii to account for it. 

We used the output of the standard pipeline processing and analyzed the data using the \textsc{XSPEC} software package \citep{1996arnaoud}. We fit the energy spectra accounting for interstellar absorption, which is modelled with an equivalent hydrogen column density ($N_{\rm H}$) by using abundances given in \cite{2000willms}. We applied Cash statistics \citep{1979cash} to obtain the  X-ray flux from the individual low-count spectra. A power-law model ({\tt tbabs$\times$powerlaw}) was used for the spectral fit during the hard state, while in the other states we used an absorbed multi-color disk blackbody to which we add a power law component to take into account residual high energy
tails, {\tt tbabs$\times$(diskbb+powerlaw)}. All observation IDs and the obtained fluxes (3-9keV) are presented in Table \ref{tab:obsswift}.

\subsection{\textit{MAXI} GSC observations}

GX\,339--4 is observed by the Monitor of All-sky X-ray Image \cite[MAXI;][]{2009Pmatsuoka} with the Gas Slit Camera (GSC). 
We downloaded publicly available data (2-6keV) covering 5 years (2018-2023) from the on demand MAXI website, \href{http://maxi.riken.jp/mxondem/}{http://maxi.riken.jp/mxondem/}. 

\subsection{MeerLICHT observations} 

MeerLICHT is an optical wide-field telescope, located at the Sutherland station of the South African Astronomical Observatory in South Africa \citep{bloemen2016}. It is equipped with a 10\,560$\times$10\,560 STA CCD detector sampling the sky at 0.56 \arcsec / pixel for a total field of view of 2.7 square degrees. MeerLICHT was designed as a prototype for the BlackGEM array \citep{groot2022}, and its express purpose is to shadow the MeerKAT array on the sky whenever possible. As a back-up/filler program to the MeerKAT-shadowing program MeerLICHT performs a high cadence survey of priority targets. The field of GX\,339--4 is one of the standard back-up fields and has been monitored since the start of full operations in early 2019 when the source was visible during the nigh-time and if possible, it was observed quasi-simultaneous with MeerKAT.  

The MeerLICHT telescope has a stricter pointing restriction towards the east (HA$>$-3hrs) due to vignetting by the dome rim. Hence, the overlap period of the optical with the radio/X-ray data is limited. Although the MeerLICHT telescope remained functioning throughout the full COVID-19 period due to its remote operations,the ground-based optical data are additionally impacted by weather. 

The MeerLICHT telescope has a six-slot filter wheel equipped with an optimized Sloan set ($u,g,r,i,z$) as well as a wide-band (440-720nm) $q$-band filter. Observations on the back-up fields are performed in the $u,q,$ and $i$-bands, which therefore make up the bulk of the data. All data are processed using the BlackBOX/ZOGY pipeline (\cite{blackbox2021}, Vreeswijk et al., in prep.) which performs standard CCD data reduction procedures as well as image subtraction for variables/transients using the ZOGY software \citep{2016zackay}. The magnitude calculation is based on PSF fitting photometry and BlackBOX uses the PSFEX \citep{2011bertin} tool for accurate PSF modeling. All MeerLICHT data are calibrated astrometrically to the Gaia DR2 \citep{gaiadr2} frame using ICRS coordinates at epoch 2015.5, and photometrically to a set of Gaia-centered photometric standard stars. All MeerLICHT magnitudes are on the AB system \citep{oke90}. Over the period of consideration here GX\,339--4 was observed 903 times in the $u,q,i$ bands in three observing seasons. A summary of the MeerLICHT observations can be found in Table \ref{tab:obsmeerlicht}. 

An extensive analysis of the optical properties of the source and how they are correlated to the multi-wavelength observations is planned to be presented in a follow up study by Alabarta et al. in prep. However, here, we consolidate all optical, radio and X-ray data into a singular, authoritative legacy data set, establishing a valuable and sustainable resource for future research and analysis by the community.

\section{Results}
\begin{sidewaysfigure*}
\vspace{-19cm}
\includegraphics[width=\textwidth]{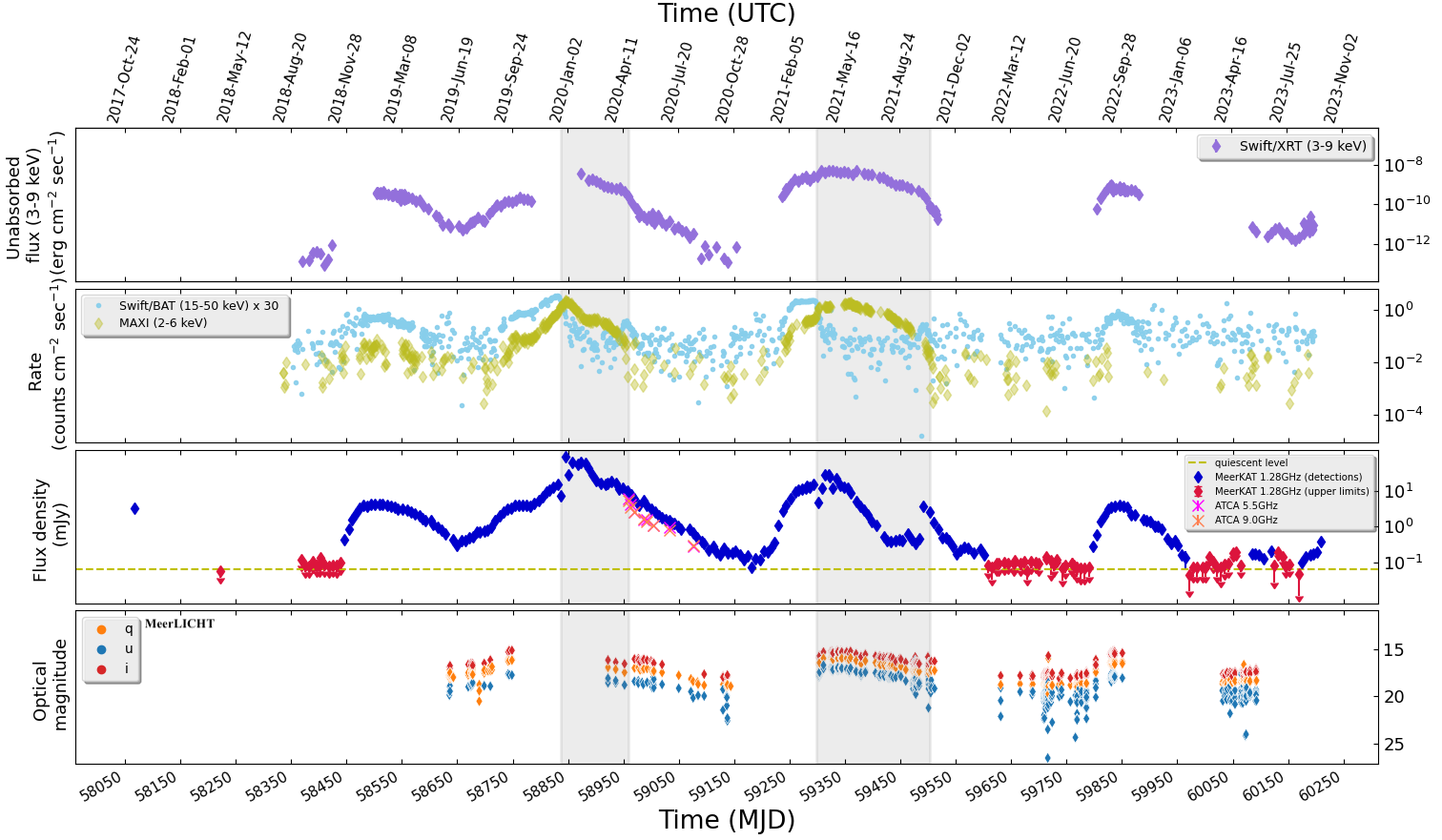}
\caption[]{The X-ray, radio and optical light-curves of GX\,339--4 over the five years of our monitoring campaign. The bottom panel shows the MeerLICHT optical magnitudes (filters $q,u,i$), while
the radio flux densities measured by MeerKAT (red:upper limits and blue:detections, dashed line: quiescent level) and ATCA (pink and orange data points) are shown in the third panel.  The second panel shows the scaled 2--6 keV (yellow) and 15--50 keV (cyan) X-ray flux from MAXI and \textit{Swift}/BAT, respectively. The top panel shows in purple the  \textit{Swift}/XRT 3-9 keV X-ray fluxes that were obtained quasi-simultaneously with the radio observations. The grey shaded background indicates the times that source was in soft X-ray state based on the spectral fitting of \textit{Swift}/XRT  3-9 keV data.} 
\label{fig:lightcurve}
\end{sidewaysfigure*}

We present the results of the densest ever quasi-simultaneous monitoring of the black hole X-ray binary, GX\,339--4, at radio, X-ray and optical wavelengths using radio data from the MeerKAT and ATCA radio telescopes, X-ray data from \textit{Swift} and MAXI, and optical data from the MeerLICHT telescope. We observed the source for a five-year period covering four outburst phases between 2018 and 2023. Figure \ref{fig:lightcurve} displays the radio, the X-ray and the optical light-curves over the five years. The bottom panel shows the optical data in AB magnitudes, the second panel from the bottom the radio flux density from MeerKAT and ATCA, and the middle panel shows the 2--6 keV and 15--50 keV X-ray flux of MAXI and \textit{Swift}/BAT, respectively. The top panel shows the  \textit{Swift}/XRT 3--9 keV X-ray flux that was obtained quasi -simultaneously with the radio observations. The grey shaded background indicates the time periods when the source was in the soft X-ray state. Table \ref{tab:summary} shows the summary of the outburst phases and their major properties.

\begin{table*}
    \begin{tabular}{|l|c|c|c|c|}
        \hline
        \textbf{Outburst} & \textbf{Duration} & \textbf{State } & \textbf{Peak Radio } & \textbf{Comments} \\
         &  & \textbf{classification}& \textbf{flux density} &  \\
         & (Days) &  & (mJy) &  \\
        \hline
        2018-2019  (MJD 58446 – 58649 & 203 & ``Hard-only" &  4.2$\pm$0.05&   no major flare\\
        \hline
        2019-2020  (MJD 58650 –59180) & 530 & ``Full/Successful" & 88.8$\pm$0.13  &   major flare \& discrete ejecta \\
        \hline
        2021  (MJD 59188 – 59607) & 419 & ``Full/Successful" & 27.9$\pm$0.03 & major flare \& discrete ejecta \\
        \hline
        2022  (MJD 59797 – 59965) & 168 & ``Hard-only" &  3.8$\pm$0.04& no major flare  \\
        \hline
    \end{tabular}
        \caption[]{Summary of the outburst phases.}
    \label{tab:summary}
\end{table*}

\subsection{The 2018-2019 (MJD 58446 -- 58649) outburst. \\ A ``hard-only" outburst or a long lived hard state?}
Following a quiescent state during the second half of 2018 \citep{2020tremou}, GX\,339--4 went into an outburst, beginning with a rapid X-ray rise at the end of November 2018 \citep[MJD 58446;][]{2018tremouatel}.

\begin{figure}
\centering
\includegraphics[width=0.48\textwidth]{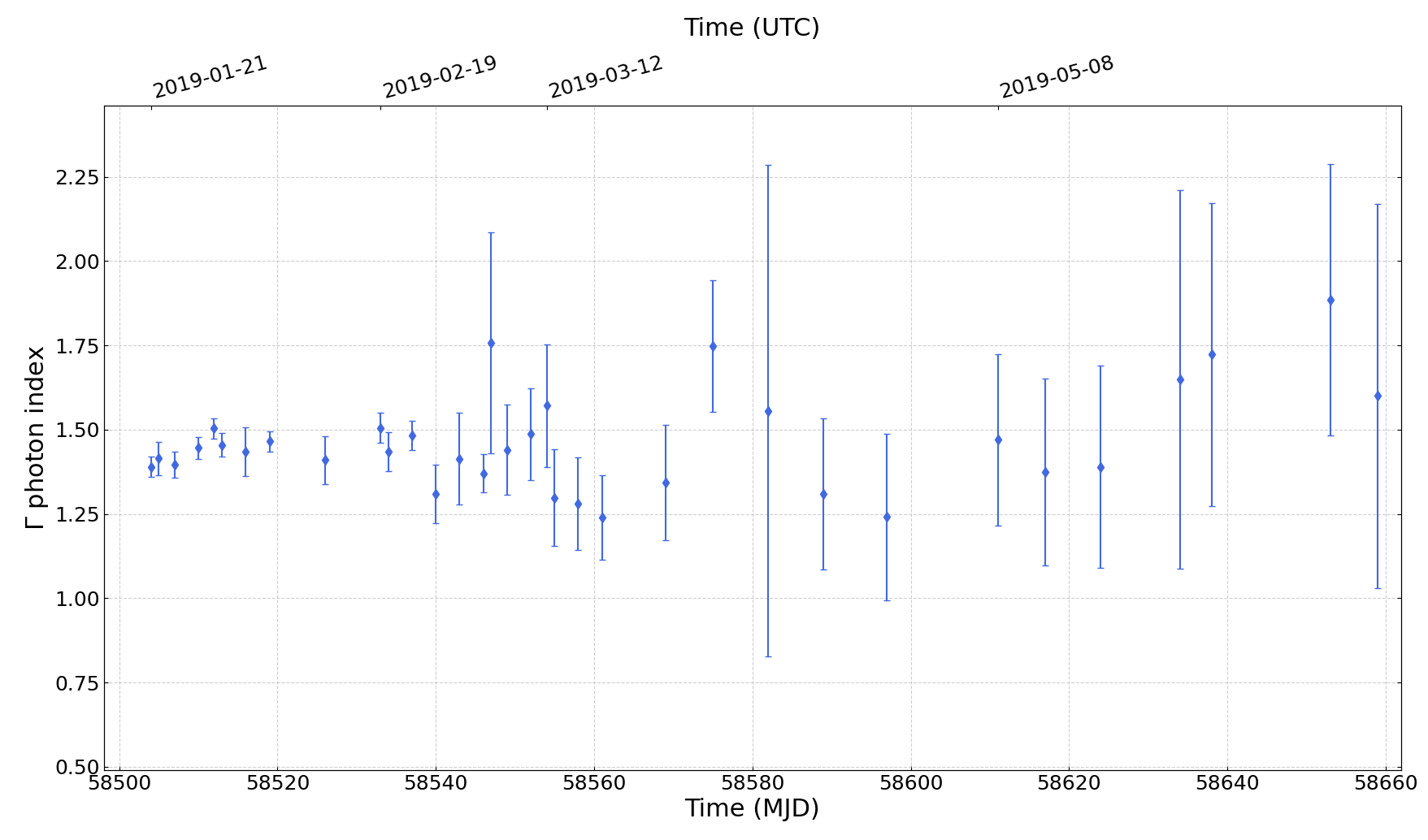}\\
\caption{Evolution of the photon index $\Gamma$ during the 2018--2019 outburst using \textit{Swift}/XRT data.} \label{fig:gamma}
\end{figure}

The 2018–2019 event was characterized as a ``hard-only" outburst, as it failed to enter the soft X-ray state characteristic of a typical ``full" outburst. The source remained in the hard X-ray state for a total of 204 days. This hard state persisted throughout 2019, before the source finally transitioned into the soft state in early 2020. This non-transition is likely because the outburst did not reach the mass accretion level required to trigger the state change. The photon index ($\Gamma$) of the hard state remained relatively flat ($\Gamma \sim 1.5$, see Figure \ref{fig:gamma}) as the outburst progressed, a value comparable to the 2017–2018 "hard-only" phase \citep[see][]{2021dehaas}. Although some steepening of $\Gamma$ was noted at the very beginning of the 2018 outburst, early \textit{Swift}/XRT data were unavailable due to the source being Sun-constrained. However, the dense radio monitoring can probe an outburst earlier than high-energy bands when those are unavailable.
``Hard-only" outbursts may sometimes be confused with re-flares/mini outbursts \citep{1995callanan, 1997chen, 2004tomosick,2012jonker,2013homan,2016patruno}, as both reach comparable X-ray flux levels that are still lower than full outbursts \citep{2004brocksopp}. Regardless, the system never reached its quiescent X-ray level \citep{2009coriat,2013corbel,2020tremou}, similar to the 2002 outburst \citep{2009coriat,2013corbel}. This continuous activity suggests the 2018–2019 phase may be a prelude or part of the subsequent 2020 outburst (see section \ref{2020outb}), implying the system underwent one of the longest hard X-ray states ever observed ($\sim 1.2$ years).

The main difference between full and hard-only outbursts is the failure of the latter to reach the critical mass accretion level. During these "hard-only" events, there may also be less efficient coupling between the jet and the accretion flow \citep{2021dehaas}. The mechanism powering renewed activity during the decay or from quiescence remains unclear \citep{2019zhang, 2021alabarta}. Similar to \textit{Swift}/XRT, observations in optical were not possible (the source was not visible in the nighttime) during the 2018-2019 outburst (see Figure \ref{fig:lightcurve}, bottom panel).

Within two months of the X-ray rise, the radio emission peaked, reaching a flux density of 4.2 mJy (MJD 58515) at 1.28 GHz. The radio flux then entered a slow decay phase, reaching a minimum flux density of 295 $\mu$Jy on MJD 58650, which is still $\sim 5$ times higher than the quiescent radio level \citep{2013corbel,2020tremou}. Following this decay, the source started rising again.

Our series of quasi-simultaneous radio and X-ray observations cover both the rise and decay of this long hard state. The radio spectral index remained relatively flat during this entire period, indicating that the compact jets were actively building up.

\subsection{The 2019--2020 (MJD 58650 --59180) outburst phase}\label{2020outb}

\begin{figure}
\centering
\includegraphics[width=0.48\textwidth]{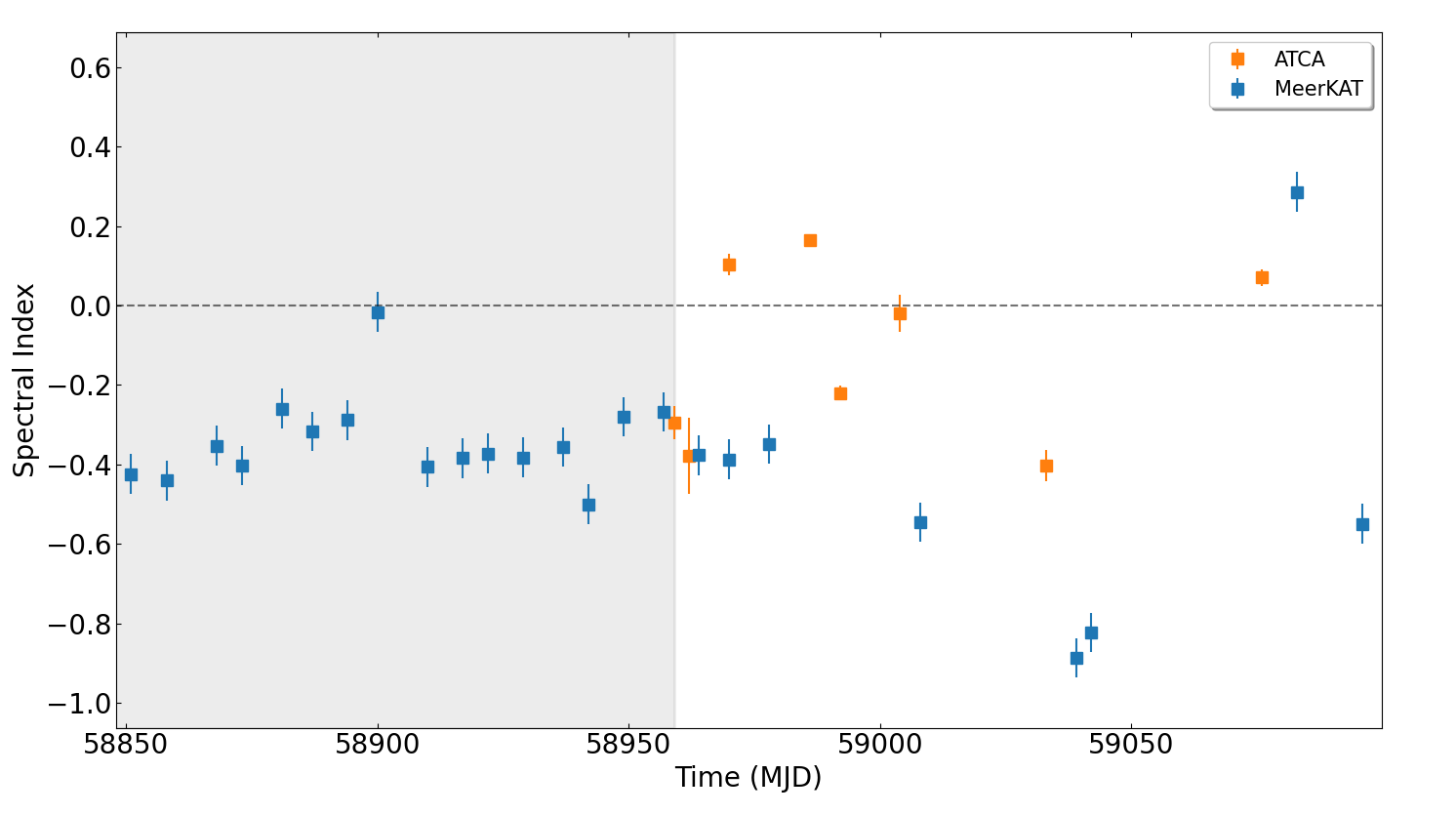}\\
\caption{Spectral index following the flare during the 2019--2020 outburst.Noting here that the spatial resolution from ATCA is higher than the one from MeerKAT. The grey shaded background indicates the time that source was in soft X-ray state.The spectral index from ATCA data turns to positive at the soft to hard transition.} \label{spixx}
\end{figure}

Following the hard state of the 2018–2019 outburst, the source underwent a critical transition phase. Although Sun constraints prevented \textit{Swift}/XRT  data acquisition during the hard-to-soft state transition in December 2019, the X-ray state change was later confirmed. Our first available \textit{Swift}/XRT  data, from early February 2020 (MJD 58887), showed that the source had successfully transitioned to the soft X-ray state, where it remained for approximately two months until mid-April 2020 (MJD 58957).
The transition from the hard to the soft state is typically characterized by the hard X-ray peak (middle panel, cyan points in Figure \ref{fig:lightcurve}) preceding the soft X-ray peak \citep{2003wenfei, 2004wenfei, 2004corbel}. This delay is due to the cooling of the corona (which produces the hard X-ray emission) and the corresponding increase in thermal emission (middle panel, yellow points in Figure \ref{fig:lightcurve}) from the accretion disk (soft X-rays). For the 2019-2020 outburst, we observed the soft X-rays peaking $\sim 20$ days (MJD 58850) after the hard X-rays (MJD 58830).
The subsequent peak of the hard X-rays at MJD 58950 (following the soft state) indicates the transition back to the hard state, marking the beginning of the decay phase and the re-build up of the steady jet.

During the hard-to-soft state transition, the accretion disk increases its thermal emission (soft X-rays), simultaneously cooling the corona (hard X-rays). The delay between the maximum flux observed in the hard and soft X-ray bands has previously been used as an indicator for the accretor's nature (black hole vs. neutron star). 
The optical flux increased rapidly during the initial rise of the 2019-2020 outburst ($\sim$1.5 optical magnitudes in all filters) following the radio flux, and the X-ray non-thermal power-law component (see Figure \ref{fig:lightcurve} and Table \ref{tab:obsmeerlicht}). Although we miss the optical behavior (Sun constraints) of the source during the hard to soft state transition, where the jet emission is suppressed or quenched and the optical flux is dominated by the thermal emission from the outer accretion disk, we could see the optical re brightening after the MJD 58950. As the system re-enters the hard state, the compact jet is re-formed and the dominant optical emission switches back to synchrotron emission from the jet, causing the optical flux to brighten again relative to the soft state though the overall luminosity is declining.

The state transition was dramatically revealed by dense radio monitoring, which successfully tracked the change even when X-ray observations were not possible.
The radio flux began a slow rise from 16 June 2019 (MJD 58650). Roughly a week prior to the main flare (MJD 58837), the radio flux density sharply decreased by a factor of $\sim 3$. This initial drop signals the ``switch-off" of the compact jet as the system moves into the soft state \citep{1999fender, 2000corbel}. The jet quenching process is frequency-dependent: the jet break frequency \citep[$\nu_{b}$;][]{2012james,2013russelld, 2014russell, 2020russellt} moves down in frequency, causing lower frequencies to become optically thin over time. This process can take a few days, and does not have to be accompanied by discrete ejection events \cite[i.e. MAXI J1836-194, ][]{2014russell}. 
The quenching was immediately followed by the brightest radio flare ever detected for GX\,339--4, reaching a flux density of 88.8 mJy at L-band at the end of 2019 (MJD 58845).
Following the major flare, the source entered the soft state, where it showed persistent radio emission similar to that seen in XTE\,J1748--288 \citep{2007brocksopp}, MAXI\,J1535--571 \citep{2019russell} and MAXI J1820+070 \citep{2020bright}. This soft state emission is not associated with ongoing core jet production. Instead, the steep spectral indices ($\alpha \lesssim -0.5$) suggest it is connected to emission from large scale downstream ejecta (Figure \ref{spixx}).
As expected \citep{1995harmon,2001fender,2004gallo}, the major radio flare was followed by an ejection event that we detected as an extended outflow at radio frequencies (5.5 and 9 GHz) observed with the ATCA, spanning from 20 April 2020 (MJD 58959) until 15 August 2020 (MJD 59076), see section \ref{flarejet}, below. 
The launch of the discrete ejections and how all those events are linked, has not been derived clearly yet \citep{2009fender, 2012james,2016rushton,2020russelld}.

The source then underwent a decay phase, reaching the quiescent radio level \citep[$\sim$70\,$\mu$Jy;][]{2020tremou} on 28 November 2020 (MJD 59181). The detection of soft state radio emission is a very interesting result that helps us to add one more LMXB, GX\,339--4, to the short list of sources that show long-lived soft state radio emission \citep{2007brocksopp, 2019russell, 2020bright, 2021carotenutoa}. The nature of the soft state radio emission can be studied through, high angular resolution radio observations where they indicate a link between the soft state radio emission and the relativistic ejections.

\subsubsection{The major radio flare and the subsequent transient jet}\label{flarejet}

\begin{figure}
\centering
\includegraphics[width=0.56\textwidth]{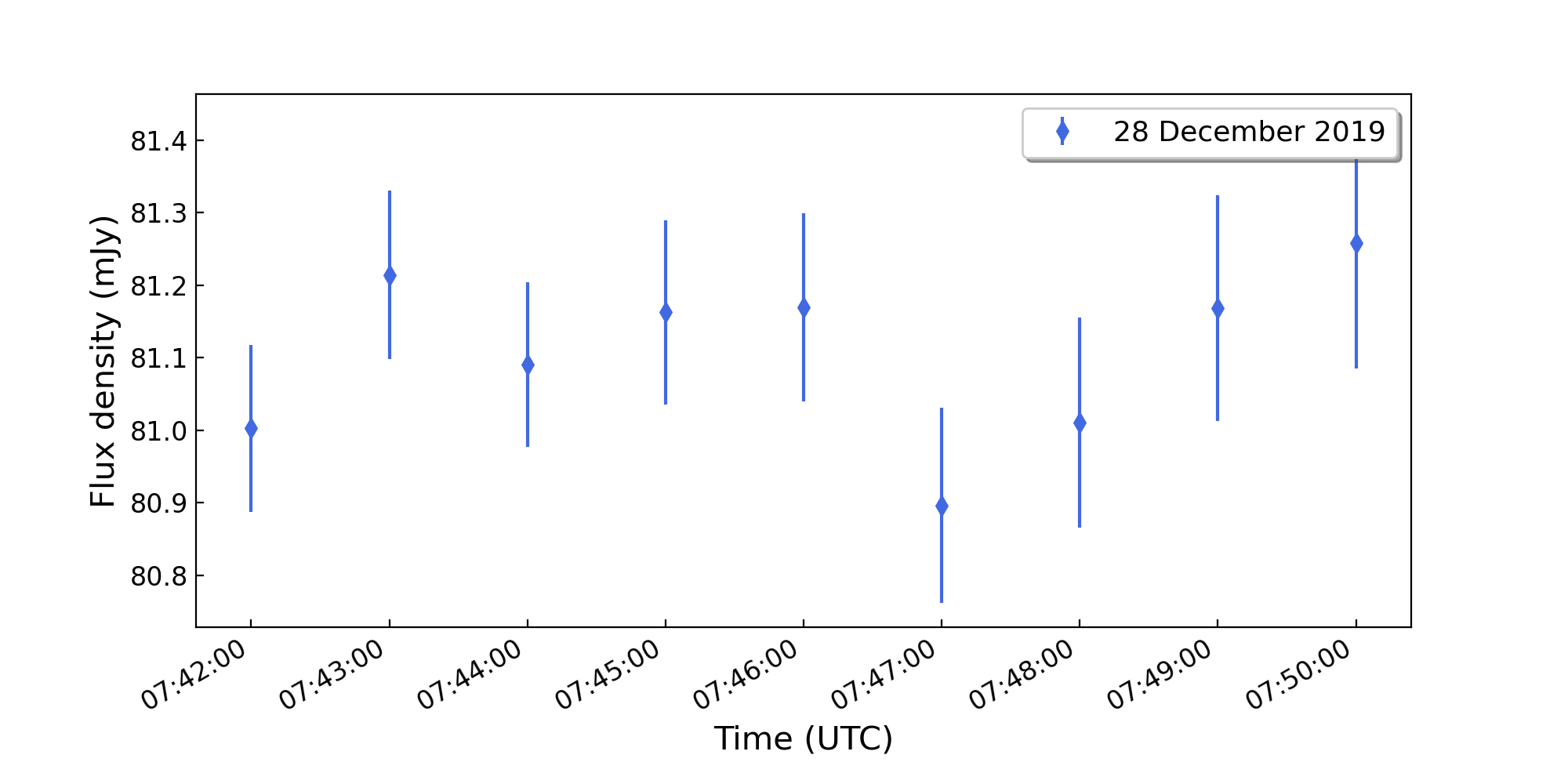}\\
\includegraphics[width=0.56\textwidth]{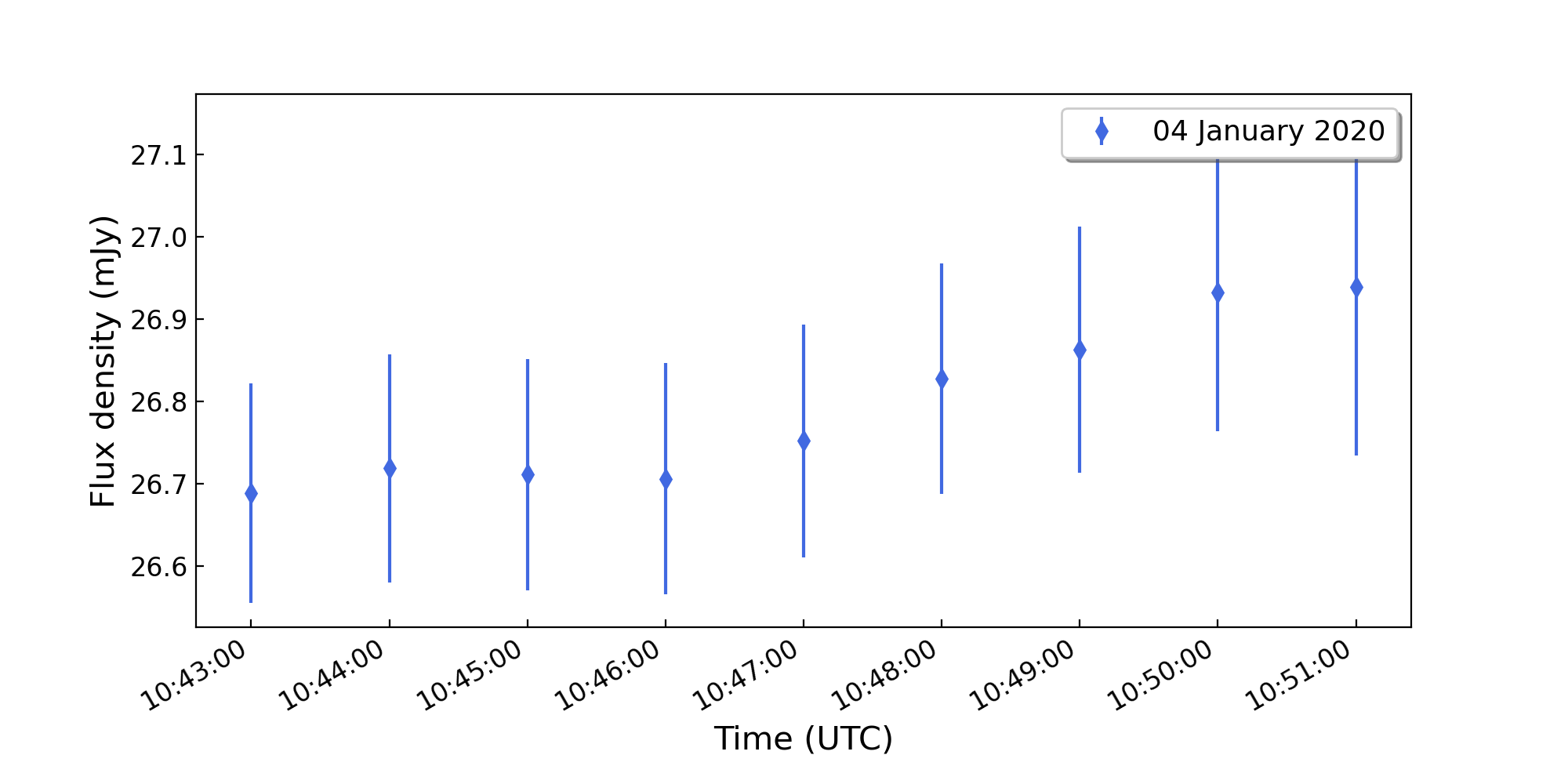}\\
\includegraphics[width=0.56\textwidth]{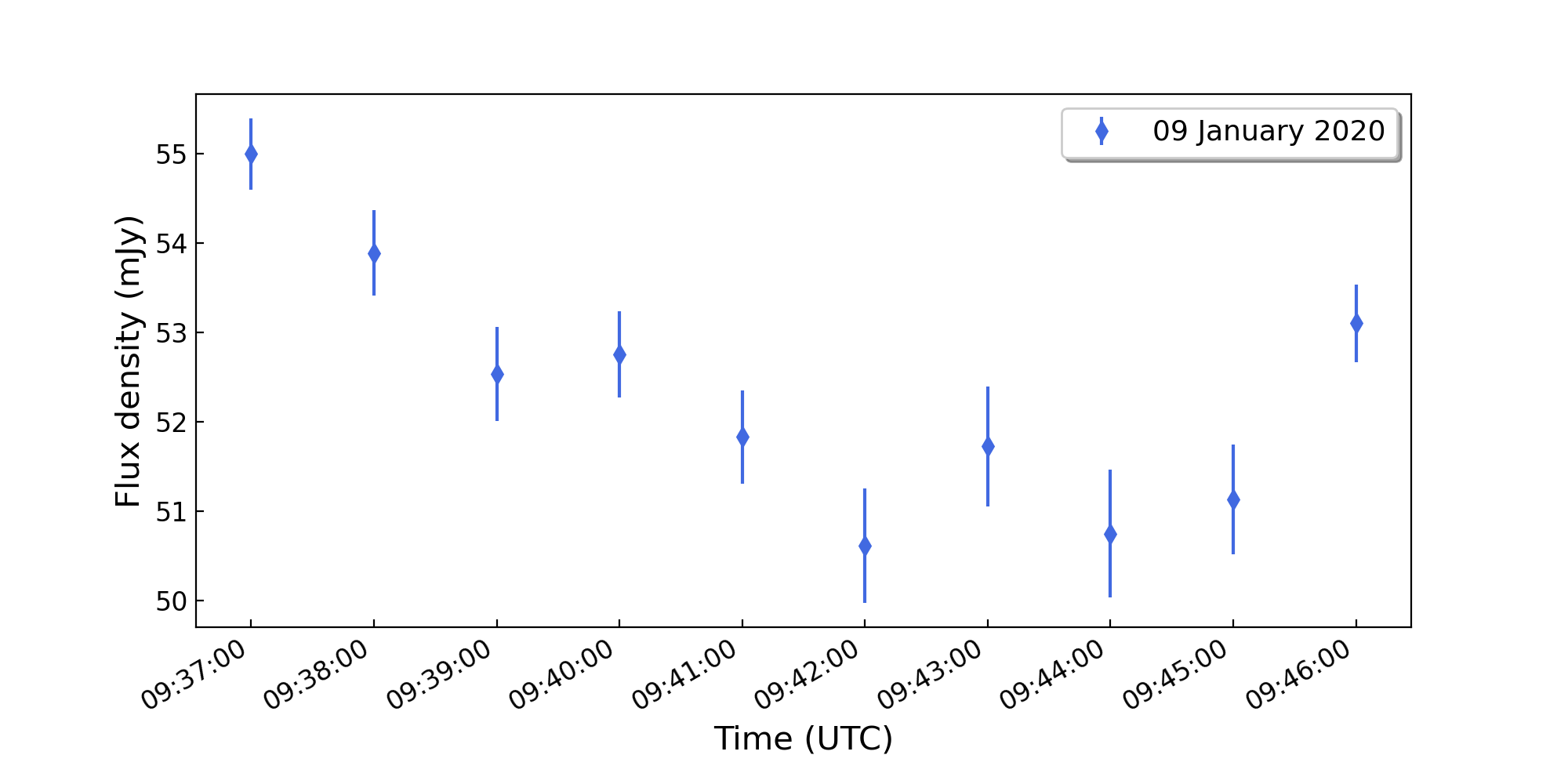}
\caption{Lightcurve of GX\,339--4 with 1 minute intervals during the time of the major flare on 2019-2020 outburst where it reached its peak flux density. \textbf{Top:} 28 December 2019, MJD 58845 (major flare). \textbf{Middle:} 03 January 2020, MJD 58851. \textbf{Bottom:} 10 January 2020, MJD 58868. The flux is decreasing by $\sim$ 4 mJy within 5 minutes.} \label{flarevar}
\end{figure}
The system's radio flux density decreased significantly on MJD 58837, indicating the quenching of the compact jet when the transition occurred. 

A week later, a strong major flare (88 mJy) was caught by MeerKAT observations coinciding with the peak of the soft X-rays (MJD 58845/2019-12-28). The L-band MeerKAT bandwidth of 856 MHz was split into four sub-bands and we imaged them individually in order to obtain a spectral index of $\alpha$ = 0.1, implying strong optically thick emission from the onset of the flare. We
divided the $\sim$10-minute scan of MJD 58845 into ten chunks of
equal temporal length, and we imaged them separately to search for any potential variability (Figure~\ref{flarevar}, top panel). The light curve does not show any significant variability, 
although the flux density is higher than the 2003 flare event that was observed at higher frequencies \citep{2004gallo}, assuming a relatively steep radio spectral index ($\alpha=-0.7$) after a major flare, such that the L-band flux density would be greater than that at higher frequencies.

The observations taken the week after the major flare showed a drop in flux density by a factor of $\sim$ 4, while afterwards the flux density increased again, reaching $\sim$ 50 mJy (Figure \ref{flarevar}) on MJD 58868, after which the decay of the outburst started. Figure \ref{flarevar} (bottom panel) shows the 10-minute long observation on MJD 58868 split into ten 1-minute intervals, when the flux density decreased by $\sim$ 4mJy within the first 5 minutes.

\begin{figure*}
\centering
\includegraphics[width=\textwidth]{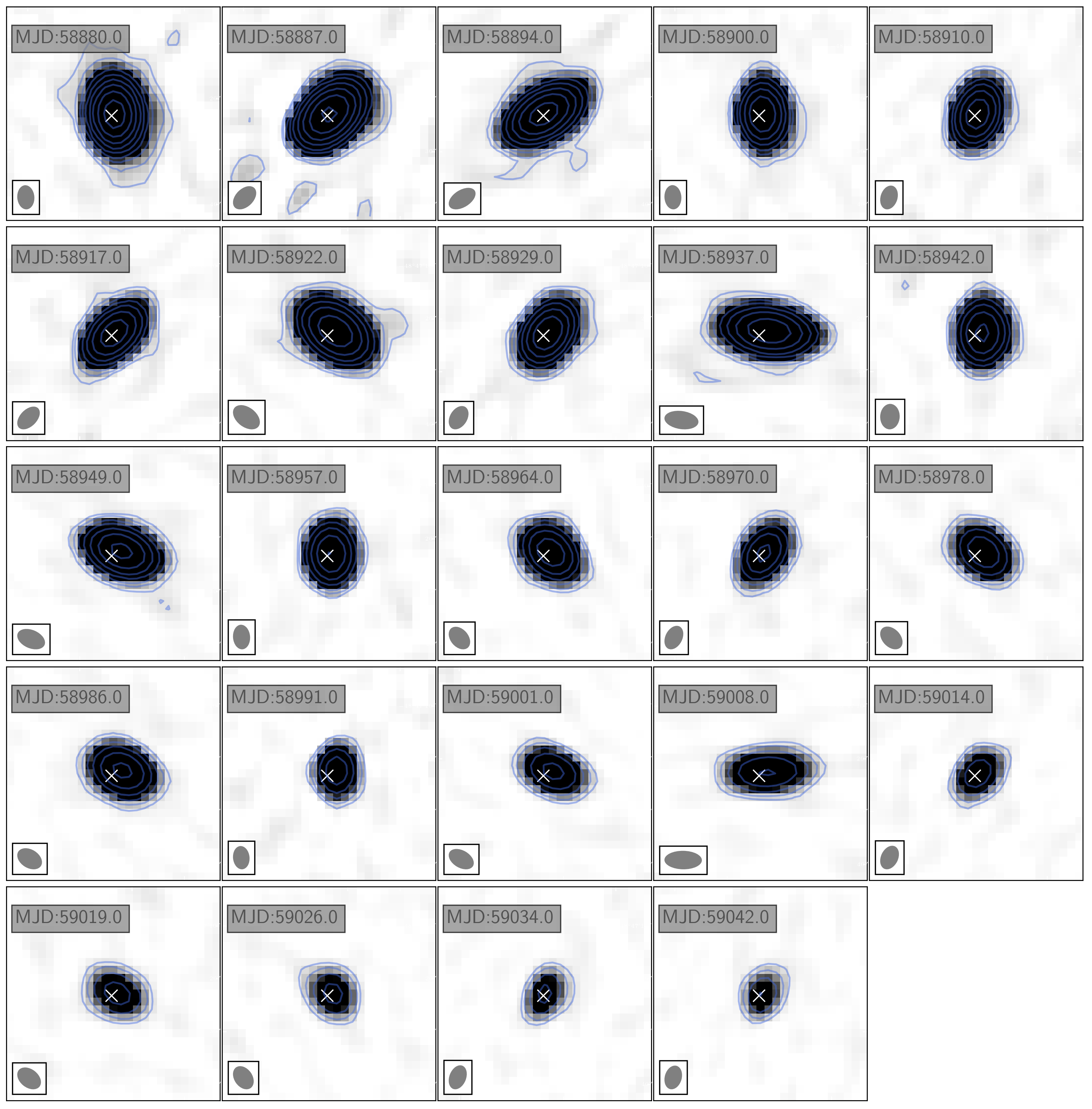}
\caption{40 $\times$ 40 arcsec cutout  MeerKAT images displaying the position shift evolution for $\sim$ 24 weeks during the 2020 outburst. Each image has an identical stretch and contrast. The images are in chronological order starting from MJD 58880 where the first shift was observed (top left) until MJD 59042 where the position shift disappears (bottom right). The contours in blue start from the 3$\times$rms noise level of $\sim$ 35 $\mu$Jy/beam and increase by factors of 2, while the white cross mark indicates the position of the GX\,339--4 core. A westward shift (along the jet direction) is noted from  MJD 58922 until MJD 59034. }\label{snaps2020}
\end{figure*}
\subsubsection{First detection of moving GX 339-4 core with MeerKAT}
\begin{figure}
 \begin{center}

\includegraphics[width=0.5\textwidth]{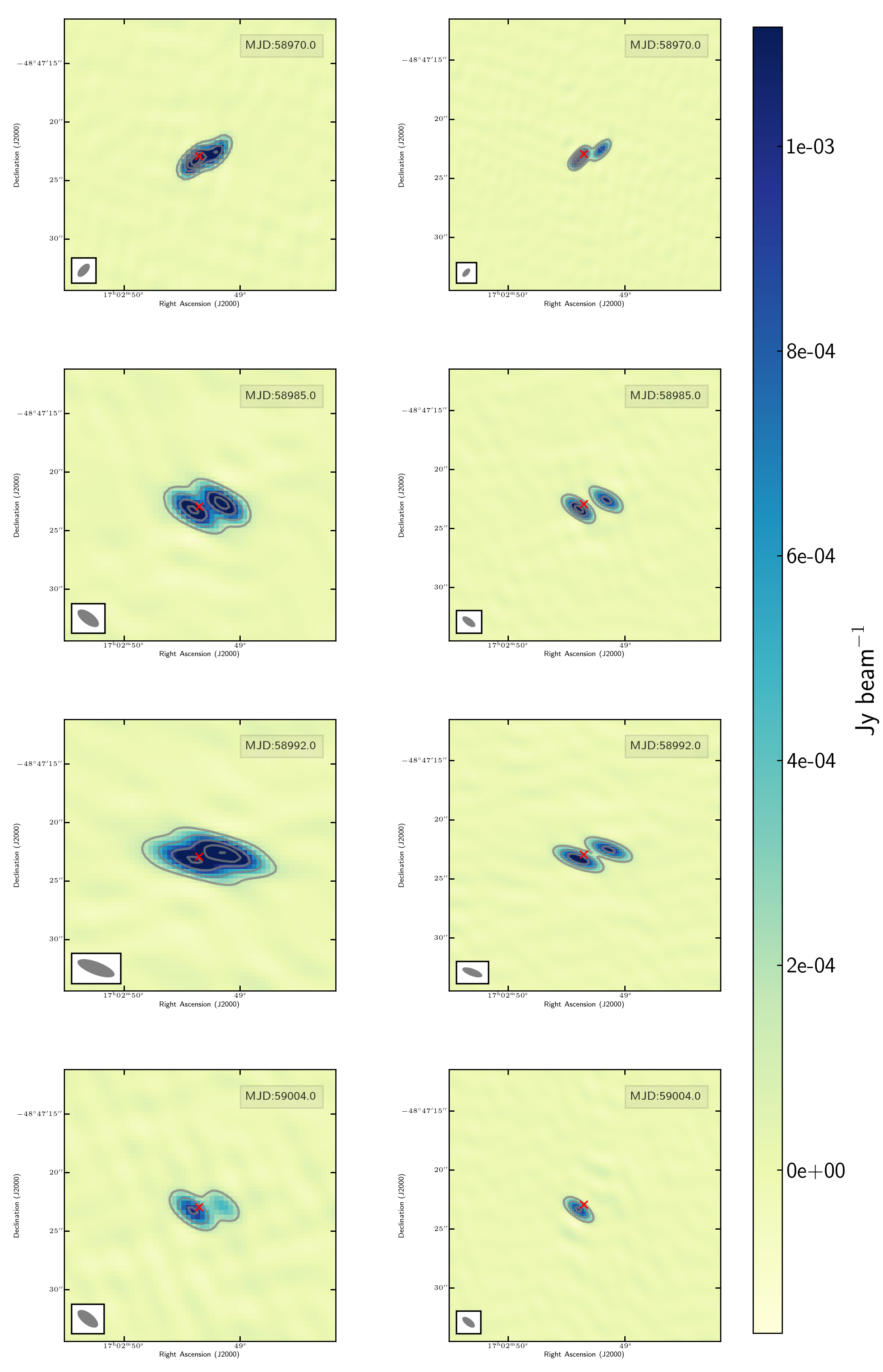}
\caption[]{ATCA radio maps at 5.5 GHz (left) and 9 GHz (right) of GX\,339--4 showing the transient ejection event between May 2020 (MJD 58970) and June 2020 (MJD 59004). The red cross symbolizes the position of the core. The grey contours start from the 3$\sigma$ significance level and increase by factors of 2. }.\label{atcamaps}
\end{center}
\end{figure}

\begin{figure}
\centering
\includegraphics[width=0.485\textwidth]{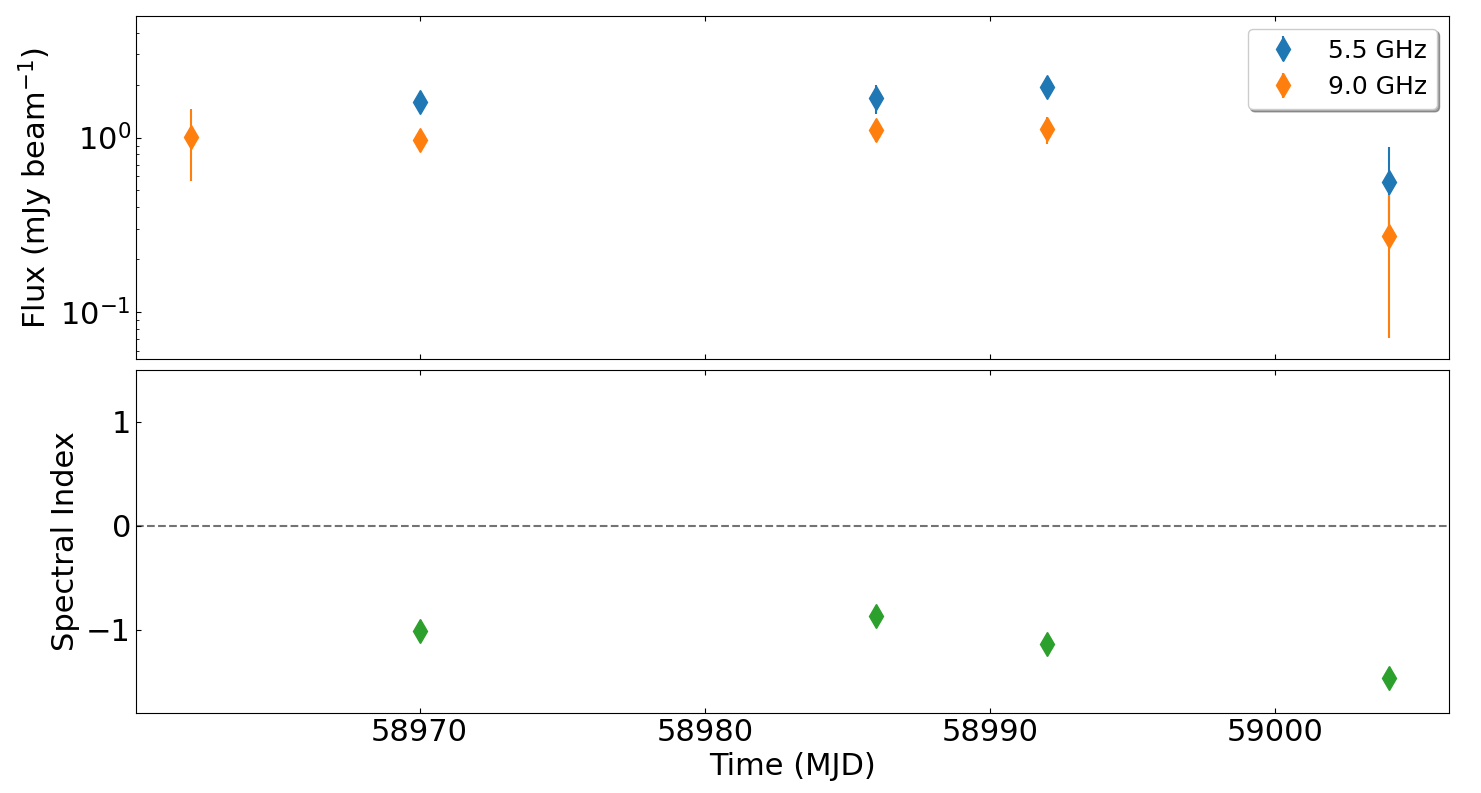}\quad
\includegraphics[width=0.49\textwidth]{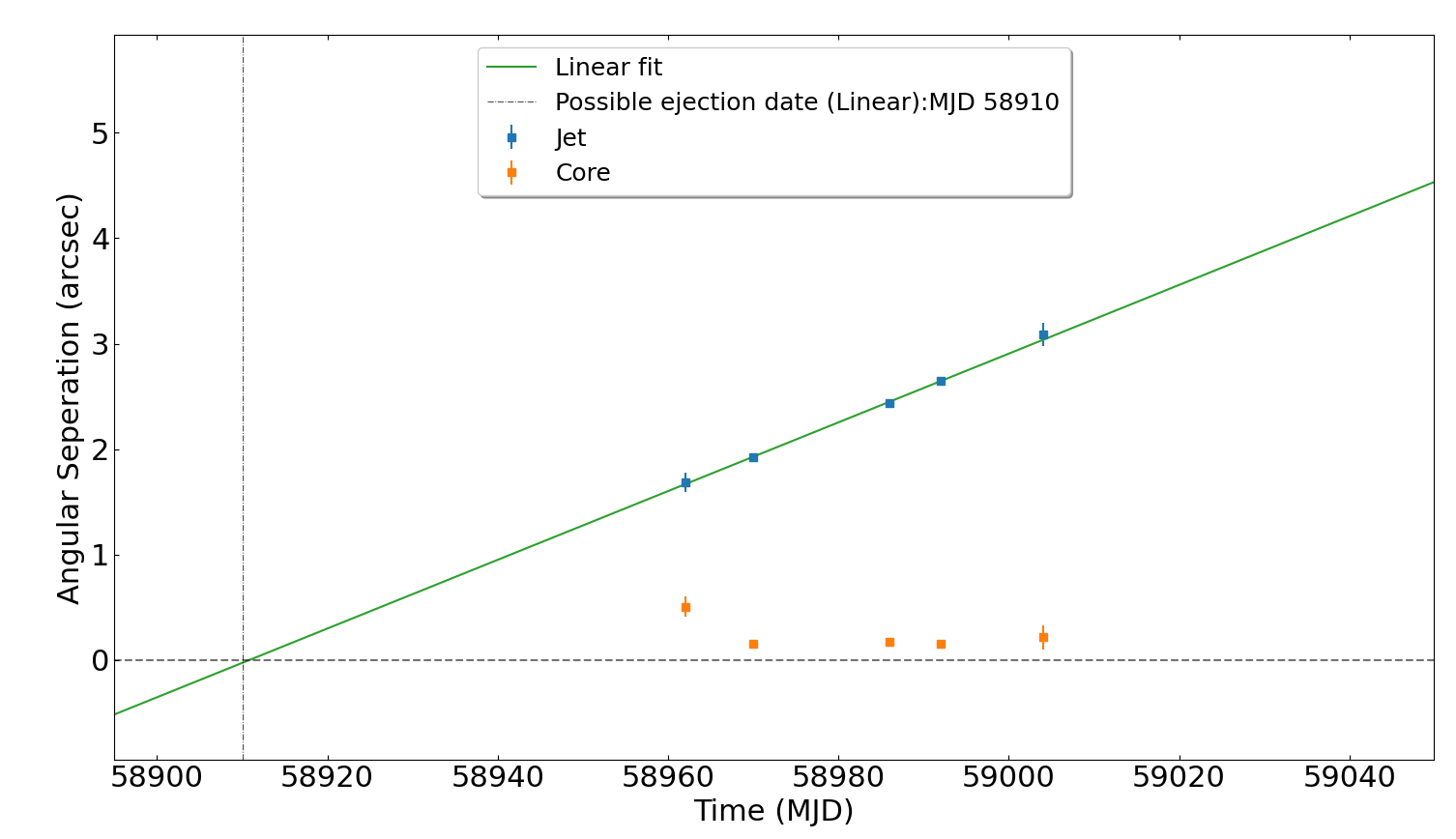}
\caption{\textbf{Top:} Light-curve and spectral indices of the  \textbf{optically thin jet} detected at 5 and 9 GHz with ATCA observations. Noting that the jet was only detected at 9 GHz on MJD 58962. \textbf{Bottom:} Angular separation of the discrete ejecta from the core of GX\,339--4 at 9 GHz (blue points). The plasmon follows a linear motion covering $32.2 \pm 0.7$ mas\,day$^{-1}$. In orange, it displays the angular separation of the core from the position of GX\,339--4. The possible ejection date is calculated at MJD 58910 $\pm$ 1 days.} \label{fig:jetlc}
\end{figure}

Starting on 14 March 2020 (MJD 58922), we observed a shift of the peak flux density to the west relative to the position of the GX\,339--4 core, which continued until 04 July 2020 (MJD 59034; see Figure \ref{snaps2020}). However, the resolution was limited and did not allow us to resolve any discrete jet component. Later, on 20 April 2020 (MJD 58959), we obtained the first ATCA observations at higher frequency than MeerKAT. Due to the array's configuration (H168/ hybrid configuration with a maximum baseline length in the North-South direction of 168 m $\sim$10 arcsec), we did not resolve any jet component at this time, but the next five ATCA observations with higher resolution ($\sim 2"$) allowed us to confirm the presence of a second component by detecting a knot to the west at 5.5 and 9 GHz (Figure \ref{atcamaps}). The ejection event followed the major flare and it was detected for 42 days (until MJD 59004) with its flux density peaking at $\sim$ 2 mJy (Figure \ref{fig:jetlc}). The jet displays a steep optically thin radio spectrum (Figure \ref{fig:jetlc}, top panel/bottom plot) ranging between $\alpha$ $\approx$ -1 and $\alpha$ $\approx$ -1.5 similar to the ejection event in 2002 \citep{1997fender, 2004gallo}. 
Tracking the jet component for 42 days allowed us to fit its linear trajectory with a ballistic motion, $R\left( t \right) = \mu \left( t-t_{\rm ejection} \right)$, where $R$ is the angular separation of the component from the core at time $t$ and $\mu$ is the proper motion. We obtained a proper motion of $\mu$ = 32.2 $\pm$ 0.7 mas\,day$^{-1}$ (Figure \ref{fig:jetlc}). For a distance of 8.4\,kpc, the projected jet velocity of GX\,339--4 is $\sim 1.56c$, meaning that the jets display apparent superluminal motion similar to the 2010 outburst \citep{2010corbel}, constraining the inclination angle to $\theta$ = 57.4 $\deg$.  For unpaired ejecta we solve for $\beta\cos\theta$ using the equation \ref{eq1} below for a distance, $d=8.4$\,kpc, and $\mu = 32.2 \pm 0.7$\,mas\,day$^{-1}$.

\begin{equation} \label{eq1}
\hspace{2.5cm}
\mu_{\rm app}=\frac{\beta \sin\theta}{1-\beta \cos\theta} \frac{c}{d}
\end{equation}

The estimated ejection date results in $t_{\rm ejection}$ = MJD 58910.1 $\pm$ 1.4 days, which implies that the ejection happened $\sim$ 65 days after the major flare triggered by the hard-to-soft X-ray state transition (MeerKAT observations, MJD 58845) and $\sim$ 12 days before we see a shift in the position towards the jet direction in MeerKAT images (see Figure \ref{snaps2020} and images at \url{https://doi.org/10.48479/4fpq-sd16}). The ejection detected by ATCA seem to be temporal inconsistent with the shift seen by the MeerKAT maps (MJD 58910) implying an earlier ejection date, or a separate ejection event following a more complicated motion rather a simple ballistic trajectory \citep{2002corbel}. MeerKAT telescope is a very good instrument at picking up low surface-brightness emission that might be resolved out by ATCA, hence it is very plausible that it could be detecting an earlier ejection event.

The trajectory of the transient jet resolved by ATCA is consistent with simple ballistic bulk motion \citep{1999rodriguez}. 

\begin{figure}
 \begin{center}
\includegraphics[width=0.495\textwidth]{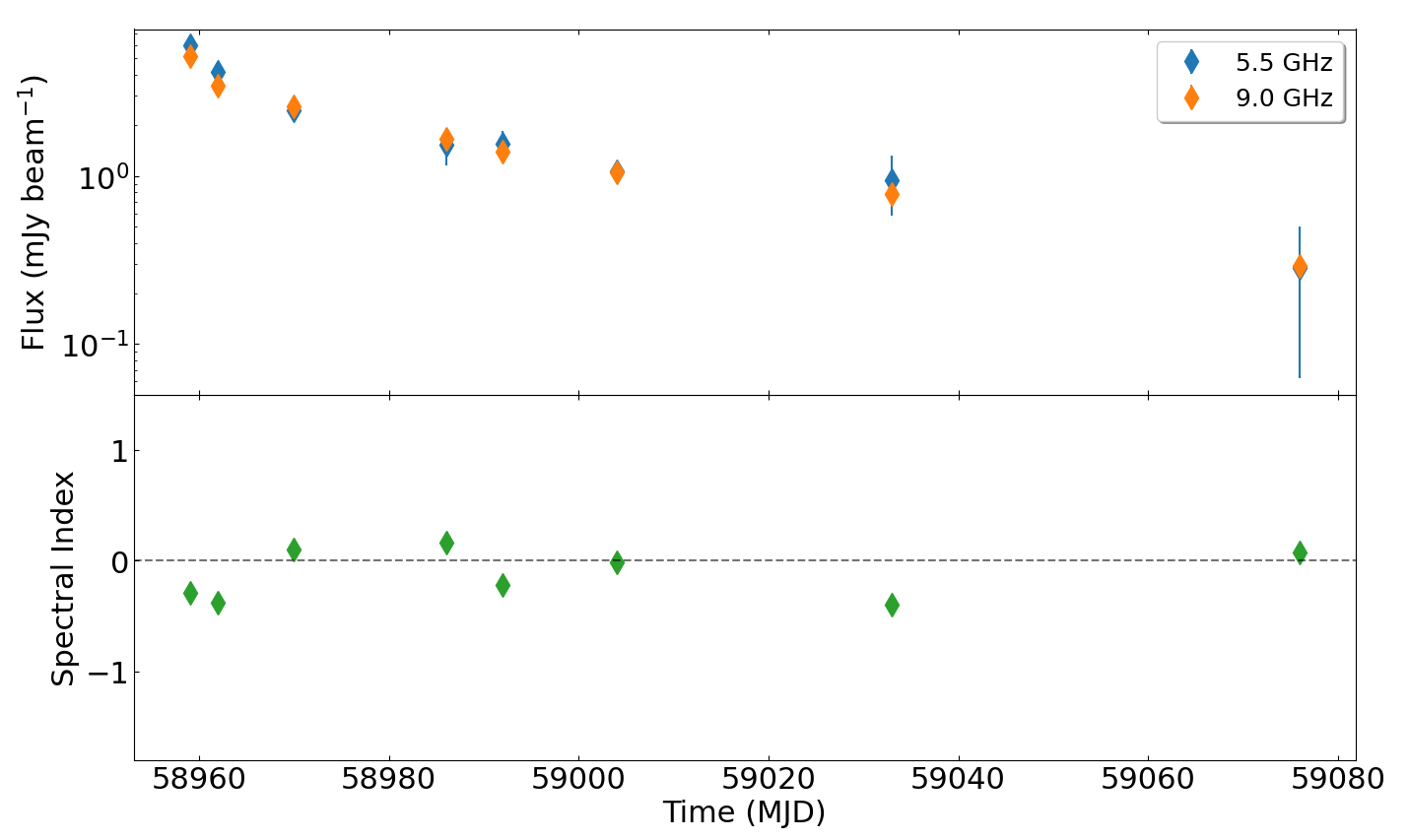}
\caption[]{\textbf{Top:} Light-curve of the core (emission from the compact jets) of GX\,339--4 at 5 (blue) and 9 (orange) GHz with ATCA observations observed after the transition from soft to hard state. \textbf{Bottom:} Using the ATCA observations, the plot shows the evolution of the core's spectral index which remains flat.}\label{corelc}
\end{center}
\end{figure}

After the source made the transition back to the hard state and the compact jet reappeared at the position of the core, the flux density of the steady jet (compact jet) was reduced by a factor of $\sim$ 6 during the 42 days that we monitored both the core and the large ejection event (Figure \ref{corelc}, top panel). The core radio spectral index determined from the ATCA observations remained flat during this time (Figure \ref{corelc}, bottom panel). 

\subsection{The 2021 (MJD 59188 -- 59607) outburst}

Following the completion of a "full" outburst in 2020, the system entered a brief quiescent phase lasting $\sim 1.5$ months (until MJD 59211). A rapid rise began on \citep[04 January 2021;][]{2021tremou}, marking the start of a new outburst on MJD 59218.

The 2021 outburst followed a full cycle progression. Initially, the source reached a plateau in hard X-rays during its peak luminosity, implying a long stay of $\sim 40$ days in the hard state, which delayed the transition to the soft state \citep[similar to the 2004 outburst;][]{2013corbel}.
The source successfully transitioned into the soft X-ray state on MJD 59300 (Hard-to-Soft Transition). It remained in this soft state for an extended period of almost 6.5 months until MJD 59489. The duration of the soft state in 2021 was notably longer compared to the 2020 outburst, even though the overall 2020 outburst lasted longer.
The hard X-ray peak observed on MJD 59489 (Figure \ref{fig:lightcurve}) indicated the system was transitioning back to the hard state \citep[Soft-to-Hard Transition][]{2021corbel}. 
The source completed its outburst at the end of 2021 (MJD 59572) and returned to quiescence. The transition into the soft state evolved rather slowly in 2021 compared to the 2020 outburst. The extended duration in the soft state suggests a prolonged period where the inner accretion disk was cooled and dominant (thermally soft).
A hard X-ray re-flare was observed on MJD 59493 at the end of the soft state. This X-ray re-flare is directly associated with the second radio flare and is linked to the build-up of the compact jets during the transition from the soft state back to the hard state \citep{2013corbel}.

In the optical, we could only track the source during the soft state where it remained at the same flux level overall and no significant variability was recorded. 

\begin{figure*}
\centering
\includegraphics[width=\textwidth]{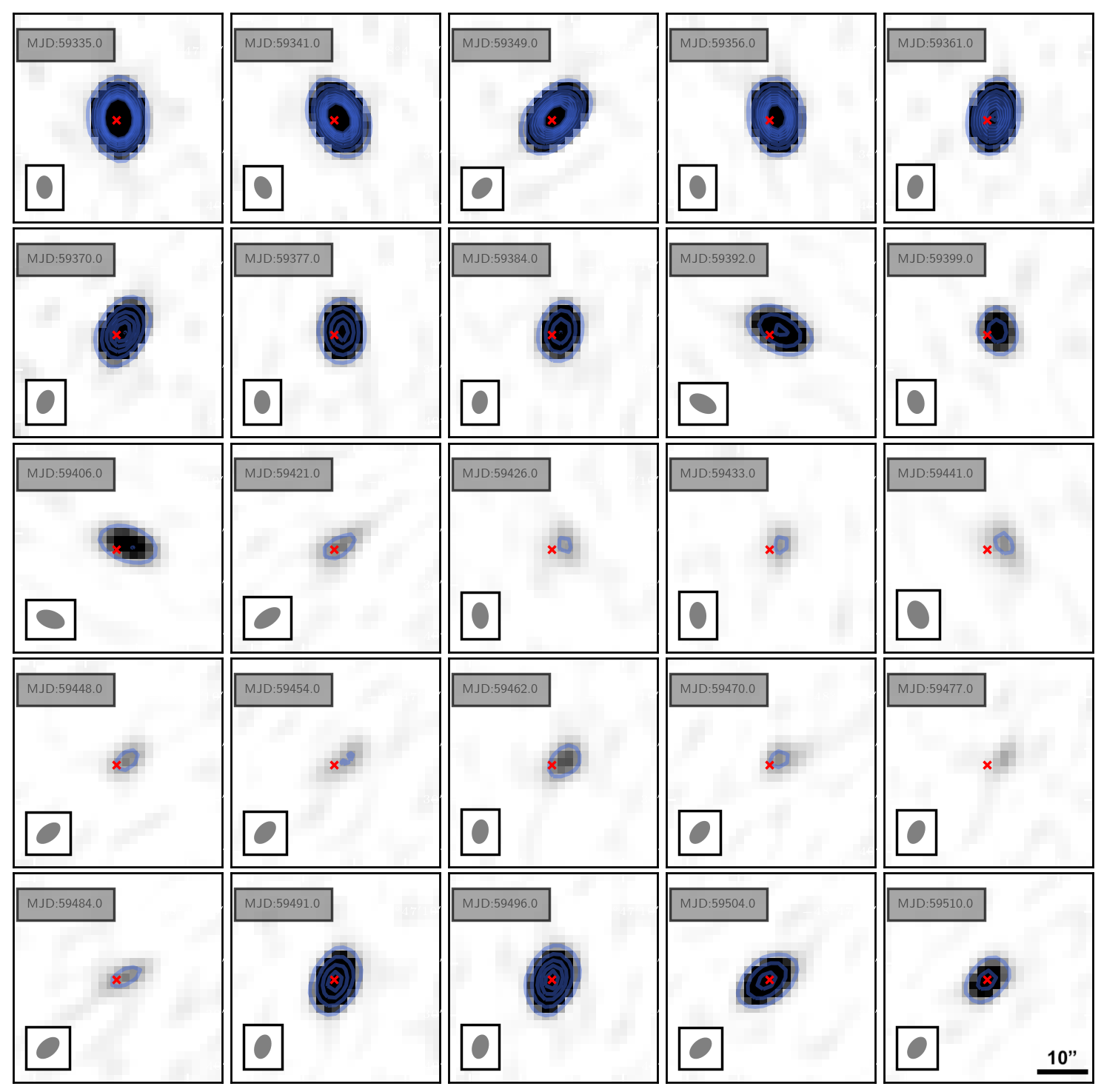}
\caption{40 $\times$ 40 arcsec cutout  MeerKAT images displaying the position shift evolution for $\sim$ 25 weeks during the 2021 outburst. Each image has an identical stretch and contrast. The images are in chronological order starting from the top left to be MJD 59335 where the first shift was observed until MJD 59510 (bottom right) where the position shift disappears. The contours in blue start from the 3$\times$rms noise level $\sim$ 35 $\mu$Jy/beam and increase by a factor of 2, while the red cross mark indicates the position of the GX\,339--4 core. The shift towards the jet direction is clear from MJD 59341 until MJD 59484. }\label{snaps2021}
\end{figure*}

\begin{figure}
\centering
\includegraphics[width=0.49\textwidth]{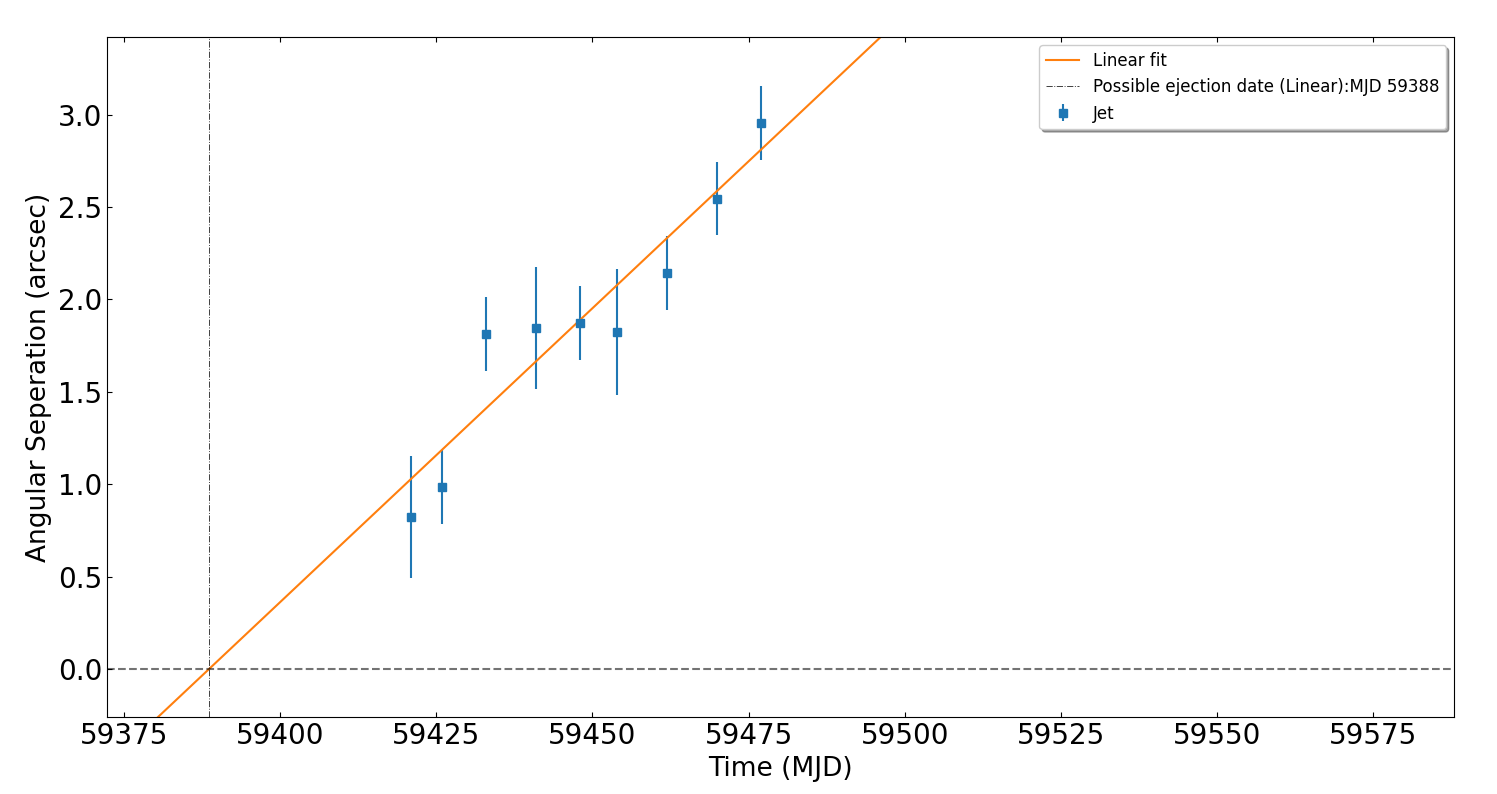}
\caption{Angular separation of the large outflow from the core of GX\,339--4 at 1.28 GHz as seen by MeerKAT during the 2021 outburst. The plasmon follows a linear motion covering 31.8 $\pm$ 4 mas/day. The possible ejection date is calculated at MJD 59388 $\pm$ 8 days.} \label{fig:jet2021}
\end{figure}

The radio emission tracked the state transitions closely:
Similar to the 2020 outburst, at the beginning of the soft state, the compact jets quenched, indicated by a flux density drop by a factor of $\sim 4$ on MJD 59301 (hard-to-soft transition). Immediately following the quenching, the flux rose again, reaching its peak radio flux of 27.9 mJy on MJD 59314. Subsequent observations showed a lower flux density, suggesting this was the decay of the first major flare.
A second radio flare, which attained the peak flux density of $\sim 4\ \text{mJy}$ for the late stages of the 2021 outburst, was noticed on MJD 59493 (Soft-to-Hard Transition). This flare is associated with the hard X-ray re-flare and is connected to the evolution from optically thin to optically thick synchrotron emission as the compact jets are re-built.
Considering the peak of the major flare on MJD 59314, a search was conducted for a potential large-scale structure (ejection event) triggered by the hard-to-soft transition. Although a jet component was challenging to be detected with the resolution of MeerKAT at L-band, a clear positional shift from the GX\,339--4 core was noted on MJD 59335, lasting until $\sim \text{MJD 59484}$ (Figure \ref{snaps2021}). This was associated with the core fading and the location of the peak flux density shifting. Although the angular separation between the first couple of epochs did not show a large positional, the ejection followed a linear track overall (Figure \ref{fig:jet2021}) with a proper motion of $32 \pm 4\ \text{mas day}^{-1}$. The inferred ejection date, calculated from the proper motion, is MJD $59388 \pm 8\ \text{days}$, which corresponds to $\sim 2$ months after the major flare. 
This measured proper motion is consistent with those seen in previous outbursts, such as the 2020 outburst, the 2002 event \citep{2004gallo}, the 2024 outburst \citep{2025mastroserio}, and is similar to what has been seen in GRS1915+105 GRS1915+105 \citep{2005james}. Unlike sources such as H1743-322, V404 Cyg \citep{2009McClintock, 2012james, 2019james,2019tetarenko} do not show consistent jet ejecta proper motions between their different outbursts.

\subsection{The 2022 (MJD 59797 -- 59965) outburst.  A ``hard-only" outburst}

Following a nearly seven-month quiescent state (from MJD 59608 to MJD 59790), GX\,339--4 entered another outburst in early August 2022 \citep{2022Kobayashi,2022tremou}. This outburst lasted for almost six months, from MJD 59797 until MJD 59965, and was characterized as a "hard only" event. 

Similar to the 2018–2019 event, the 2022 outburst did not follow the typical "full" outburst progression. It completely failed to enter the intermediate soft state, instead remaining in the hard X-ray state for approximately 168 days. 
X-ray observations (\textit{Swift}/XRT) confirmed the hard state throughout both the rise and the decay phases. As the outburst progressed, the photon index ($\Gamma$) remained flat ($\Gamma \sim 1.5$), a value comparable to the previous "hard only" outbursts observed for this source. The fact that the source remained in the hard X-ray state for its entire duration suggests the mass accretion rate did not reach the high level necessary to cause the inner accretion disk to significantly cool and transition to the soft state.

During the outburst, the radio flux density of the source reached a maximum of $\sim 3.8\ \text{mJy}$ on 24 September 2022 (MJD 59846), after which its brightness began to decrease. This radio emission is associated with the compact jet that is characteristically present during the hard state of black hole X-ray binaries.
A series of quasi-simultaneous radio observations were conducted using the MeerKAT radio telescope during this hard state.
Following the decay, the source returned to quiescence for another $\sim 9$ months until a new outburst began to rise in early October 2023 \citep[][Nyamai et al. in prep.]{2023Alabarta,2024nyamai}. Minor weekly variability was noted during this quiescent phase, primarily resulting in non-detections (upper limits) but also including a couple of radio detections close to the quiescent level (see Figure \ref{fig:lightcurve}).

\subsection{On the Radio/X-ray correlation}\label{corrdisc}

\begin{figure*}
\begin{center}

\includegraphics[width=\textwidth]{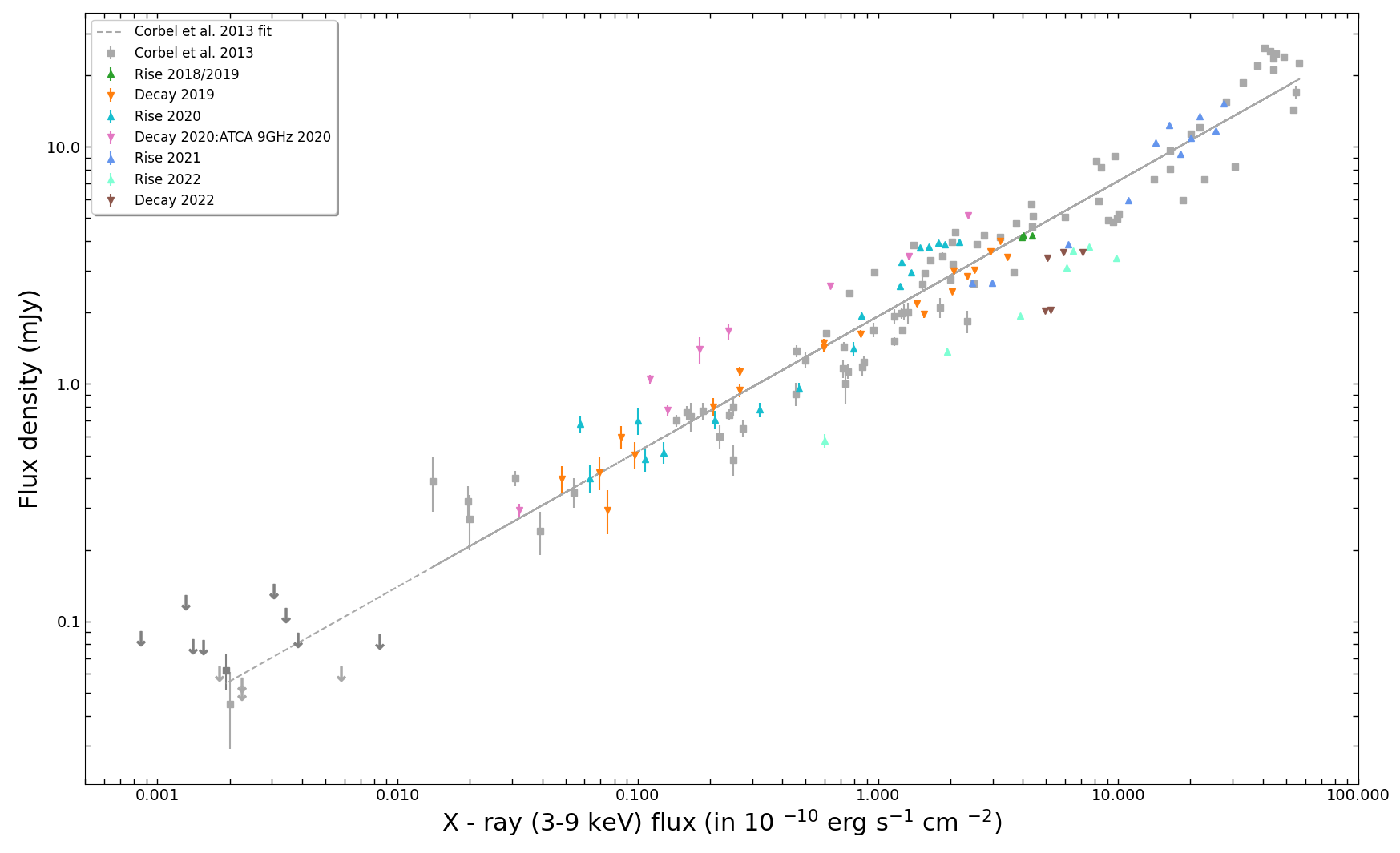}
\caption[]{1.28 GHz radio emission from GX\,339--4 in the hard state versus the \textit{Swift}/XRT unabsorbed 3--9 keV X-ray flux. The  grey points show the past studies from \cite{2013corbel} at 9 GHz radio frequency as well as their fit in the gray dashed line. The colored points display the correlation using data presented in the current study split by rise and decay for each outburst cycle. See Table \ref{tab:corr} below for details. The MeerKAT data taken during the decays of the 2020 and the 2021 outbursts is not part of the correlation plot due to the unknown contribution of the the unresolved large-scale transient ejecta that persist into the hard state decay.}

\label{fig:lxlrphase}
\end{center}
\end{figure*}

In Figure \ref{fig:lxlrphase}, we place our quasi-simultaneous radio
and X-ray measurements from this study on the X-ray--radio plane. We choose observations taken within 24 hours from the radio observations, to compare our results with the existing large sample \citep[a series of radio and X-ray observations of the system;][]{2013corbel}. Table \ref{tab:corr} indicates the exact dates that represent each phase of the outburst and the data points of the Figure \ref{fig:lxlrphase}. We note the difference in frequency of our MeerKAT L-band data compared to the standard 5 GHz radio/X-ray correlation, however we assume a flat radio spectral index that is usually present during the hard state. All the data shown in Figure \ref{fig:lxlrphase} are taken in the hard-state of the source.  

Due to Sun-constraints some of the phases were not fully covered due to the lack of \textit{Swift}/XRT observations. We also did not fully cover the rise of the 2018--2019 outburst and the decay of the 2021 outburst.

\begin{figure*}
\includegraphics[width=\textwidth]{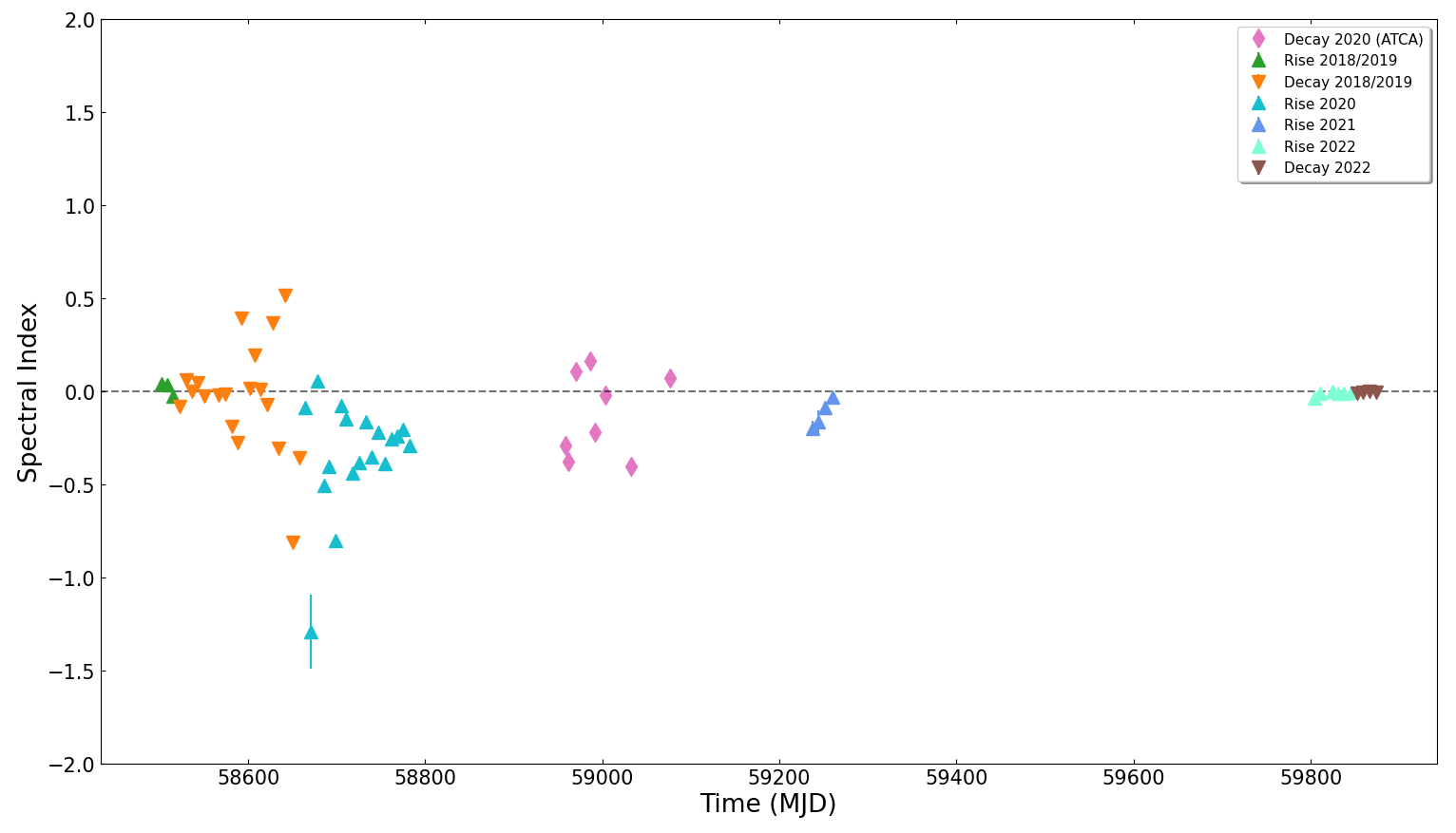}
\caption[]{The radio spectral index calculated from the MeerKAT L-band data (during the hard state of the source) split into four sub-bands and imaged individually. We note here that the sub-band calibration has not been properly evaluated and it is challenging when the source is not bright. Hence we are aware that our estimates may include a few percent of calibration errors. We choose the dates where the signal to noise is enough to allow us the in-band measurements.}
\label{fig:spix}
\end{figure*}

During the 2018--2019 outburst and the 2022 ``hard only" outburst, the slope of the radio/X-ray correlation remains consistent with the fit from the GX\,339--4 data presented in \cite{2013corbel} and we cannot conclude significant flattening as has been discussed by \cite{2021dehaas}. However, we note that we only have two data points from the 2018--2019 rise, which limits our sampling. The radio spectral index remains relatively flat during both the rise and decay of the 2018--2019 and the 2022 outbursts, where the compact jets are increasing in flux (Figure \ref{fig:spix}). 

As the 2019--2020 outburst progresses the spectral index changes from flat to optically thin. Interestingly, the radio spectral index during the decay of 2020 and 2021 remains optically thin, as the radio emission is dominated by the large scale jet emission. 
Although, the radio spectral index may be an important indicator for radiative behaviors between the rising and decaying phases but the limited radio band may not be fully representative of the whole jet spectrum \citep{2022barnier}.

The limited angular resolution of the L-band MeerKAT did not allow us to resolve the inferred transient plasmon from the core. Hence, the contribution of the large scale outflow to the total radio emission detected by MeerKAT $\sim$ 5 $\arcsec$ resolution is significant and unknown. Therefore, we exclude from the correlation plot (Figure \ref{fig:lxlrphase}) both 2020 (April - October 2020) and 2021 (May - December 2021) decays. However, we plot in Figure \ref{fig:lxlrphase} in pink the 9 GHz radio flux density (higher frequency ATCA observations) as a function of the 3--9 keV \textit{Swift}/XRT X-ray flux. Those observations were taken during the decay of the 2020 outburst and the high angular resolution allows us to resolve the compact jet from the transient ejecta and measure the flux density of the core (compact jet; see Figure \ref{atcamaps}).

\begin{table*}
\begin{tabular}{|| m{14.3cm} | m{2cm}|| }
  \hline
  \textbf{Time} & \textbf{Phase} \\ \hline
  \textbf{MJD} 58502, 58509, 58515 & Rise 2018/2019  \\ \hline
  \textbf{MJD} 58523, 58530, 58537, 58543, 58551, 58567, 58574, 58582, 58588, 58593, 58602, 58607, 58614, 58621, 58628, 58634, 58642, 58650, 58658  & Decay 2019  \\ \hline
  \textbf{MJD} 58664, 58671, 58678, 58686, 58691, 58699, 58705, 58711, 58718, 58726, 58733, 58740, 58747, 58755, 58762, 58768, 58775, 58782 & Rise 2020\\\hline
  \textbf{MJD} 58970,58978,58986,58992,59001,59008,59014,59019,59026,59034,59039,59042,59049,59054, 59069, 59076, 59091, 59096,59104,59118,59132,59139, 59153 & Decay 2020 \\\hline
  \textbf{MJD} 59239, 59244, 59251, 59258, 59265, 59273, 59279, 59281,59286, 59293 & Rise 2021 \\\hline
  \textbf{MJD} 59491, 59504, 59510, 59518 & Decay 2021 \\\hline
  \textbf{MJD} 59804, 59811, 59818, 59823, 59831, 59839, 59846 & Rise 2022 \\\hline
  \textbf{MJD} 59851, 59859, 59867, 59873 & Decay 2022 \\
  \hline
  \end{tabular}
  \caption[]{A list of the MJD dates that used for the compilation of the Radio/X-ray correlation plot as shown in  Figure \ref{fig:lxlrphase}.}
  \label{tab:corr}
\end{table*}

\section{Discussion}
We have presented weekly quasi simultaneous radio, X-ray and optical observations of the recurrent black hole X-ray binary, GX\,339--4, covering a period of five years (2018 -- 2023). This monitoring represents the densest and longest-standing campaign of this source using some of the main optical, radio and X-ray facilities currently available such as MeerLICHT, MeerKAT, ATCA and \textit{Swift} and MAXI. Over the five years, the source underwent two ``full" outbursts, in which it transitioned from the hard X-ray spectra state to the soft state and back to the hard state before decaying to quiescence.

At the beginning of our monitoring (2018) the source was in quiescence \citep[see][]{2020tremou} and a new outburst began in late 2018, which progressed through mid-2019. However, the source never entered into the intermediate or soft state and it did not completely return to quiescence before re-brightening in late 2019 to undergo the 2020 ``full" outburst.  The 2018--2019 outburst was fainter and lasted only $\sim$ 200 days compared to the following ones that exceeded 370 days. This type of outburst is generally referred to as a ``hard-only" outburst \citep{1994harmon,2000hynes,2001brocksopp,2004brocksopp,2010brocksopp,2002belloni,2004aref,2005sturner,2013curran,2021motta} and the cause that prevents the source from entering the soft state is not yet well understood. ``Hard-only" outbursts are shorter in duration and on-average fainter compared to the ``full" outbursts \citep{2016tetarenko}. This suggests that the lower accretion rates may be insufficient to sustain a full outburst, since less material from the outer disk flows to the inner regions \citep{2021dehaas}. However, outbursts that leave the hard state and enter only the intermediate state without transitioning to the soft state have also been observed in the past \citep[e.g.:][]{2002zand,2002wijnands,2009capitanio,2012ferrigno,2012reis,2013soleri,2013zhou,2014curran}. The last outburst discussed in this work (the 2022 outburst) followed the same ``hard-only" track as the 2018--2019 outburst, although it lasted for almost a month less. However, the maximum flux densities were similar for both outbursts.  

The following outburst started in June 2020 (MJD 58650) and finished in November 2021 (MJD 59180). This outburst was the brightest and the longest in duration ($\sim$ 1.5 years) among the four outbursts that we discuss in this work, and seems to be comparable to the 2008--2009 outburst \citep{2013corbel,2000corbel}. Although, the soft state lasted less long than that of the 2021 outburst, we were able to observe the quenching of the compact jet by a factor of $\sim$ 4 prior to a very bright major flare ($\sim$ 88\,mJy) and monitor the evolution of the following large scale outflow for almost a month. 
 
The transient jet followed a linear motion travelling away from the core covering $32.2 \pm 0.7$\,mas\,day$^{-1}$. ATCA data at higher frequencies (4cm wavelength) were essential in allowing us to track this ejection. The radio spectral index of the outflow was optically thin ($\alpha \approx -1$) during this month, and the peak flux density reached $\sim 2$\,mJy at 5.5\,GHz. 
The source transitioned from the soft to the hard state on $\sim$ MJD 58960 and the outburst decayed until November 2021 (MJD 59180) when the source entered the quiescent state. 

It is interesting to note that the system remained in quiescence for only $\sim$ 20 days (until MJD 59200). The next outburst that occurred during 2021 was shorter compared to the previous outburst, lasting for $\sim$ 1 year. The 2021 outburst was also a ``full" outburst where the source transitioned from the hard to soft state and then back to the hard state until reaching quiescence. Compared to the 2020 outburst, the rise of 2021 was faster, however, the system remained in the hard state for a long time after the peak of the hard X-rays before entering the soft state, similar to the 2004 outburst \citep[$\sim$ 5 months,][]{2013corbel}. This time, even with the resolution of MeerKAT we detected the large-scale outflow, as shown in Figure \ref{snaps2021}. The detection may seem marginal ($\sim 5 \sigma$), however the previous detected position of the outflow to the west supports this detection. The angular separation of the large outflow from the core of GX\,339--4 reached a maximum of $\sim 3$\,arcsec. The plasmon followed a linear motion covering $31.81 \pm 4$\,mas\,day$^{-1}$. The inferred ejection date is calculated to be at MJD 59388.64 $\pm$ 8 days, $\sim$ 2 months after the major flare that was triggered at the transition state from the hard to the soft state.

\cite{2013corbel} presented the longest-term campaign of 88 quasi-simultaneous radio and X-ray observations of GX\,339--4 during its hard state, covering a total of seven outbursts over a 15-year period and allowing the confirmation of the non-linear coupling between the jet and the inner accretion flow ($L_{\rm X} \propto L_{\rm Rad}^{0.62\pm0.01}$). Our dense monitoring allowed us to add 
90 more data points to the correlation, from which 82 come from quasi simultaneous observations of MeerKAT and \textit{Swift}/XRT, while the rest of them (8) are from ATCA radio data. We sampled the radio/X-ray correlation during rises and decays (hard state) when the \textit{Swift}/XRT observations were possible. In both ``full" outbursts, we note the quenching of the compact jet before the source entered the soft state (less radio bright comparing to the entire rise phase). Although we observed with MeerKAT the source during the decays of the two ``full" outbursts (2019-2020 and 2021), we excluded those points from the correlation plot due to the insufficient angular resolution. The large-scale outflow was not resolved and its contribution to the total radio flux cannot be clearly derived.

The ThunderKAT X-ray binary monitoring program was operational for five years. GX\,339--4 is an ideal candidate to probe the mechanisms of accretion and the connection with compact jets during the evolution of outbursts. Hence, it has been the key X-ray binary source in the ThunderKAT monitoring program, which has been detecting numerous radio jets and large scale outflows from black hole X-ray binary systems \citep[e.g.][]{2022williams,2022eijnden,2022rhodes,2022carotenuto,2022zhang,2021carotenutob,2021carotenutoa,2021motta,2021monageng,2020bright,2020tremou,2020williams,2019russell}.

\section{Summary}

In this work, we present the longest weekly quasi-simultaneous radio, X-ray and optical monitoring of a low mass X-ray binary, the GX\,339--4. We discuss the progression of four outbursts, two ``hard-only" and two ``full" outbursts, and present the detection a main radio flare showing no short-term radio variability and the subsequent detection of a discrete outflow with apparent superluminal motion of a projected speed $1.56c$ at an inclination angle of an upper limit $\theta$ = 57.4 $\deg$. We note that the proper motions measured from different ejection events show a consistent motion. 

We add more data points to the well sampled radio/X-ray correlation showing that the correlation index remains consistent. However, we highlight the importance of not only the densely sampled observations but also the need of high angular resolution in order to resolve large outflows and distinguish between the compact and discrete ejections. Finally, we release 252 epochs of radio data taken by the MeerKAT radio telescope at the phase centre of the low mass X-ray binary GX\,339--4.

\section*{Data availability}
ThunderKAT raw data are available in the SARAO archive (\url{https://archive.sarao.ac.za/}). Calibrated radio maps are available at \url{https://doi.org/10.48479/4fpq-sd16}. The un-calibrated ATCA visibility data are publicly available at the ATNF archive at \url{https://atoa.atnf.csiro.au}. \textit{Swift}/XRT data are available in \url{https://www.swift.ac.uk/swift_portal/}. Online tables \ref{tab:obs}, \ref{tab:obsswift}, \ref{tab:obsmeerlicht} provide the full list of fluxes and observations at radio, X-ray and optical. respectively.

\section*{Acknowledgments}
The authors wish to acknowledge the referee for their constructive report that helped us improve the manuscript. 
We thank the staff at the South African Radio Astronomy Observatory (SARAO) for scheduling these observations. The MeerKAT telescope is operated by the South African Radio Astronomy Observatory (SARAO), which is a facility of the National Research Foundation, an agency of the Department of Science and Innovation. We thank the SARAO staff for conducting these observations. We acknowledge the use of the ilifu cloud computing facility – \url{www.ilifu.ac.za}, a partnership between the University of Cape Town, the University of the Western Cape, the University of Stellenbosch, Sol Plaatje University, the Cape Peninsula University of Technology and the South African Radio Astronomy Observatory. The ilifu facility is supported by contributions from the Inter-University Institute for Data Intensive Astronomy (IDIA – a partnership between the University of Cape Town, the University of Pretoria and the University of the Western Cape), the Computational Biology division at UCT and the Data Intensive Research Initiative of South Africa (DIRISA). 
We acknowledge the use of the Nançay Data Center (CDN – Centre de Données de Nançay) facility. The CDN is hosted by the Observatoire Radioastronomique de Nançay (ORN) in partnership with the Observatoire de Paris, the Université d’Orléans, the Observatoire des Sciences de l’Univers d’Orléans (OSUC) and the French Centre National de la Recherche Scientifique (CNRS). The CDN is supported by the Région Centre-Val de Loire (département du Cher). The ORN is operated by the Observatoire de Paris, associated with the CNRS.
ATCA is part of the ATNF which is funded by the Australian Government for operation as a National Facility managed by CSIRO. We acknowledge the Gomeroi people as the traditional owners of the ATCA observatory site. JvdE acknowledges a Warwick Astrophysics prize post-doctoral fellowship made possible thanks to a generous philanthropic donation and was supported by a Lee Hysan Junior Research Fellowship awarded by St Hilda's College, Oxford during part of this work. PJG is partly supported by NRF SARChI Grant 111692. DMR and PS are supported by Tamkeen under the NYU Abu Dhabi Research Institute grant CASS. The MeerLICHT telescope is operated by a consortium consisting of Radboud University, the University of Cape Town, the South African Astronomical Observatory, the University of Oxford, the University of Manchester and the University of Amsterdam. 






\bibliographystyle{mnras}
\interlinepenalty=10000

\bibliography{gx339_4-q_new.bib}

@ARTICLE{2024nyamai,
       author = {{Nyamai}, Miriam M. and {Tremou}, Lilia and {Fender}, Rob and {X-KAT Team}},
        title = "{Recent radio brightening of the black-hole X-ray binary GX 339-4 as observed with MeerKAT}",
      journal = {The Astronomer's Telegram},
         year = 2024,
        month = jan,
       volume = {16421},
        pages = {1},
       adsurl = {https://ui.adsabs.harvard.edu/abs/2024ATel16421....1N}
}

@ARTICLE{2022driessen,
       author = {{Driessen}, L.~N. and {Stappers}, B.~W. and {Tremou}, E. and {Fender}, R.~P. and {Woudt}, P.~A. and {Armstrong}, R. and {Bloemen}, S. and {Groot}, P. and {Heywood}, I. and {Horesh}, A. and {van der Horst}, A.~J. and {Koerding}, E. and {McBride}, V.~A. and {Miller-Jones}, J.~C.~A. and {Mooley}, K.~P. and {Rowlinson}, A. and {Wijers}, R.~A.~M.~J.},
        title = "{21 new long-term variables in the GX 339-4 field: two years of MeerKAT monitoring}",
      journal = {\mnras},
         year = 2022,
        month = jun,
       volume = {512},
       number = {4},
        pages = {5037-5066},
          doi = {10.1093/mnras/stac756},
archivePrefix = {arXiv},
       eprint = {2203.09806},
 primaryClass = {astro-ph.HE},
       adsurl = {https://ui.adsabs.harvard.edu/abs/2022MNRAS.512.5037D}
}

@ARTICLE{2022casa,
       author = {{CASA Team} and {Bean}, Ben and {Bhatnagar}, Sanjay and {Castro}, Sandra and {Donovan Meyer}, Jennifer and {Emonts}, Bjorn and {Garcia}, Enrique and {Garwood}, Robert and {Golap}, Kumar and {Gonzalez Villalba}, Justo and {Harris}, Pamela and {Hayashi}, Yohei and {Hoskins}, Josh and {Hsieh}, Mingyu and {Jagannathan}, Preshanth and {Kawasaki}, Wataru and {Keimpema}, Aard and {Kettenis}, Mark and {Lopez}, Jorge and {Marvil}, Joshua and {Masters}, Joseph and {McNichols}, Andrew and {Mehringer}, David and {Miel}, Renaud and {Moellenbrock}, George and {Montesino}, Federico and {Nakazato}, Takeshi and {Ott}, Juergen and {Petry}, Dirk and {Pokorny}, Martin and {Raba}, Ryan and {Rau}, Urvashi and {Schiebel}, Darrell and {Schweighart}, Neal and {Sekhar}, Srikrishna and {Shimada}, Kazuhiko and {Small}, Des and {Steeb}, Jan-Willem and {Sugimoto}, Kanako and {Suoranta}, Ville and {Tsutsumi}, Takahiro and {van Bemmel}, Ilse M. and {Verkouter}, Marjolein and {Wells}, Akeem and {Xiong}, Wei and {Szomoru}, Arpad and {Griffith}, Morgan and {Glendenning}, Brian and {Kern}, Jeff},
        title = "{CASA, the Common Astronomy Software Applications for Radio Astronomy}",
      journal = {\pasp},
         year = 2022,
        month = nov,
       volume = {134},
       number = {1041},
          eid = {114501},
        pages = {114501},
          doi = {10.1088/1538-3873/ac9642},
archivePrefix = {arXiv},
       eprint = {2210.02276},
 primaryClass = {astro-ph.IM},
       adsurl = {https://ui.adsabs.harvard.edu/abs/2022PASP..134k4501C}
}

@ARTICLE{1979cash,
       author = {{Cash}, W.},
        title = "{Parameter estimation in astronomy through application of the likelihood ratio.}",
      journal = {\apj},
         year = 1979,
        month = mar,
       volume = {228},
        pages = {939-947},
          doi = {10.1086/156922},
       adsurl = {https://ui.adsabs.harvard.edu/abs/1979ApJ...228..939C}
}

@ARTICLE{2009coriat,
       author = {{Coriat}, M. and {Corbel}, S. and {Buxton}, M.~M. and {Bailyn}, C.~D. and {Tomsick}, J.~A. and {K{\"o}rding}, E. and {Kalemci}, E.},
        title = "{The infrared/X-ray correlation of GX 339-4: probing hard X-ray emission in accreting black holes}",
      journal = {\mnras},
         year = 2009,
        month = nov,
       volume = {400},
       number = {1},
        pages = {123-133},
          doi = {10.1111/j.1365-2966.2009.15461.x},
archivePrefix = {arXiv},
       eprint = {0909.3283},
 primaryClass = {astro-ph.HE},
       adsurl = {https://ui.adsabs.harvard.edu/abs/2009MNRAS.400..123C}
}

@ARTICLE{2005james,
       author = {{Miller-Jones}, J.~C.~A. and {McCormick}, D.~G. and {Fender}, R.~P. and {Spencer}, R.~E. and {Muxlow}, T.~W.~B. and {Pooley}, G.~G.},
        title = "{Multiple relativistic outbursts of GRS1915+105: radio emission and internal shocks}",
      journal = {\mnras},
         year = 2005,
        month = nov,
       volume = {363},
       number = {3},
        pages = {867-881},
          doi = {10.1111/j.1365-2966.2005.09488.x},
archivePrefix = {arXiv},
       eprint = {astro-ph/0508230},
 primaryClass = {astro-ph},
       adsurl = {https://ui.adsabs.harvard.edu/abs/2005MNRAS.363..867M}
}

@ARTICLE{2007brocksopp,
       author = {{Brocksopp}, C. and {Miller-Jones}, J.~C.~A. and {Fender}, R.~P. and {Stappers}, B.~W.},
        title = "{A highly polarized radio jet during the 1998 outburst of the black hole transient XTE J1748-288}",
      journal = {\mnras},
         year = 2007,
        month = jul,
       volume = {378},
       number = {3},
        pages = {1111-1117},
          doi = {10.1111/j.1365-2966.2007.11846.x},
archivePrefix = {arXiv},
       eprint = {0705.1125},
 primaryClass = {astro-ph},
       adsurl = {https://ui.adsabs.harvard.edu/abs/2007MNRAS.378.1111B}
}

@ARTICLE{2004gallo,
       author = {{Gallo}, E. and {Corbel}, S. and {Fender}, R.~P. and {Maccarone}, T.~J. and {Tzioumis}, A.~K.},
        title = "{A transient large-scale relativistic radio jet from GX 339-4}",
      journal = {\mnras},
         year = 2004,
        month = jan,
       volume = {347},
       number = {3},
        pages = {L52-L56},
          doi = {10.1111/j.1365-2966.2004.07435.x},
archivePrefix = {arXiv},
       eprint = {astro-ph/0311452},
 primaryClass = {astro-ph},
       adsurl = {https://ui.adsabs.harvard.edu/abs/2004MNRAS.347L..52G}
}

@ARTICLE{2015smirnov,
       author = {{Smirnov}, O.~M. and {Tasse}, C.},
        title = "{Radio interferometric gain calibration as a complex optimization problem}",
      journal = {\mnras},
         year = 2015,
        month = may,
       volume = {449},
       number = {3},
        pages = {2668-2684},
          doi = {10.1093/mnras/stv418},
archivePrefix = {arXiv},
       eprint = {1502.06974},
 primaryClass = {astro-ph.IM},
       adsurl = {https://ui.adsabs.harvard.edu/abs/2015MNRAS.449.2668S}
}

@ARTICLE{2022williams,
       author = {{Williams}, D.~R.~A. and {Motta}, S.~E. and {Fender}, R. and {Miller-Jones}, J.~C.~A. and {Neilsen}, J. and {Allison}, J.~R. and {Bright}, J. and {Heywood}, I. and {Jacob}, P.~F.~L. and {Rhodes}, L. and {Tremou}, E. and {Woudt}, P. and {van den Eijnden}, J. and {Carotenuto}, F. and {Green}, D.~A. and {Titterington}, D. and {van der Horst}, A.~J. and {Saikia}, P.},
        title = "{Radio observations of the Black Hole X-ray Binary EXO 1846-031 re-awakening from a 34-year slumber}",
      journal = {\mnras},
         year = 2022,
        month = sep,
          doi = {10.1093/mnras/stac2700},
       adsurl = {https://ui.adsabs.harvard.edu/abs/2022MNRAS.tmp.2530W}
}

@ARTICLE{2022rhodes,
       author = {{Rhodes}, L. and {Fender}, R.~P. and {Motta}, S. and {van den Eijnden}, J. and {Williams}, D.~R.~A. and {Bright}, J. and {Sivakoff}, G.~R.},
        title = "{Long-term radio monitoring of the neutron star X-ray binary Swift J1858.6-0814}",
      journal = {\mnras},
         year = 2022,
        month = jun,
       volume = {513},
       number = {2},
        pages = {2708-2718},
          doi = {10.1093/mnras/stac954},
archivePrefix = {arXiv},
       eprint = {2204.01598},
 primaryClass = {astro-ph.HE},
       adsurl = {https://ui.adsabs.harvard.edu/abs/2022MNRAS.513.2708R}
}

@ARTICLE{2022carotenuto,
       author = {{Carotenuto}, F. and {Tetarenko}, A.~J. and {Corbel}, S.},
        title = "{Modelling the kinematics of the decelerating jets from the black hole X-ray binary MAXI J1348-630}",
      journal = {\mnras},
         year = 2022,
        month = apr,
       volume = {511},
       number = {4},
        pages = {4826-4841},
          doi = {10.1093/mnras/stac329},
archivePrefix = {arXiv},
       eprint = {2202.01514},
 primaryClass = {astro-ph.HE},
       adsurl = {https://ui.adsabs.harvard.edu/abs/2022MNRAS.511.4826C}
}

@ARTICLE{2000sunyaev,
       author = {{Sunyaev}, R. and {Revnivtsev}, M.},
        title = "{Fourier power spectra at high frequencies: a way to distinguish a neutron star from a black hole}",
      journal = {\aap},
         year = 2000,
        month = jun,
       volume = {358},
        pages = {617-623},
archivePrefix = {arXiv},
       eprint = {astro-ph/0003308},
 primaryClass = {astro-ph},
       adsurl = {https://ui.adsabs.harvard.edu/abs/2000A&A...358..617S}
}

@ARTICLE{1998zdziarski,
       author = {{Zdziarski}, Andrzej A. and {Poutanen}, Juri and {Mikolajewska}, Joanna and {Gierlinski}, Marek and {Ebisawa}, Ken and {Johnson}, W. Neil},
        title = "{Broad-band X-ray/gamma-ray spectra and binary parameters of GX 339-4 and their astrophysical implications}",
      journal = {\mnras},
         year = 1998,
        month = dec,
       volume = {301},
       number = {2},
        pages = {435-450},
          doi = {10.1046/j.1365-8711.1998.02021.x},
archivePrefix = {arXiv},
       eprint = {astro-ph/9807300},
 primaryClass = {astro-ph},
       adsurl = {https://ui.adsabs.harvard.edu/abs/1998MNRAS.301..435Z}
}

@ARTICLE{1999wilms,
       author = {{Wilms}, J{\"o}rn and {Nowak}, Michael A. and {Dove}, James B. and {Fender}, Robert P. and {Di Matteo}, Tiziana},
        title = "{Low-Luminosity States of the Black Hole Candidate GX 339-4. I. ASCA and Simultaneous Radio/RXTE Observations}",
      journal = {\apj},
         year = 1999,
        month = sep,
       volume = {522},
       number = {1},
        pages = {460-475},
          doi = {10.1086/307622},
archivePrefix = {arXiv},
       eprint = {astro-ph/9904123},
 primaryClass = {astro-ph},
       adsurl = {https://ui.adsabs.harvard.edu/abs/1999ApJ...522..460W}
}

@ARTICLE{2002cowley,
       author = {{Cowley}, A.~P. and {Schmidtke}, P.~C. and {Hutchings}, J.~B. and {Crampton}, David},
        title = "{Optical Observations of the Black Hole Candidate GX 339-4 (V821 Arae)}",
      journal = {\aj},
         year = 2002,
        month = mar,
       volume = {123},
       number = {3},
        pages = {1741-1749},
          doi = {10.1086/339028},
       adsurl = {https://ui.adsabs.harvard.edu/abs/2002AJ....123.1741C}
}

@ARTICLE{2002corbel,
       author = {{Corbel}, S. and {Fender}, R.~P. and {Tzioumis}, A.~K. and {Tomsick}, J.~A. and {Orosz}, J.~A. and {Miller}, J.~M. and {Wijnands}, R. and {Kaaret}, P.},
        title = "{Large-Scale, Decelerating, Relativistic X-ray Jets from the Microquasar XTE J1550-564}",
      journal = {Science},
         year = 2002,
        month = oct,
       volume = {298},
       number = {5591},
        pages = {196-199},
          doi = {10.1126/science.1075857},
archivePrefix = {arXiv},
       eprint = {astro-ph/0210224},
 primaryClass = {astro-ph},
       adsurl = {https://ui.adsabs.harvard.edu/abs/2002Sci...298..196C}
}

@INCOLLECTION{2006fender,
       author = {{Fender}, Rob},
        title = "{Jets from X-ray binaries}",
    booktitle = {Compact stellar X-ray sources},
         year = 2006,
       volume = {39},
        pages = {381-419},
       adsurl = {https://ui.adsabs.harvard.edu/abs/2006csxs.book..381F}
}

@ARTICLE{1988hjellming,
       author = {{Hjellming}, R.~M. and {Johnston}, K.~J.},
        title = "{Radio Emission from Conical Jets Associated with X-Ray Binaries}",
      journal = {\apj},
         year = 1988,
        month = may,
       volume = {328},
        pages = {600},
          doi = {10.1086/166318},
       adsurl = {https://ui.adsabs.harvard.edu/abs/1988ApJ...328..600H}
}

@ARTICLE{1979blandford,
       author = {{Blandford}, R.~D. and {K{\"o}nigl}, A.},
        title = "{Relativistic jets as compact radio sources.}",
      journal = {\apj},
         year = 1979,
        month = aug,
       volume = {232},
        pages = {34-48},
          doi = {10.1086/157262},
       adsurl = {https://ui.adsabs.harvard.edu/abs/1979ApJ...232...34B}
}

@ARTICLE{2001markoff,
       author = {{Markoff}, S. and {Falcke}, H. and {Fender}, R.},
        title = "{A jet model for the broadband spectrum of XTE J1118+480. Synchrotron emission from radio to X-rays in the Low/Hard spectral state}",
      journal = {\aap},
         year = 2001,
        month = jun,
       volume = {372},
        pages = {L25-L28},
          doi = {10.1051/0004-6361:20010420},
archivePrefix = {arXiv},
       eprint = {astro-ph/0010560},
 primaryClass = {astro-ph},
       adsurl = {https://ui.adsabs.harvard.edu/abs/2001A&A...372L..25M}
}

@ARTICLE{1992mirabel,
       author = {{Mirabel}, I.~F. and {Rodriguez}, L.~F. and {Cordier}, B. and {Paul}, J. and {Lebrun}, F.},
        title = "{A double-sided radio jet from the compact Galactic Centre annihilator 1E1740.7-2942}",
      journal = {\nat},
         year = 1992,
        month = jul,
       volume = {358},
       number = {6383},
        pages = {215-217},
          doi = {10.1038/358215a0},
       adsurl = {https://ui.adsabs.harvard.edu/abs/1992Natur.358..215M}
}

@ARTICLE{1973pringle,
       author = {{Pringle}, J.~E. and {Rees}, M.~J. and {Pacholczyk}, A.~G.},
        title = "{Accretion onto Massive Black Holes}",
      journal = {\aap},
         year = 1973,
        month = dec,
       volume = {29},
        pages = {179},
       adsurl = {https://ui.adsabs.harvard.edu/abs/1973A&A....29..179P}
}

@ARTICLE{2019zhang,
       author = {{Zhang}, G. -B. and {Bernardini}, F. and {Russell}, D.~M. and {Gelfand}, J.~D. and {Lasota}, J. -P. and {Qasim}, A. Al and {AlMannaei}, A. and {Koljonen}, K.~I.~I. and {Shaw}, A.~W. and {Lewis}, F. and {Tomsick}, J.~A. and {Plotkin}, R.~M. and {Miller-Jones}, J.~C.~A. and {Maitra}, D. and {Homan}, J. and {Charles}, P.~A. and {Kobel}, P. and {Perez}, D. and {Doran}, R.},
        title = "{Bright Mini-outburst Ends the 12 yr Long Activity of the Black Hole Candidate Swift J1753.5-0127}",
      journal = {\apj},
         year = 2019,
        month = may,
       volume = {876},
       number = {1},
          eid = {5},
        pages = {5},
          doi = {10.3847/1538-4357/ab12dd},
archivePrefix = {arXiv},
       eprint = {1903.09455},
 primaryClass = {astro-ph.HE},
       adsurl = {https://ui.adsabs.harvard.edu/abs/2019ApJ...876....5Z}
}

@ARTICLE{2009fender,
       author = {{Fender}, R.~P. and {Homan}, J. and {Belloni}, T.~M.},
        title = "{Jets from black hole X-ray binaries: testing, refining and extending empirical models for the coupling to X-rays}",
      journal = {\mnras},
         year = 2009,
        month = jul,
       volume = {396},
       number = {3},
        pages = {1370-1382},
          doi = {10.1111/j.1365-2966.2009.14841.x},
archivePrefix = {arXiv},
       eprint = {0903.5166},
 primaryClass = {astro-ph.HE},
       adsurl = {https://ui.adsabs.harvard.edu/abs/2009MNRAS.396.1370F}
}

@ARTICLE{2005homan,
       author = {{Homan}, Jeroen and {Buxton}, Michelle and {Markoff}, Sera and {Bailyn}, Charles D. and {Nespoli}, Elisa and {Belloni}, Tomaso},
        title = "{Multiwavelength Observations of the 2002 Outburst of GX 339-4: Two Patterns of X-Ray-Optical/Near-Infrared Behavior}",
      journal = {\apj},
         year = 2005,
        month = may,
       volume = {624},
       number = {1},
        pages = {295-306},
          doi = {10.1086/428722},
archivePrefix = {arXiv},
       eprint = {astro-ph/0501349},
 primaryClass = {astro-ph},
       adsurl = {https://ui.adsabs.harvard.edu/abs/2005ApJ...624..295H}
}

@ARTICLE{1991hughes,
       author = {{Hughes}, P.~A. and {Aller}, H.~D. and {Aller}, M.~F.},
        title = "{Synchrotron Emission from Shocked Relativistic Jets. III. Models for the Centimeter Wave Band Quiescent and Burst Emission from 3C 279 and OT 081}",
      journal = {\apj},
         year = 1991,
        month = jun,
       volume = {374},
        pages = {57},
          doi = {10.1086/170096},
       adsurl = {https://ui.adsabs.harvard.edu/abs/1991ApJ...374...57H}
}

@ARTICLE{2018camilo1,
       author = {{Camilo}, Fernando},
        title = "{African star joins the radio astronomy firmament}",
      journal = {Nature Astronomy},
         year = 2018,
        month = jul,
       volume = {2},
        pages = {594-594},
          doi = {10.1038/s41550-018-0516-y},
       adsurl = {https://ui.adsabs.harvard.edu/abs/2018NatAs...2..594C}
}

@ARTICLE{1973markert,
       author = {{Markert}, T.~H. and {Canizares}, C.~R. and {Clark}, G.~W. and {Lewin}, W.~H.~G. and {Schnopper}, H.~W. and {Sprott}, G.~F.},
        title = "{Observations of the Highly Variable X-Ray Source GX 339-4}",
      journal = {\apjl},
         year = 1973,
        month = sep,
       volume = {184},
        pages = {L67},
          doi = {10.1086/181290},
       adsurl = {https://ui.adsabs.harvard.edu/abs/1973ApJ...184L..67M}
}

@ARTICLE{2012buxton,
       author = {{Buxton}, Michelle M. and {Bailyn}, Charles D. and {Capelo}, Holly L. and {Chatterjee}, Ritaban and {Din{\c{c}}er}, Tolga and {Kalemci}, Emrah and {Tomsick}, John A.},
        title = "{Optical and Near-infrared Monitoring of the Black Hole X-Ray Binary GX 339-4 during 2002-2010}",
      journal = {\aj},
         year = 2012,
        month = jun,
       volume = {143},
       number = {6},
          eid = {130},
        pages = {130},
          doi = {10.1088/0004-6256/143/6/130},
archivePrefix = {arXiv},
       eprint = {1203.5700},
 primaryClass = {astro-ph.GA},
       adsurl = {https://ui.adsabs.harvard.edu/abs/2012AJ....143..130B}
}

@ARTICLE{1994sood,
       author = {{Sood}, R. and {Campbell-Wilson}, D.},
        title = "{GX 339-4}",
      journal = {\iaucirc},
         year = 1994,
        month = jun,
       volume = {6006},
        pages = {1},
       adsurl = {https://ui.adsabs.harvard.edu/abs/1994IAUC.6006....1S}
}

@ARTICLE{2017heida,
       author = {{Heida}, M. and {Jonker}, P.~G. and {Torres}, M.~A.~P. and {Chiavassa}, A.},
        title = "{The Mass Function of GX 339-4 from Spectroscopic Observations of Its Donor Star}",
      journal = {\apj},
         year = 2017,
        month = sep,
       volume = {846},
       number = {2},
          eid = {132},
        pages = {132},
          doi = {10.3847/1538-4357/aa85df},
archivePrefix = {arXiv},
       eprint = {1708.04667},
 primaryClass = {astro-ph.HE},
       adsurl = {https://ui.adsabs.harvard.edu/abs/2017ApJ...846..132H}
}

@ARTICLE{2022zhang,
       author = {{Zhang}, X. and {Yu}, W. and {Motta}, S.~E. and {Fender}, R. and {Woudt}, P. and {Miller-Jones}, J.~C.~A. and {Sivakoff}, G.~R.},
        title = "{MeerKAT radio detection of the Galactic black hole candidate Swift J1842.5-1124 during its 2020 outburst}",
      journal = {\mnras},
         year = 2022,
        month = feb,
       volume = {510},
       number = {1},
        pages = {1258-1263},
          doi = {10.1093/mnras/stab3463},
archivePrefix = {arXiv},
       eprint = {2112.02202},
 primaryClass = {astro-ph.HE},
       adsurl = {https://ui.adsabs.harvard.edu/abs/2022MNRAS.510.1258Z}
}

@ARTICLE{2022eijnden,
       author = {{van den Eijnden}, J. and {Heywood}, I. and {Fender}, R. and {Mohamed}, S. and {Sivakoff}, G.~R. and {Saikia}, P. and {Russell}, T.~D. and {Motta}, S. and {Miller-Jones}, J.~C.~A. and {Woudt}, P.~A.},
        title = "{MeerKAT discovery of radio emission from the Vela X-1 bow shock}",
      journal = {\mnras},
         year = 2022,
        month = feb,
       volume = {510},
       number = {1},
        pages = {515-530},
          doi = {10.1093/mnras/stab3395},
archivePrefix = {arXiv},
       eprint = {2111.10159},
 primaryClass = {astro-ph.HE},
       adsurl = {https://ui.adsabs.harvard.edu/abs/2022MNRAS.510..515V}
}

@ARTICLE{2021carotenutob,
       author = {{Carotenuto}, F. and {Corbel}, S. and {Tremou}, E. and {Russell}, T.~D. and {Tzioumis}, A. and {Fender}, R.~P. and {Woudt}, P.~A. and {Motta}, S.~E. and {Miller-Jones}, J.~C.~A. and {Tetarenko}, A.~J. and {Sivakoff}, G.~R.},
        title = "{The hybrid radio/X-ray correlation of the black hole transient MAXI J1348-630}",
      journal = {\mnras},
         year = 2021,
        month = jul,
       volume = {505},
       number = {1},
        pages = {L58-L63},
          doi = {10.1093/mnrasl/slab049},
archivePrefix = {arXiv},
       eprint = {2105.06006},
 primaryClass = {astro-ph.HE},
       adsurl = {https://ui.adsabs.harvard.edu/abs/2021MNRAS.505L..58C}
}

@ARTICLE{2021carotenutoa,
       author = {{Carotenuto}, F. and {Corbel}, S. and {Tremou}, E. and {Russell}, T.~D. and {Tzioumis}, A. and {Fender}, R.~P. and {Woudt}, P.~A. and {Motta}, S.~E. and {Miller-Jones}, J.~C.~A. and {Chauhan}, J. and {Tetarenko}, A.~J. and {Sivakoff}, G.~R. and {Heywood}, I. and {Horesh}, A. and {van der Horst}, A.~J. and {Koerding}, E. and {Mooley}, K.~P.},
        title = "{The black hole transient MAXI J1348-630: evolution of the compact and transient jets during its 2019/2020 outburst}",
      journal = {\mnras},
         year = 2021,
        month = jun,
       volume = {504},
       number = {1},
        pages = {444-468},
          doi = {10.1093/mnras/stab864},
archivePrefix = {arXiv},
       eprint = {2103.12190},
 primaryClass = {astro-ph.HE},
       adsurl = {https://ui.adsabs.harvard.edu/abs/2021MNRAS.504..444C}
}

@ARTICLE{2021motta,
       author = {{Motta}, S.~E. and {Kajava}, J.~J.~E. and {Giustini}, M. and {Williams}, D.~R.~A. and {Del Santo}, M. and {Fender}, R. and {Green}, D.~A. and {Heywood}, I. and {Rhodes}, L. and {Segreto}, A. and {Sivakoff}, G. and {Woudt}, P.~A.},
        title = "{Observations of a radio-bright, X-ray obscured GRS 1915+105}",
      journal = {\mnras},
         year = 2021,
        month = may,
       volume = {503},
       number = {1},
        pages = {152-161},
          doi = {10.1093/mnras/stab511},
archivePrefix = {arXiv},
       eprint = {2101.01187},
 primaryClass = {astro-ph.HE},
       adsurl = {https://ui.adsabs.harvard.edu/abs/2021MNRAS.503..152M}
}

@ARTICLE{2021monageng,
       author = {{Monageng}, I.~M. and {Motta}, S.~E. and {Fender}, R. and {Yu}, W. and {Woudt}, P.~A. and {Tremou}, E. and {Miller-Jones}, J.~C.~A. and {van der Horst}, A.~J.},
        title = "{Radio flaring and dual radio loud/quiet behaviour in the new candidate black hole X-ray binary MAXI J1631-472}",
      journal = {\mnras},
         year = 2021,
        month = mar,
       volume = {501},
       number = {4},
        pages = {5776-5781},
          doi = {10.1093/mnras/stab043},
archivePrefix = {arXiv},
       eprint = {2101.01569},
 primaryClass = {astro-ph.HE},
       adsurl = {https://ui.adsabs.harvard.edu/abs/2021MNRAS.501.5776M}
}

@ARTICLE{2020bright,
       author = {{Bright}, J.~S. and {Fender}, R.~P. and {Motta}, S.~E. and {Williams}, D.~R.~A. and {Moldon}, J. and {Plotkin}, R.~M. and {Miller-Jones}, J.~C.~A. and {Heywood}, I. and {Tremou}, E. and {Beswick}, R. and {Sivakoff}, G.~R. and {Corbel}, S. and {Buckley}, D.~A.~H. and {Homan}, J. and {Gallo}, E. and {Tetarenko}, A.~J. and {Russell}, T.~D. and {Green}, D.~A. and {Titterington}, D. and {Woudt}, P.~A. and {Armstrong}, R.~P. and {Groot}, P.~J. and {Horesh}, A. and {van der Horst}, A.~J. and {K{\"o}rding}, E.~G. and {McBride}, V.~A. and {Rowlinson}, A. and {Wijers}, R.~A.~M.~J.},
        title = "{An extremely powerful long-lived superluminal ejection from the black hole MAXI J1820+070}",
      journal = {Nature Astronomy},
         year = 2020,
        month = mar,
       volume = {4},
        pages = {697-703},
          doi = {10.1038/s41550-020-1023-5},
archivePrefix = {arXiv},
       eprint = {2003.01083},
 primaryClass = {astro-ph.HE},
       adsurl = {https://ui.adsabs.harvard.edu/abs/2020NatAs...4..697B}
}

@ARTICLE{2020williams,
       author = {{Williams}, D.~R.~A. and {Motta}, S.~E. and {Fender}, R. and {Bright}, J. and {Heywood}, I. and {Tremou}, E. and {Woudt}, P. and {Buckley}, D.~A.~H. and {Corbel}, S. and {Coriat}, M. and {Joseph}, T. and {Rhodes}, L. and {Sivakoff}, G.~R. and {van der Horst}, A.~J.},
        title = "{The 2018 outburst of BHXB H1743-322 as seen with MeerKAT}",
      journal = {\mnras},
         year = 2020,
        month = jan,
       volume = {491},
       number = {1},
        pages = {L29-L33},
          doi = {10.1093/mnrasl/slz152},
archivePrefix = {arXiv},
       eprint = {1910.00349},
 primaryClass = {astro-ph.HE},
       adsurl = {https://ui.adsabs.harvard.edu/abs/2020MNRAS.491L..29W}
}

@ARTICLE{2019russell,
       author = {{Russell}, T.~D. and {Tetarenko}, A.~J. and {Miller-Jones}, J.~C.~A. and {Sivakoff}, G.~R. and {Parikh}, A.~S. and {Rapisarda}, S. and {Wijnands}, R. and {Corbel}, S. and {Tremou}, E. and {Altamirano}, D. and {Baglio}, M.~C. and {Ceccobello}, C. and {Degenaar}, N. and {van den Eijnden}, J. and {Fender}, R. and {Heywood}, I. and {Krimm}, H.~A. and {Lucchini}, M. and {Markoff}, S. and {Russell}, D.~M. and {Soria}, R. and {Woudt}, P.~A.},
        title = "{Disk-Jet Coupling in the 2017/2018 Outburst of the Galactic Black Hole Candidate X-Ray Binary MAXI J1535-571}",
      journal = {\apj},
         year = 2019,
        month = oct,
       volume = {883},
       number = {2},
          eid = {198},
        pages = {198},
          doi = {10.3847/1538-4357/ab3d36},
archivePrefix = {arXiv},
       eprint = {1906.00998},
 primaryClass = {astro-ph.HE},
       adsurl = {https://ui.adsabs.harvard.edu/abs/2019ApJ...883..198R}
}

@ARTICLE{2021tremou,
       author = {{Tremou}, Evangelia and {Corbel}, Stephane and {Fender}, Rob and {Woudt}, Patrick and {Miller-Jones}, James and {Motta}, Sara and {Sivakoff}, Gregory R.},
        title = "{MeerKAT observations revealed a rapid rising outburst of the recurrent black hole GX339-4}",
      journal = {The Astronomer's Telegram},
         year = 2021,
        month = jan,
       volume = {14336},
        pages = {1},
       adsurl = {https://ui.adsabs.harvard.edu/abs/2021ATel14336....1T}
}

@ARTICLE{2002zand,
       author = {{in't Zand}, J.~J.~M. and {Miller}, J.~M. and {Oosterbroek}, T. and {Parmar}, A.~N.},
        title = "{Broad-band X-ray measurements of the black hole candidate XTE J1908+094}",
      journal = {\aap},
         year = 2002,
        month = nov,
       volume = {394},
        pages = {553-560},
          doi = {10.1051/0004-6361:20021123},
archivePrefix = {arXiv},
       eprint = {astro-ph/0205535},
 primaryClass = {astro-ph},
       adsurl = {https://ui.adsabs.harvard.edu/abs/2002A&A...394..553I}
}

@ARTICLE{2013zhou,
       author = {{Zhou}, J.~N. and {Liu}, Q.~Z. and {Chen}, Y.~P. and {Li}, J. and {Qu}, J.~L. and {Zhang}, S. and {Gao}, H.~Q. and {Zhang}, Z.},
        title = "{The last three outbursts of H1743-322 observed by RXTE in its latest service phase}",
      journal = {\mnras},
         year = 2013,
        month = may,
       volume = {431},
       number = {3},
        pages = {2285-2293},
          doi = {10.1093/mnras/stt326},
archivePrefix = {arXiv},
       eprint = {1302.5520},
 primaryClass = {astro-ph.HE},
       adsurl = {https://ui.adsabs.harvard.edu/abs/2013MNRAS.431.2285Z}
}

@ARTICLE{2022tremou,
       author = {{Tremou}, Evangelia and {Corbel}, Stephane and {Fender}, Rob and {Woudt}, Patrick and {Miller-Jones}, James and {Motta}, Sara and {Sivakoff}, Gregory R.},
        title = "{Radio detection of the new outburst phase of the recurrent black hole GX 339-4: MeerKAT observations}",
      journal = {The Astronomer's Telegram},
         year = 2022,
        month = aug,
       volume = {15580},
        pages = {1},
       adsurl = {https://ui.adsabs.harvard.edu/abs/2022ATel15580....1T}
}

@ARTICLE{2014curran,
       author = {{Curran}, P.~A. and {Coriat}, M. and {Miller-Jones}, J.~C.~A. and {Armstrong}, R.~P. and {Edwards}, P.~G. and {Sivakoff}, G.~R. and {Woudt}, P. and {Altamirano}, D. and {Belloni}, T.~M. and {Corbel}, S. and {Fender}, R.~P. and {K{\"o}rding}, E.~G. and {Krimm}, H.~A. and {Markoff}, S. and {Migliari}, S. and {Russell}, D.~M. and {Stevens}, J. and {Tzioumis}, T.},
        title = "{The evolving polarized jet of black hole candidate Swift J1745-26}",
      journal = {\mnras},
         year = 2014,
        month = feb,
       volume = {437},
       number = {4},
        pages = {3265-3273},
          doi = {10.1093/mnras/stt2125},
archivePrefix = {arXiv},
       eprint = {1309.4926},
 primaryClass = {astro-ph.HE},
       adsurl = {https://ui.adsabs.harvard.edu/abs/2014MNRAS.437.3265C}
}

@ARTICLE{2013soleri,
       author = {{Soleri}, P. and {Mu{\~n}oz-Darias}, T. and {Motta}, S. and {Belloni}, T. and {Casella}, P. and {M{\'e}ndez}, M. and {Altamirano}, D. and {Linares}, M. and {Wijnands}, R. and {Fender}, R. and {van der Klis}, M.},
        title = "{A complex state transition from the black hole candidate Swift J1753.5-0127}",
      journal = {\mnras},
         year = 2013,
        month = feb,
       volume = {429},
       number = {2},
        pages = {1244-1257},
          doi = {10.1093/mnras/sts405},
archivePrefix = {arXiv},
       eprint = {1211.3537},
 primaryClass = {astro-ph.HE},
       adsurl = {https://ui.adsabs.harvard.edu/abs/2013MNRAS.429.1244S}
}

@ARTICLE{2012reis,
       author = {{Reis}, R.~C. and {Miller}, J.~M. and {Reynolds}, M.~T. and {Fabian}, A.~C. and {Walton}, D.~J.},
        title = "{Suzaku Observation of the Black Hole Candidate Maxi J1836-194 in a Hard/Intermediate Spectral State}",
      journal = {\apj},
         year = 2012,
        month = may,
       volume = {751},
       number = {1},
          eid = {34},
        pages = {34},
          doi = {10.1088/0004-637X/751/1/34},
archivePrefix = {arXiv},
       eprint = {1111.6665},
 primaryClass = {astro-ph.HE},
       adsurl = {https://ui.adsabs.harvard.edu/abs/2012ApJ...751...34R}
}

@ARTICLE{2012ferrigno,
       author = {{Ferrigno}, C. and {Bozzo}, E. and {Del Santo}, M. and {Capitanio}, F.},
        title = "{The first outburst of the black-hole candidate MAXI J1836-194 observed by INTEGRAL, Swift, and RXTE}",
      journal = {\aap},
         year = 2012,
        month = jan,
       volume = {537},
          eid = {L7},
        pages = {L7},
          doi = {10.1051/0004-6361/201118474},
archivePrefix = {arXiv},
       eprint = {1112.1240},
 primaryClass = {astro-ph.GA},
       adsurl = {https://ui.adsabs.harvard.edu/abs/2012A&A...537L...7F}
}

@ARTICLE{2009capitanio,
       author = {{Capitanio}, F. and {Belloni}, T. and {Del Santo}, M. and {Ubertini}, P.},
        title = "{A failed outburst of H1743-322}",
      journal = {\mnras},
         year = 2009,
        month = sep,
       volume = {398},
       number = {3},
        pages = {1194-1200},
          doi = {10.1111/j.1365-2966.2009.15196.x},
archivePrefix = {arXiv},
       eprint = {0906.1137},
 primaryClass = {astro-ph.HE},
       adsurl = {https://ui.adsabs.harvard.edu/abs/2009MNRAS.398.1194C}
}

@ARTICLE{2002wijnands,
       author = {{Wijnands}, Rudy and {Miller}, Jon M.},
        title = "{The New X-Ray Transient SAX J1711.6-3808: Decoupling between Its 3-20 keV Luminosity and Its State Transitions}",
      journal = {\apj},
         year = 2002,
        month = jan,
       volume = {564},
       number = {2},
        pages = {974-980},
          doi = {10.1086/324329},
archivePrefix = {arXiv},
       eprint = {astro-ph/0105182},
 primaryClass = {astro-ph},
       adsurl = {https://ui.adsabs.harvard.edu/abs/2002ApJ...564..974W}
}

@ARTICLE{2010brocksopp,
       author = {{Brocksopp}, C. and {Jonker}, P.~G. and {Maitra}, D. and {Krimm}, H.~A. and {Pooley}, G.~G. and {Ramsay}, G. and {Zurita}, C.},
        title = "{Disentangling jet and disc emission from the 2005 outburst of XTE J1118+480}",
      journal = {\mnras},
         year = 2010,
        month = may,
       volume = {404},
       number = {2},
        pages = {908-916},
          doi = {10.1111/j.1365-2966.2010.16323.x},
archivePrefix = {arXiv},
       eprint = {1001.1965},
 primaryClass = {astro-ph.HE},
       adsurl = {https://ui.adsabs.harvard.edu/abs/2010MNRAS.404..908B}
}

@ARTICLE{2004brocksopp,
       author = {{Brocksopp}, C. and {Bandyopadhyay}, R.~M. and {Fender}, R.~P.},
        title = "{``Soft X-ray transient'' outbursts which are not soft}",
      journal = {\na},
         year = 2004,
        month = may,
       volume = {9},
       number = {4},
        pages = {249-264},
          doi = {10.1016/j.newast.2003.11.002},
archivePrefix = {arXiv},
       eprint = {astro-ph/0311152},
 primaryClass = {astro-ph},
       adsurl = {https://ui.adsabs.harvard.edu/abs/2004NewA....9..249B}
}

@ARTICLE{2001brocksopp,
       author = {{Brocksopp}, C. and {Jonker}, P.~G. and {Fender}, R.~P. and {Groot}, P.~J. and {van der Klis}, M. and {Tingay}, S.~J.},
        title = "{The 1997 hard-state outburst of the X-ray transient GS 1354-64/BW Cir}",
      journal = {\mnras},
         year = 2001,
        month = may,
       volume = {323},
       number = {2},
        pages = {517-528},
          doi = {10.1046/j.1365-8711.2001.04193.x},
archivePrefix = {arXiv},
       eprint = {astro-ph/0011145},
 primaryClass = {astro-ph},
       adsurl = {https://ui.adsabs.harvard.edu/abs/2001MNRAS.323..517B}
}

@ARTICLE{1994harmon,
       author = {{Harmon}, B.~A. and {Wilson}, C.~A. and {Paciesas}, W.~S. and {Pendleton}, G.~N. and {Briggs}, M.~S. and {Rubin}, B.~C. and {Finger}, M.~H. and {Fishman}, G.~J. and {Brock}, M.~N. and {Wilson}, R.~B. and {Meegan}, C.~A.},
        title = "{Observation of GX 339-4 Hard State Outbursts in 1991 and 1992}",
      journal = {\apjl},
         year = 1994,
        month = apr,
       volume = {425},
        pages = {L17},
          doi = {10.1086/187300},
       adsurl = {https://ui.adsabs.harvard.edu/abs/1994ApJ...425L..17H}
}

@ARTICLE{2000hynes,
       author = {{Hynes}, R.~I. and {Mauche}, C.~W. and {Haswell}, C.~A. and {Shrader}, C.~R. and {Cui}, W. and {Chaty}, S.},
        title = "{The X-Ray Transient XTE J1118+480: Multiwavelength Observations of a Low-State Minioutburst}",
      journal = {\apjl},
         year = 2000,
        month = aug,
       volume = {539},
       number = {1},
        pages = {L37-L40},
          doi = {10.1086/312836},
archivePrefix = {arXiv},
       eprint = {astro-ph/0005398},
 primaryClass = {astro-ph},
       adsurl = {https://ui.adsabs.harvard.edu/abs/2000ApJ...539L..37H}
}

@ARTICLE{2002belloni,
       author = {{Belloni}, T. and {Colombo}, A.~P. and {Homan}, J. and {Campana}, S. and {van der Klis}, M.},
        title = "{A low/hard state outburst of XTE J1550-564}",
      journal = {\aap},
         year = 2002,
        month = jul,
       volume = {390},
        pages = {199-204},
          doi = {10.1051/0004-6361:20020703},
archivePrefix = {arXiv},
       eprint = {astro-ph/0205231},
 primaryClass = {astro-ph},
       adsurl = {https://ui.adsabs.harvard.edu/abs/2002A&A...390..199B}
}

@ARTICLE{2004aref,
       author = {{Aref'ev}, V.~A. and {Revnivtsev}, M.~G. and {Lutovinov}, A.~A. and {Sunyaev}, R.~A.},
        title = "{Broadband X-ray Spectrum of XTE J1550-564 during the Outburst of 2003}",
      journal = {Astronomy Letters},
         year = 2004,
        month = oct,
       volume = {30},
        pages = {669-674},
          doi = {10.1134/1.1808053},
archivePrefix = {arXiv},
       eprint = {astro-ph/0404460},
 primaryClass = {astro-ph},
       adsurl = {https://ui.adsabs.harvard.edu/abs/2004AstL...30..669A}
}

@ARTICLE{2005sturner,
       author = {{Sturner}, S.~J. and {Shrader}, C.~R.},
        title = "{XTE J1550-564: INTEGRAL Observations of a Failed Outburst}",
      journal = {\apj},
         year = 2005,
        month = jun,
       volume = {625},
       number = {2},
        pages = {923-930},
          doi = {10.1086/429815},
archivePrefix = {arXiv},
       eprint = {astro-ph/0502498},
 primaryClass = {astro-ph},
       adsurl = {https://ui.adsabs.harvard.edu/abs/2005ApJ...625..923S}
}

@ARTICLE{2013curran,
       author = {{Curran}, P.~A. and {Chaty}, S.},
        title = "{Near-infrared and optical observations of the failed outbursts of black hole binary XTE J1550-564}",
      journal = {\aap},
         year = 2013,
        month = sep,
       volume = {557},
          eid = {A45},
        pages = {A45},
          doi = {10.1051/0004-6361/201321865},
archivePrefix = {arXiv},
       eprint = {1308.2053},
 primaryClass = {astro-ph.HE},
       adsurl = {https://ui.adsabs.harvard.edu/abs/2013A&A...557A..45C}
}

@ARTICLE{2013russelld,
       author = {{Russell}, D.~M. and {Russell}, T.~D. and {Miller-Jones}, J.~C.~A. and {O'Brien}, K. and {Soria}, R. and {Sivakoff}, G.~R. and {Slaven-Blair}, T. and {Lewis}, F. and {Markoff}, S. and {Homan}, J. and {Altamirano}, D. and {Curran}, P.~A. and {Rupen}, M.~P. and {Belloni}, T.~M. and {Cadolle Bel}, M. and {Casella}, P. and {Corbel}, S. and {Dhawan}, V. and {Fender}, R.~P. and {Gallo}, E. and {Gandhi}, P. and {Heinz}, S. and {K{\"o}rding}, E.~G. and {Krimm}, H.~A. and {Maitra}, D. and {Migliari}, S. and {Remillard}, R.~A. and {Sarazin}, C.~L. and {Shahbaz}, T. and {Tudose}, V.},
        title = "{An Evolving Compact Jet in the Black Hole X-Ray Binary MAXI J1836-194}",
      journal = {\apjl},
         year = 2013,
        month = may,
       volume = {768},
       number = {2},
          eid = {L35},
        pages = {L35},
          doi = {10.1088/2041-8205/768/2/L35},
archivePrefix = {arXiv},
       eprint = {1304.3510},
 primaryClass = {astro-ph.HE},
       adsurl = {https://ui.adsabs.harvard.edu/abs/2013ApJ...768L..35R}
}

@ARTICLE{2020russelld,
       author = {{Russell}, D.~M. and {Casella}, P. and {Kalemci}, E. and {Vahdat Motlagh}, A. and {Saikia}, P. and {Pirbhoy}, S.~F. and {Maitra}, D.},
        title = "{The appearance of a compact jet in the soft-intermediate state of 4U 1543-47}",
      journal = {\mnras},
         year = 2020,
        month = jun,
       volume = {495},
       number = {1},
        pages = {182-191},
          doi = {10.1093/mnras/staa1182},
archivePrefix = {arXiv},
       eprint = {2002.08399},
 primaryClass = {astro-ph.HE},
       adsurl = {https://ui.adsabs.harvard.edu/abs/2020MNRAS.495..182R}
}

@ARTICLE{2012james,
       author = {{Miller-Jones}, J.~C.~A. and {Sivakoff}, G.~R. and {Altamirano}, D. and {Coriat}, M. and {Corbel}, S. and {Dhawan}, V. and {Krimm}, H.~A. and {Remillard}, R.~A. and {Rupen}, M.~P. and {Russell}, D.~M. and {Fender}, R.~P. and {Heinz}, S. and {K{\"o}rding}, E.~G. and {Maitra}, D. and {Markoff}, S. and {Migliari}, S. and {Sarazin}, C.~L. and {Tudose}, V.},
        title = "{Disc-jet coupling in the 2009 outburst of the black hole candidate H1743-322}",
      journal = {\mnras},
         year = 2012,
        month = mar,
       volume = {421},
       number = {1},
        pages = {468-485},
          doi = {10.1111/j.1365-2966.2011.20326.x},
archivePrefix = {arXiv},
       eprint = {1201.1678},
 primaryClass = {astro-ph.HE},
       adsurl = {https://ui.adsabs.harvard.edu/abs/2012MNRAS.421..468M}
}

@ARTICLE{2004corbel,
       author = {{Corbel}, S. and {Fender}, R.~P. and {Tomsick}, J.~A. and {Tzioumis}, A.~K. and {Tingay}, S.},
        title = "{On the Origin of Radio Emission in the X-Ray States of XTE J1650-500 during the 2001-2002 Outburst}",
      journal = {\apj},
         year = 2004,
        month = dec,
       volume = {617},
       number = {2},
        pages = {1272-1283},
          doi = {10.1086/425650},
archivePrefix = {arXiv},
       eprint = {astro-ph/0409154},
 primaryClass = {astro-ph},
       adsurl = {https://ui.adsabs.harvard.edu/abs/2004ApJ...617.1272C}
}

@ARTICLE{2021dehaas,
       author = {{de Haas}, S.~E.~M. and {Russell}, T.~D. and {Degenaar}, N. and {Markoff}, S. and {Tetarenko}, A.~J. and {Tetarenko}, B.~E. and {van den Eijnden}, J. and {Miller-Jones}, J.~C.~A. and {Parikh}, A.~S. and {Plotkin}, R.~M. and {Sivakoff}, G.~R.},
        title = "{Disc-jet coupling changes as a possible indicator for outbursts from GX 339-4 remaining within the X-ray hard state}",
      journal = {\mnras},
         year = 2021,
        month = mar,
       volume = {502},
       number = {1},
        pages = {521-540},
          doi = {10.1093/mnras/staa3853},
archivePrefix = {arXiv},
       eprint = {2012.05206},
 primaryClass = {astro-ph.HE},
       adsurl = {https://ui.adsabs.harvard.edu/abs/2021MNRAS.502..521D}
}

@ARTICLE{2022barnier,
       author = {{Barnier}, S. and {Petrucci}, P. -O. and {Ferreira}, J. and {Marcel}, G. and {Belmont}, R. and {Clavel}, M. and {Corbel}, S. and {Coriat}, M. and {Espinasse}, M. and {Henri}, G. and {Malzac}, J. and {Rodriguez}, J.},
        title = "{Clues on jet behavior from simultaneous radio-X-ray fits of GX 339-4}",
      journal = {\aap},
         year = 2022,
        month = jan,
       volume = {657},
          eid = {A11},
        pages = {A11},
          doi = {10.1051/0004-6361/202141182},
archivePrefix = {arXiv},
       eprint = {2109.02895},
 primaryClass = {astro-ph.HE},
       adsurl = {https://ui.adsabs.harvard.edu/abs/2022A&A...657A..11B}
}

@ARTICLE{2012jonker,
       author = {{Jonker}, P.~G. and {Miller-Jones}, J.~C.~A. and {Homan}, J. and {Tomsick}, J. and {Fender}, R.~P. and {Kaaret}, P. and {Markoff}, S. and {Gallo}, E.},
        title = "{The black hole candidate MAXI J1659-152 in and towards quiescence in X-ray and radio}",
      journal = {\mnras},
         year = 2012,
        month = jul,
       volume = {423},
       number = {4},
        pages = {3308-3315},
          doi = {10.1111/j.1365-2966.2012.21116.x},
archivePrefix = {arXiv},
       eprint = {1204.4832},
 primaryClass = {astro-ph.HE},
       adsurl = {https://ui.adsabs.harvard.edu/abs/2012MNRAS.423.3308J}
}

@ARTICLE{1995callanan,
       author = {{Callanan}, Paul J. and {Garcia}, Michael R. and {McClintock}, Jeffrey E. and {Zhao}, Ping and {Remillard}, Ronald A. and {Bailyn}, Charles D. and {Orosz}, Jerome A. and {Harmon}, B. Alan and {Paciesas}, William S.},
        title = "{Observations of the X-Ray Nova GRO J0422+32. I. Outburst and the Decay to Quiescence}",
      journal = {\apj},
         year = 1995,
        month = mar,
       volume = {441},
        pages = {786},
          doi = {10.1086/175402},
       adsurl = {https://ui.adsabs.harvard.edu/abs/1995ApJ...441..786C}
}

@ARTICLE{1997chen,
       author = {{Chen}, Wan and {Shrader}, C.~R. and {Livio}, Mario},
        title = "{The Properties of X-Ray and Optical Light Curves of X-Ray Novae}",
      journal = {\apj},
         year = 1997,
        month = dec,
       volume = {491},
       number = {1},
        pages = {312-338},
          doi = {10.1086/304921},
archivePrefix = {arXiv},
       eprint = {astro-ph/9707138},
 primaryClass = {astro-ph},
       adsurl = {https://ui.adsabs.harvard.edu/abs/1997ApJ...491..312C}
}

@ARTICLE{2013homan,
       author = {{Homan}, Jeroen and {Fridriksson}, Joel K. and {Jonker}, Peter G. and {Russell}, David M. and {Gallo}, Elena and {Kuulkers}, Erik and {Rea}, Nanda and {Altamirano}, Diego},
        title = "{The X-Ray Properties of the Black Hole Transient MAXI J1659-152 in Quiescence}",
      journal = {\apj},
         year = 2013,
        month = sep,
       volume = {775},
       number = {1},
          eid = {9},
        pages = {9},
          doi = {10.1088/0004-637X/775/1/9},
archivePrefix = {arXiv},
       eprint = {1308.2580},
 primaryClass = {astro-ph.HE},
       adsurl = {https://ui.adsabs.harvard.edu/abs/2013ApJ...775....9H}
}

@ARTICLE{2004tomosick,
       author = {{Tomsick}, John A. and {Kalemci}, Emrah and {Kaaret}, Philip},
        title = "{Detection of Low-Hard State Spectral and Timing Signatures from the Black Hole X-Ray Transient XTE J1650-500 at Low X-Ray Luminosities}",
      journal = {\apj},
         year = 2004,
        month = jan,
       volume = {601},
       number = {1},
        pages = {439-449},
          doi = {10.1086/380484},
archivePrefix = {arXiv},
       eprint = {astro-ph/0307458},
 primaryClass = {astro-ph},
       adsurl = {https://ui.adsabs.harvard.edu/abs/2004ApJ...601..439T}
}

@ARTICLE{2003wenfei,
       author = {{Yu}, Wenfei and {Klein-Wolt}, Marc and {Fender}, Rob and {van der Klis}, Michiel},
        title = "{Hard X-Ray Flares Preceding Soft X-Ray Outbursts in Aquila X-1: A Link between Neutron Star and Black Hole State Transitions}",
      journal = {\apjl},
         year = 2003,
        month = may,
       volume = {589},
       number = {1},
        pages = {L33-L36},
          doi = {10.1086/375714},
archivePrefix = {arXiv},
       eprint = {astro-ph/0304134},
 primaryClass = {astro-ph},
       adsurl = {https://ui.adsabs.harvard.edu/abs/2003ApJ...589L..33Y}
}

@ARTICLE{2004wenfei,
       author = {{Yu}, Wenfei and {van der Klis}, Michiel and {Fender}, Rob},
        title = "{The Correlation between Hard X-Ray Peak Flux and Soft X-Ray Peak Flux in the Outburst Rise of Low-Mass X-Ray Binaries}",
      journal = {\apjl},
         year = 2004,
        month = aug,
       volume = {611},
       number = {2},
        pages = {L121-L124},
          doi = {10.1086/423953},
archivePrefix = {arXiv},
       eprint = {astro-ph/0407515},
 primaryClass = {astro-ph},
       adsurl = {https://ui.adsabs.harvard.edu/abs/2004ApJ...611L.121Y}
}

@ARTICLE{2021alabarta,
       author = {{Alabarta}, K. and {Altamirano}, D. and {M{\'e}ndez}, M. and {C{\'u}neo}, V.~A. and {Vincentelli}, F.~M. and {Castro-Segura}, N. and {Garc{\'\i}a}, F. and {Luff}, B. and {Veledina}, A.},
        title = "{Failed-transition outbursts in black hole low-mass X-ray binaries}",
      journal = {\mnras},
         year = 2021,
        month = nov,
       volume = {507},
       number = {4},
        pages = {5507-5522},
          doi = {10.1093/mnras/stab2241},
archivePrefix = {arXiv},
       eprint = {2107.10035},
 primaryClass = {astro-ph.HE},
       adsurl = {https://ui.adsabs.harvard.edu/abs/2021MNRAS.507.5507A}
}

@ARTICLE{2016rushton,
       author = {{Rushton}, A.~P. and {Shaw}, A.~W. and {Fender}, R.~P. and {Altamirano}, D. and {Gandhi}, P. and {Uttley}, P. and {Charles}, P.~A. and {Kolehmainen}, M. and {Anderson}, G.~E. and {Rumsey}, C. and {Titterington}, D.~J.},
        title = "{Disc-jet quenching of the galactic black hole Swift J1753.5-0127}",
      journal = {\mnras},
         year = 2016,
        month = nov,
       volume = {463},
       number = {1},
        pages = {628-634},
          doi = {10.1093/mnras/stw2020},
archivePrefix = {arXiv},
       eprint = {1608.02886},
 primaryClass = {astro-ph.HE},
       adsurl = {https://ui.adsabs.harvard.edu/abs/2016MNRAS.463..628R}
}

@ARTICLE{2020russellt,
       author = {{Russell}, T.~D. and {Lucchini}, M. and {Tetarenko}, A.~J. and {Miller-Jones}, J.~C.~A. and {Sivakoff}, G.~R. and {Krau{\ss}}, F. and {Mulaudzi}, W. and {Baglio}, M.~C. and {Russell}, D.~M. and {Altamirano}, D. and {Ceccobello}, C. and {Corbel}, S. and {Degenaar}, N. and {van den Eijnden}, J. and {Fender}, R. and {Heinz}, S. and {Koljonen}, K.~I.~I. and {Maitra}, D. and {Markoff}, S. and {Migliari}, S. and {Parikh}, A.~S. and {Plotkin}, R.~M. and {Rupen}, M. and {Sarazin}, C. and {Soria}, R. and {Wijnands}, R.},
        title = "{Rapid compact jet quenching in the Galactic black hole candidate X-ray binary MAXI J1535-571}",
      journal = {\mnras},
         year = 2020,
        month = nov,
       volume = {498},
       number = {4},
        pages = {5772-5785},
          doi = {10.1093/mnras/staa2650},
archivePrefix = {arXiv},
       eprint = {2008.11216},
 primaryClass = {astro-ph.HE},
       adsurl = {https://ui.adsabs.harvard.edu/abs/2020MNRAS.498.5772R}
}

@ARTICLE{2014russell,
       author = {{Russell}, T.~D. and {Soria}, R. and {Miller-Jones}, J.~C.~A. and {Curran}, P.~A. and {Markoff}, S. and {Russell}, D.~M. and {Sivakoff}, G.~R.},
        title = "{The accretion-ejection coupling in the black hole candidate X-ray binary MAXI J1836-194}",
      journal = {\mnras},
         year = 2014,
        month = apr,
       volume = {439},
       number = {2},
        pages = {1390-1402},
          doi = {10.1093/mnras/stt2498},
archivePrefix = {arXiv},
       eprint = {1312.5822},
 primaryClass = {astro-ph.HE},
       adsurl = {https://ui.adsabs.harvard.edu/abs/2014MNRAS.439.1390R}
}

@ARTICLE{1999fender,
       author = {{Fender}, Robert and {Corbel}, St{\'e}phane and {Tzioumis}, Tasso and {McIntyre}, Vince and {Campbell-Wilson}, Duncan and {Nowak}, Mike and {Sood}, Ravi and {Hunstead}, Richard and {Harmon}, Alan and {Durouchoux}, Philippe and {Heindl}, William},
        title = "{Quenching of the Radio Jet during the X-Ray High State of GX 339-4}",
      journal = {\apjl},
         year = 1999,
        month = jul,
       volume = {519},
       number = {2},
        pages = {L165-L168},
          doi = {10.1086/312128},
archivePrefix = {arXiv},
       eprint = {astro-ph/9905121},
 primaryClass = {astro-ph},
       adsurl = {https://ui.adsabs.harvard.edu/abs/1999ApJ...519L.165F}
}

@ARTICLE{2000corbel,
       author = {{Corbel}, S. and {Fender}, R.~P. and {Tzioumis}, A.~K. and {Nowak}, M. and {McIntyre}, V. and {Durouchoux}, P. and {Sood}, R.},
        title = "{Coupling of the X-ray and radio emission in the black hole candidate and compact jet source GX 339-4}",
      journal = {\aap},
         year = 2000,
        month = jul,
       volume = {359},
        pages = {251-268},
archivePrefix = {arXiv},
       eprint = {astro-ph/0003460},
 primaryClass = {astro-ph},
       adsurl = {https://ui.adsabs.harvard.edu/abs/2000A&A...359..251C}
}

@ARTICLE{2021corbel,
       author = {{Corbel}, S. and {Tzioumis}, T. and {Tremou}, L. and {Carotenuto}, F.},
        title = "{GX 339-4 in transition back to the hard state with the compact jets in formation}",
      journal = {The Astronomer's Telegram},
         year = 2021,
        month = oct,
       volume = {14953},
        pages = {1},
       adsurl = {https://ui.adsabs.harvard.edu/abs/2021ATel14953....1C}
}

@ARTICLE{1995harmon,
       author = {{Harmon}, B.~A. and {Wilson}, C.~A. and {Zhang}, S.~N. and {Paciesas}, W.~S. and {Fishman}, G.~J. and {Hjellming}, R.~M. and {Rupen}, M.~P. and {Scott}, D.~M. and {Briggs}, M.~S. and {Rubin}, B.~C.},
        title = "{Correlations between X-ray outbursts and relativistic ejections in the X-ray transient GRO J1655 - 40}",
      journal = {\nat},
         year = 1995,
        month = apr,
       volume = {374},
       number = {6524},
        pages = {703-706},
          doi = {10.1038/374703a0},
       adsurl = {https://ui.adsabs.harvard.edu/abs/1995Natur.374..703H}
}

@ARTICLE{1999rodriguez,
       author = {{Rodr{\'\i}guez}, L.~F. and {Mirabel}, I.~F.},
        title = "{Repeated Relativistic Ejections in GRS 1915+105}",
      journal = {\apj},
         year = 1999,
        month = jan,
       volume = {511},
       number = {1},
        pages = {398-404},
          doi = {10.1086/306642},
archivePrefix = {arXiv},
       eprint = {astro-ph/9808341},
 primaryClass = {astro-ph},
       adsurl = {https://ui.adsabs.harvard.edu/abs/1999ApJ...511..398R}
}

@ARTICLE{2000willms,
       author = {{Wilms}, J. and {Allen}, A. and {McCray}, R.},
        title = "{On the Absorption of X-Rays in the Interstellar Medium}",
      journal = {\apj},
         year = 2000,
        month = oct,
       volume = {542},
       number = {2},
        pages = {914-924},
          doi = {10.1086/317016},
archivePrefix = {arXiv},
       eprint = {astro-ph/0008425},
 primaryClass = {astro-ph},
       adsurl = {https://ui.adsabs.harvard.edu/abs/2000ApJ...542..914W}
}

@ARTICLE{2018tremouatel,
       author = {{Tremou}, Evangelia and {Corbel}, Stephane and {Fender}, Rob and {Woudt}, Patrick and {Miller-Jones}, James and {Girard}, Julien},
        title = "{MeerKAT observations revealed a rising outburst of the recurrent black hole GX 339-4}",
      journal = {The Astronomer's Telegram},
         year = 2018,
        month = dec,
       volume = {12287},
        pages = {1},
       adsurl = {https://ui.adsabs.harvard.edu/abs/2018ATel12287....1T}
}

@ARTICLE{2020tremou,
       author = {{Tremou}, E. and {Corbel}, S. and {Fender}, R.~P. and {Woudt}, P.~A. and {Miller-Jones}, J.~C.~A. and {Motta}, S.~E. and {Heywood}, I. and {Armstrong}, R.~P. and {Groot}, P. and {Horesh}, A. and {van der Horst}, A.~J. and {Koerding}, E. and {Mooley}, K.~P. and {Rowlinson}, A. and {Wijers}, R.~A.~M.~J.},
        title = "{Radio \& X-ray detections of GX 339-4 in quiescence using MeerKAT and Swift}",
      journal = {\mnras},
         year = 2020,
        month = mar,
       volume = {493},
       number = {1},
        pages = {L132-L137},
          doi = {10.1093/mnrasl/slaa019},
archivePrefix = {arXiv},
       eprint = {2002.01522},
 primaryClass = {astro-ph.HE},
       adsurl = {https://ui.adsabs.harvard.edu/abs/2020MNRAS.493L.132T}
}

@ARTICLE{2018tasse,
       author = {{Tasse}, C. and {Hugo}, B. and {Mirmont}, M. and {Smirnov}, O. and
        {Atemkeng}, M. and {Bester}, L. and {Hardcastle}, M.~J. and
        {Lakhoo}, R. and {Perkins}, S. and {Shimwell}, T.},
        title = "{Faceting for direction-dependent spectral deconvolution}",
      journal = {\aap},
         year = 2018,
        month = Apr,
       volume = {611},
          doi = {10.1051/0004-6361/201731474},
       adsurl = {https://ui.adsabs.harvard.edu/#abs/2018A&A...611A..87T}
}

@ARTICLE{2015saikia,
       author = {{Saikia}, Payaswini and {K{\"o}rding}, Elmar and {Falcke}, Heino},
        title = "{The Fundamental Plane of black hole activity in the optical band}",
      journal = {\mnras},
         year = 2015,
        month = jul,
       volume = {450},
       number = {3},
        pages = {2317-2326},
          doi = {10.1093/mnras/stv731},
archivePrefix = {arXiv},
       eprint = {1504.00363},
 primaryClass = {astro-ph.HE},
       adsurl = {https://ui.adsabs.harvard.edu/abs/2015MNRAS.450.2317S}
}

@ARTICLE{2018saikia,
       author = {{Saikia}, Payaswini and {K{\"o}rding}, Elmar and {Coppejans}, Deanne L. and {Falcke}, Heino and {Williams}, David and {Baldi}, Ranieri D. and {Mchardy}, Ian and {Beswick}, Rob},
        title = "{15-GHz radio emission from nearby low-luminosity active galactic nuclei}",
      journal = {\aap},
         year = 2018,
        month = sep,
       volume = {616},
          eid = {A152},
        pages = {A152},
          doi = {10.1051/0004-6361/201833233},
archivePrefix = {arXiv},
       eprint = {1805.06696},
 primaryClass = {astro-ph.HE},
       adsurl = {https://ui.adsabs.harvard.edu/abs/2018A&A...616A.152S}
}

@ARTICLE{2004falcke,
       author = {{Falcke}, H. and {K{\"o}rding}, E. and {Markoff}, S.},
        title = "{A scheme to unify low-power accreting black holes. Jet-dominated accretion flows and the radio/X-ray correlation}",
      journal = {\aap},
         year = 2004,
        month = feb,
       volume = {414},
        pages = {895-903},
          doi = {10.1051/0004-6361:20031683},
archivePrefix = {arXiv},
       eprint = {astro-ph/0305335},
 primaryClass = {astro-ph},
       adsurl = {https://ui.adsabs.harvard.edu/abs/2004A&A...414..895F}
}

@ARTICLE{2011russell,
       author = {{Russell}, D.~M. and {Miller-Jones}, J.~C.~A. and {Maccarone}, T.~J. and {Yang}, Y.~J. and {Fender}, R.~P. and {Lewis}, F.},
        title = "{Testing the Jet Quenching Paradigm with an Ultradeep Observation of a Steadily Soft State Black Hole}",
      journal = {\apjl},
         year = 2011,
        month = sep,
       volume = {739},
       number = {1},
          eid = {L19},
        pages = {L19},
          doi = {10.1088/2041-8205/739/1/L19},
archivePrefix = {arXiv},
       eprint = {1106.0723},
 primaryClass = {astro-ph.HE},
       adsurl = {https://ui.adsabs.harvard.edu/abs/2011ApJ...739L..19R}
}

@ARTICLE{2019tetarenko,
       author = {{Tetarenko}, A.~J. and {Sivakoff}, G.~R. and {Miller-Jones}, J.~C.~A. and {Bremer}, M. and {Mooley}, K.~P. and {Fender}, R.~P. and {Rumsey}, C. and {Bahramian}, A. and {Altamirano}, D. and {Heinz}, S. and {Maitra}, D. and {Markoff}, S.~B. and {Migliari}, S. and {Rupen}, M.~P. and {Russell}, D.~M. and {Russell}, T.~D. and {Sarazin}, C.~L.},
        title = "{Tracking the variable jets of V404 Cygni during its 2015 outburst}",
      journal = {\mnras},
         year = 2019,
        month = jan,
       volume = {482},
       number = {3},
        pages = {2950-2972},
          doi = {10.1093/mnras/sty2853},
archivePrefix = {arXiv},
       eprint = {1810.05709},
 primaryClass = {astro-ph.HE},
       adsurl = {https://ui.adsabs.harvard.edu/abs/2019MNRAS.482.2950T}
}

@ARTICLE{2010corbel,
       author = {{Corbel}, S. and {Broderick}, J. and {Calvelo}, D. and {Kaaret}, P. and {Brocksopp}, C. and {Tomsick}, J. and {Orosz}, J. and {Coriat}, M. and {Fender}, R. and {Tzioumis}, T.},
        title = "{Witnessing with the ATCA the interaction of a relativistic jet from GX 339-4 with the interstellar medium}",
      journal = {The Astronomer's Telegram},
         year = 2010,
        month = jul,
       volume = {2745},
        pages = {1},
       adsurl = {https://ui.adsabs.harvard.edu/abs/2010ATel.2745....1C}
}

@ARTICLE{2016patruno,
       author = {{Patruno}, A. and {Maitra}, D. and {Curran}, P.~A. and {D'Angelo}, C. and {Fridriksson}, J.~K. and {Russell}, D.~M. and {Middleton}, M. and {Wijnands}, R.},
        title = "{The Reflares and Outburst Evolution in the Accreting Millisecond Pulsar SAX J1808.4-3658: A Disk Truncated Near Co-Rotation?}",
      journal = {\apj},
         year = 2016,
        month = feb,
       volume = {817},
       number = {2},
          eid = {100},
        pages = {100},
          doi = {10.3847/0004-637X/817/2/100},
archivePrefix = {arXiv},
       eprint = {1504.05048},
 primaryClass = {astro-ph.HE},
       adsurl = {https://ui.adsabs.harvard.edu/abs/2016ApJ...817..100P}
}

@ARTICLE{2001shahbaz,
       author = {{Shahbaz}, T. and {Fender}, R. and {Charles}, P.~A.},
        title = "{VLT optical observations of V821 Ara(=GX339-4) in an extended ``off'' state}",
      journal = {\aap},
         year = 2001,
        month = sep,
       volume = {376},
        pages = {L17-L21},
          doi = {10.1051/0004-6361:20011042},
archivePrefix = {arXiv},
       eprint = {astro-ph/0107455},
 primaryClass = {astro-ph},
       adsurl = {https://ui.adsabs.harvard.edu/abs/2001A&A...376L..17S}
}

@ARTICLE{2016parker,
       author = {{Parker}, M.~L. and {Tomsick}, J.~A. and {Kennea}, J.~A. and {Miller}, J.~M. and {Harrison}, F.~A. and {Barret}, D. and {Boggs}, S.~E. and {Christensen}, F.~E. and {Craig}, W.~W. and {Fabian}, A.~C. and {F{\"u}rst}, F. and {Grinberg}, V. and {Hailey}, C.~J. and {Romano}, P. and {Stern}, D. and {Walton}, D.~J. and {Zhang}, W.~W.},
        title = "{NuSTAR and Swift Observations of the Very High State in GX 339-4: Weighing the Black Hole with X-Rays}",
      journal = {\apjl},
         year = 2016,
        month = apr,
       volume = {821},
       number = {1},
          eid = {L6},
        pages = {L6},
          doi = {10.3847/2041-8205/821/1/L6},
archivePrefix = {arXiv},
       eprint = {1603.03777},
 primaryClass = {astro-ph.HE},
       adsurl = {https://ui.adsabs.harvard.edu/abs/2016ApJ...821L...6P}
}

@ARTICLE{2003hynes,
       author = {{Hynes}, R.~I. and {Steeghs}, D. and {Casares}, J. and {Charles}, P.~A. and {O'Brien}, K.},
        title = "{Dynamical Evidence for a Black Hole in GX 339-4}",
      journal = {\apjl},
         year = 2003,
        month = feb,
       volume = {583},
       number = {2},
        pages = {L95-L98},
          doi = {10.1086/368108},
archivePrefix = {arXiv},
       eprint = {astro-ph/0301127},
 primaryClass = {astro-ph},
       adsurl = {https://ui.adsabs.harvard.edu/abs/2003ApJ...583L..95H}
}

@ARTICLE{2011shidatsu,
       author = {{Shidatsu}, Megumi and {Ueda}, Yoshihiro and {Tazaki}, Fumie and {Yoshikawa}, Tatsuhito and {Nagayama}, Takahiro and {Nagata}, Tetsuya and {Oi}, Nagisa and {Yamaoka}, Kazutaka and {Takahashi}, Hiromitsu and {Kubota}, Aya and {Cottam}, Jean and {Remillard}, Ronald and {Negoro}, Hitoshi},
        title = "{X-Ray and Near-Infrared Observations of GX 339-4 in the Low/Hard State with Suzaku and IRSF}",
      journal = {\pasj},
         year = 2011,
        month = nov,
       volume = {63},
        pages = {S785-S801},
          doi = {10.1093/pasj/63.sp3.S785},
archivePrefix = {arXiv},
       eprint = {1105.3586},
 primaryClass = {astro-ph.HE},
       adsurl = {https://ui.adsabs.harvard.edu/abs/2011PASJ...63S.785S}
}

@ARTICLE{2008munoz,
       author = {{Mu{\~n}oz-Darias}, T. and {Casares}, J. and {Mart{\'\i}nez-Pais}, I.~G.},
        title = "{On the masses and evolutionary status of the black hole binary GX 339-4: a twin system of XTE J1550-564?}",
      journal = {\mnras},
         year = 2008,
        month = apr,
       volume = {385},
       number = {4},
        pages = {2205-2209},
          doi = {10.1111/j.1365-2966.2008.12987.x},
archivePrefix = {arXiv},
       eprint = {0801.3268},
 primaryClass = {astro-ph},
       adsurl = {https://ui.adsabs.harvard.edu/abs/2008MNRAS.385.2205M}
}

@ARTICLE{1992callanan,
       author = {{Callanan}, P.~J. and {Charles}, P.~A. and {Honey}, W.~B. and {Thorstensen}, J.~R.},
        title = "{The 14.8-h orbital period of GX 339-4.}",
      journal = {\mnras},
         year = 1992,
        month = nov,
       volume = {259},
        pages = {395-400},
          doi = {10.1093/mnras/259.2.395},
       adsurl = {https://ui.adsabs.harvard.edu/abs/1992MNRAS.259..395C}
}

@ARTICLE{2017fender,
       author = {{Fender}, R. and {Woudt}, P.~A. and {Armstrong}, R. and {Groot}, P. and
         {McBride}, V. and {Miller-Jones}, J. and {Mooley}, K. and
         {Stappers}, B. and {Wijers}, R. and {Bietenholz}, M. and {Blyth}, S. and
         {Bottcher}, M. and {Buckley}, D. and {Charles}, P. and {Chomiuk}, L. and
         {Coppejans}, D. and {Corbel}, S. and {Coriat}, M. and {Daigne}, F. and
         {de Blok}, W.~J.~G. and {Falcke}, H. and {Girard}, J. and
         {Heywood}, I. and {Horesh}, A. and {Horrell}, J. and {Jonker}, P. and
         {Joseph}, T. and {Kamble}, A. and {Knigge}, C. and {Koerding}, E. and
         {Kotze}, M. and {Kouveliotou}, C. and {Lynch}, C. and {Maccarone}, T. and
         {Meintjes}, P. and {Migliari}, S. and {Murphy}, T. and {Nagayama}, T. and
         {Nelemans}, G. and {Nicholson}, G. and {O'Brien}, T. and
         {Oodendaal}, A. and {Oozeer}, N. and {Osborne}, J. and
         {Perez-Torres}, M. and {Ratcliffe}, S. and {Ribeiro}, V. and {Rol}, E. and
         {Rushton}, A. and {Scaife}, A. and {Schurch}, M. and {Sivakoff}, G. and
         {Staley}, T. and {Steeghs}, D. and {Stewart}, I. and {Swinbank}, J. and
         {van der Heyden}, K. and {van der Horst}, A. and {van Soelen}, B. and
         {Vergani}, S. and {Warner}, B. and {Wiersema}, K.},
        title = "{ThunderKAT: The MeerKAT Large Survey Project for Image-Plane Radio Transients}",
      journal = {arXiv e-prints},
         year = "2017",
        month = "Nov",
          eid = {arXiv:1711.04132},
        pages = {arXiv:1711.04132},
archivePrefix = {arXiv},
       eprint = {1711.04132},
 primaryClass = {astro-ph.HE},
       adsurl = {https://ui.adsabs.harvard.edu/abs/2017arXiv171104132F}
}

@ARTICLE{1995narayan,
       author = {{Narayan}, Ramesh and {Yi}, Insu},
        title = "{Advection-dominated Accretion: Underfed Black Holes and Neutron Stars}",
      journal = {\apj},
         year = 1995,
        month = oct,
       volume = {452},
        pages = {710},
          doi = {10.1086/176343},
archivePrefix = {arXiv},
       eprint = {astro-ph/9411059},
 primaryClass = {astro-ph},
       adsurl = {https://ui.adsabs.harvard.edu/abs/1995ApJ...452..710N}
}

@ARTICLE{2004fender,
       author = {{Fender}, R.~P. and {Belloni}, T.~M. and {Gallo}, E.},
        title = "{Towards a unified model for black hole X-ray binary jets}",
      journal = {\mnras},
         year = 2004,
        month = dec,
       volume = {355},
       number = {4},
        pages = {1105-1118},
          doi = {10.1111/j.1365-2966.2004.08384.x},
archivePrefix = {arXiv},
       eprint = {astro-ph/0409360},
 primaryClass = {astro-ph},
       adsurl = {https://ui.adsabs.harvard.edu/abs/2004MNRAS.355.1105F}
}

@INCOLLECTION{2010belloni,
       author = {{Belloni}, T.~M.},
        title = "{States and Transitions in Black Hole Binaries}",
    booktitle = {Lecture Notes in Physics, Berlin Springer Verlag},
         year = 2010,
       editor = {{Belloni}, Tomaso},
       volume = {794},
        pages = {53},
          doi = {10.1007/978-3-540-76937-8_3},
       adsurl = {https://ui.adsabs.harvard.edu/abs/2010LNP...794...53B}
}

@ARTICLE{2000dhawan,
       author = {{Dhawan}, V. and {Mirabel}, I.~F. and {Rodr{\'\i}guez}, L.~F.},
        title = "{AU-Scale Synchrotron Jets and Superluminal Ejecta in GRS 1915+105}",
      journal = {\apj},
         year = 2000,
        month = nov,
       volume = {543},
       number = {1},
        pages = {373-385},
          doi = {10.1086/317088},
archivePrefix = {arXiv},
       eprint = {astro-ph/0006086},
 primaryClass = {astro-ph},
       adsurl = {https://ui.adsabs.harvard.edu/abs/2000ApJ...543..373D}
}

@ARTICLE{2005markoff,
       author = {{Markoff}, Sera and {Nowak}, Michael A. and {Wilms}, J{\"o}rn},
        title = "{Going with the Flow: Can the Base of Jets Subsume the Role of Compact Accretion Disk Coronae?}",
      journal = {\apj},
         year = 2005,
        month = dec,
       volume = {635},
       number = {2},
        pages = {1203-1216},
          doi = {10.1086/497628},
archivePrefix = {arXiv},
       eprint = {astro-ph/0509028},
 primaryClass = {astro-ph},
       adsurl = {https://ui.adsabs.harvard.edu/abs/2005ApJ...635.1203M}
}

@ARTICLE{2012ponti,
       author = {{Ponti}, G. and {Fender}, R.~P. and {Begelman}, M.~C. and {Dunn}, R.~J.~H. and {Neilsen}, J. and {Coriat}, M.},
        title = "{Ubiquitous equatorial accretion disc winds in black hole soft states}",
      journal = {\mnras},
         year = 2012,
        month = may,
       volume = {422},
       number = {1},
        pages = {L11-L15},
          doi = {10.1111/j.1745-3933.2012.01224.x},
archivePrefix = {arXiv},
       eprint = {1201.4172},
 primaryClass = {astro-ph.HE},
       adsurl = {https://ui.adsabs.harvard.edu/abs/2012MNRAS.422L..11P}
}

@ARTICLE{2003corbel,
       author = {{Corbel}, S. and {Nowak}, M.~A. and {Fender}, R.~P. and
         {Tzioumis}, A.~K. and {Markoff}, S.},
        title = "{Radio/X-ray correlation in the low/hard state of GX 339-4}",
      journal = {\aap},
         year = "2003",
        month = "Mar",
       volume = {400},
        pages = {1007-1012},
          doi = {10.1051/0004-6361:20030090},
archivePrefix = {arXiv},
       eprint = {astro-ph/0301436},
 primaryClass = {astro-ph},
       adsurl = {https://ui.adsabs.harvard.edu/abs/2003A&A...400.1007C}
}

@ARTICLE{2013corbel,
       author = {{Corbel}, S. and {Coriat}, M. and {Brocksopp}, C. and {Tzioumis}, A.~K. and {Fender}, R.~P. and {Tomsick}, J.~A. and {Buxton}, M.~M. and {Bailyn}, C.~D.},
        title = "{The `universal' radio/X-ray flux correlation: the case study of the black hole GX 339-4}",
      journal = {\mnras},
         year = 2013,
        month = jan,
       volume = {428},
       number = {3},
        pages = {2500-2515},
          doi = {10.1093/mnras/sts215},
archivePrefix = {arXiv},
       eprint = {1211.1600},
 primaryClass = {astro-ph.HE},
       adsurl = {https://ui.adsabs.harvard.edu/abs/2013MNRAS.428.2500C}
}

@ARTICLE{2011coriat,
       author = {{Coriat}, M. and {Corbel}, S. and {Prat}, L. and
         {Miller-Jones}, J.~C.~A. and {Cseh}, D. and {Tzioumis}, A.~K. and
         {Brocksopp}, C. and {Rodriguez}, J. and {Fender}, R.~P. and
         {Sivakoff}, G.~R.},
        title = "{Radiatively efficient accreting black holes in the hard state: the case study of H1743-322}",
      journal = {\mnras},
         year = "2011",
        month = "Jun",
       volume = {414},
       number = {1},
        pages = {677-690},
          doi = {10.1111/j.1365-2966.2011.18433.x},
archivePrefix = {arXiv},
       eprint = {1101.5159},
 primaryClass = {astro-ph.HE},
       adsurl = {https://ui.adsabs.harvard.edu/abs/2011MNRAS.414..677C}
}

@ARTICLE{2003fender,
       author = {{Fender}, R.~P. and {Gallo}, E. and {Jonker}, P.~G.},
        title = "{Jet-dominated states: an alternative to advection across black hole event horizons in `quiescent' X-ray binaries}",
      journal = {\mnras},
         year = "2003",
        month = "Aug",
       volume = {343},
       number = {4},
        pages = {L99-L103},
          doi = {10.1046/j.1365-8711.2003.06950.x},
archivePrefix = {arXiv},
       eprint = {astro-ph/0306614},
 primaryClass = {astro-ph},
       adsurl = {https://ui.adsabs.harvard.edu/abs/2003MNRAS.343L..99F}
}

@ARTICLE{2003gallo,
       author = {{Gallo}, E. and {Fender}, R.~P. and {Pooley}, G.~G.},
        title = "{A universal radio-X-ray correlation in low/hard state black hole binaries}",
      journal = {\mnras},
         year = "2003",
        month = "Sep",
       volume = {344},
       number = {1},
        pages = {60-72},
          doi = {10.1046/j.1365-8711.2003.06791.x},
archivePrefix = {arXiv},
       eprint = {astro-ph/0305231},
 primaryClass = {astro-ph},
       adsurl = {https://ui.adsabs.harvard.edu/abs/2003MNRAS.344...60G}
}

@ARTICLE{1998hannikainen,
       author = {{Hannikainen}, D.~C. and {Hunstead}, R.~W. and {Campbell-Wilson}, D. and
         {Sood}, R.~K.},
        title = "{MOST radio monitoring of GX 339-4}",
      journal = {\aap},
         year = "1998",
        month = "Sep",
       volume = {337},
        pages = {460-464},
archivePrefix = {arXiv},
       eprint = {astro-ph/9805332},
 primaryClass = {astro-ph},
       adsurl = {https://ui.adsabs.harvard.edu/abs/1998A&A...337..460H}
}

@ARTICLE{2004hynes,
       author = {{Hynes}, R.~I. and {Steeghs}, D. and {Casares}, J. and {Charles}, P.~A. and
         {O'Brien}, K.},
        title = "{The Distance and Interstellar Sight Line to GX 339-4}",
      journal = {\apj},
         year = "2004",
        month = "Jul",
       volume = {609},
       number = {1},
        pages = {317-324},
          doi = {10.1086/421014},
archivePrefix = {arXiv},
       eprint = {astro-ph/0402408},
 primaryClass = {astro-ph},
       adsurl = {https://ui.adsabs.harvard.edu/abs/2004ApJ...609..317H}
}

@INPROCEEDINGS{2011bertin,
       author = {{Bertin}, E.},
        title = "{Automated Morphometry with SExtractor and PSFEx}",
    booktitle = {Astronomical Data Analysis Software and Systems XX},
         year = 2011,
       editor = {{Evans}, I.~N. and {Accomazzi}, A. and {Mink}, D.~J. and {Rots}, A.~H.},
       series = {Astronomical Society of the Pacific Conference Series},
       volume = {442},
        month = jul,
        pages = {435},
       adsurl = {https://ui.adsabs.harvard.edu/abs/2011ASPC..442..435B}
}


\onecolumn

\newpage
\appendix

\section{Some extra material}

\begin{center}

\begin{longtable}{|l|l|l|l|l||l|l|l|}
\caption{Radio (MeerKAT \& ATCA) observations of GX 339--4. Epochs noted with a star sign ``*" represent upper limits during the quiescence phase of the source resulted in non detections. The complete table is provided in the supplementary material.} \\

\hline \multicolumn{1}{|c|}{\textbf{UTC}} & \multicolumn{1}{|c|}{\textbf{MJD}} & \multicolumn{1}{c|}{\textbf{Telescope}} & \multicolumn{1}{c|}{\textbf{Frequency}} & \multicolumn{1}{c|}{\textbf{S$_{\nu}$ core}} & \multicolumn{1}{c|}{\textbf{$\alpha_{core}$}} & \multicolumn{1}{c|}{\textbf{S$_{\nu}$ ejecta}} & \multicolumn{1}{c|}{\textbf{$\alpha_{ejecta}$}} \\ 
\multicolumn{1}{|c|}{\textbf{--}} &\multicolumn{1}{|c|}{\textbf{--}} & \multicolumn{1}{c|}{\textbf{--}} & \multicolumn{1}{c|}{\textbf{ (GHz) }} & \multicolumn{1}{c|}{\textbf{($\mu$Jy)}} & \multicolumn{1}{c|}{--} & \multicolumn{1}{c|}{\textbf{($\mu$Jy)}} & \multicolumn{1}{c|}{--} \\ \hline 
\endfirsthead

\multicolumn{8}{c}%
{{\bfseries \tablename\ \thetable{} -- continued from previous page}} \\
\hline \multicolumn{1}{|c|}{\textbf{UTC}} & \multicolumn{1}{|c|}{\textbf{MJD}} & \multicolumn{1}{c|}{\textbf{Telescope}} & \multicolumn{1}{c|}{\textbf{Frequency}} & \multicolumn{1}{c|}{\textbf{S$_{\nu}$ core}} & \multicolumn{1}{c|}{\textbf{$\alpha_{core}$}} & \multicolumn{1}{c|}{\textbf{S$_{\nu}$ ejecta}} & \multicolumn{1}{c|}{\textbf{$\alpha_{ejecta}$}} \\ 
\multicolumn{1}{|c|}{\textbf{--}} &\multicolumn{1}{|c|}{\textbf{--}} & \multicolumn{1}{c|}{\textbf{--}} & \multicolumn{1}{c|}{\textbf{ (GHz) }} & \multicolumn{1}{c|}{\textbf{($\mu$Jy)}} & \multicolumn{1}{c|}{--} & \multicolumn{1}{c|}{\textbf{($\mu$Jy)}} & \multicolumn{1}{c|}{--} \\ \hline 
\endhead

\endfoot

\hline
\endlastfoot
*2018-09-08 &			58369& MeerKAT &   1.28&   120.20~$\pm$~40.1    & 		&	&                                         \\
*2018-09-14& 			58375& MeerKAT &   1.28&   78.60~$\pm$~26.2     & 		&	&                                        \\
...& 			...& .. &   ...&   ...    & 		&	&                                        \\
2023-09-23    & 60210  &	MeerKAT &   1.28&   369.14    ~$\pm$ ~28.0                 & 		&    &                                   
\label{tab:obs}
\end{longtable}
\end{center}

\begin{center}
\begin{longtable}{|l|l|l|l|c||l|l|}
\caption{\textit{Swift}/XRT  observations of GX 339--4. The complete table is provided in the supplementary material.} \\

\hline \multicolumn{1}{|c|}{\textbf{UTC}} & \multicolumn{1}{|c|}{\textbf{MJD}} & \multicolumn{1}{c|}{\textbf{ObsID}} & \multicolumn{1}{c|}{\textbf{Exposure}} & \multicolumn{1}{c|}{\textbf{Spectral}} & \multicolumn{1}{c|}{\textbf{Unabsorbed flux}}   \\ 
\multicolumn{1}{|c|}{} &\multicolumn{1}{|c|}{} & \multicolumn{1}{c|}{} & \multicolumn{1}{c|}{\textbf{ time (s)}} & \multicolumn{1}{c|}{\textbf{State}} & \multicolumn{1}{c|}{\textbf{3-9 keV (erg cm$^{-2}$ sec$^{-1}$)}}   \\ \hline 
\endfirsthead

\multicolumn{6}{c}%
{{\bfseries \tablename\ \thetable{} -- continued from previous page}} \\
\hline \multicolumn{1}{|c|}{\textbf{UTC}} & \multicolumn{1}{|c|}{\textbf{MJD}} & \multicolumn{1}{c|}{\textbf{ObsID}} & \multicolumn{1}{c|}{\textbf{Exposure}} & \multicolumn{1}{c|}{\textbf{Spectral}} & \multicolumn{1}{c|}{\textbf{Unabsorbed flux}}   \\ 
\multicolumn{1}{|c|}{} &\multicolumn{1}{|c|}{} & \multicolumn{1}{c|}{} & \multicolumn{1}{c|}{\textbf{ time (s)}} & \multicolumn{1}{c|}{\textbf{State}} & \multicolumn{1}{c|}{\textbf{3-9 keV (erg cm$^{-2}$ sec$^{-1}$)}} \\ \hline
\endhead

\hline \multicolumn{6}{|c|}{{Continued on next page}} \\ \hline
\endfoot

\hline
\endlastfoot
2019-01-21	&	58504 &32898182     & 969.6    & HS &  4.03E-10 ~$\pm$~  4.03E-11      \\   
2019-01-22	&	58505 &32898183     & 1821.0   & HS &  3.37E-10 ~$\pm$~  3.37E-11      \\   
...	&	... &...     & ...   & ... &  ... \\
2023-09-08 	 &	 60195  &14052167&	1300	&	HS &	9.20E-12 ~$\pm$~  9.20E-13      
\label{tab:obsswift}
\end{longtable}%
\end{center}

\begin{center}
\begin{longtable}{|c|c|c|c|c||c|}
\caption{MeerLicht observations of GX 339--4. The complete table is provided in the supplementary material.} \\

\hline \multicolumn{1}{|c|}{\textbf{UTC}} & \multicolumn{1}{|c|}{\textbf{MJD}} & \multicolumn{1}{c|}{\textbf{Filter}} & \multicolumn{1}{c|}{\textbf{Magnitude}} & \multicolumn{1}{c|}{\textbf{Seeing}} & \multicolumn{1}{c|}{\textbf{Airmass}}   \\  \hline 
\endfirsthead 

\multicolumn{6}{c}%
{{\bfseries \tablename\ \thetable{} -- continued from previous page}} \\
\hline \multicolumn{1}{|c|}{\textbf{UTC}} & \multicolumn{1}{|c|}{\textbf{MJD}} & \multicolumn{1}{c|}{\textbf{Filter}} & \multicolumn{1}{c|}{\textbf{Magnitude}} & \multicolumn{1}{c|}{\textbf{Seeing}} & \multicolumn{1}{c|}{\textbf{Airmass}}   \\ \hline 
\endhead 

\hline \multicolumn{6}{|c|}{{Continued on next page}} \\ \hline
\endfoot

\hline
\endlastfoot

  2019-06-01/20:11:18 & 58635.841& q& 18.02 ~$\pm$~  0.01&	3.001  & 1.22  \\       
  2019-06-01/20:12:51 & 58635.842& u& 19.85 ~$\pm$~  0.31&	3.354  & 1.22  \\       
  ... & ...& ...&...&	...  &...  \\       
  2023-05-28/23:13:56 &  60092.968& u& 20.43 ~$\pm$~  0.54&	3.821  & 1.04    
\label{tab:obsmeerlicht}
\end{longtable}%
\end{center}




\bsp	
\label{lastpage}
\end{document}